%% file: main.tex
\documentclass[final,10pt, 3p, twocolumn, times]{elsarticle}

\usepackage{amssymb}

\biboptions{sort&compress}

\usepackage[dvipsnames]{xcolor}
\usepackage{fancyhdr}
\usepackage[utf8]{inputenc}
\usepackage[us,12hr]{datetime}
\usepackage[hyphens]{url}
\usepackage{microtype}
\usepackage[keeplastbox, ancient]{flushend}
\usepackage[binary-units]{siunitx}
\usepackage{setspace}
\usepackage{multirow}
\usepackage{booktabs}
\usepackage{subfig}
\usepackage{tikz}
\newcommand*\mycirc[1]
{\tikz[baseline=(char.base)]{
  \node[shape=circle,draw,inner sep=0.5pt] (char) {#1};}}
\usepackage{xcolor}

\newcommand{\breakTable}[2][c]{%
  \begin{tabular}[#1]{@{}c@{}}#2\end{tabular}}

\usepackage{amsmath}
\usepackage{hyperref} 
\newif\ifcameraready
\camerareadytrue

\newcommand{\versionnum}[0]{8}

\ifcameraready
\else
  \fancypagestyle{firststyle}
  {
    \fancyhead[C]{\textcolor{BrickRed}{\emph{Version \versionnum~---~\today, \ampmtime}} \\
    \normalsize{CONFIDENTIAL DRAFT VERSION. Please do NOT distribute.}}
  }
  \fancypagestyle{allpages}{
    \fancyhead[C]{\normalsize{CONFIDENTIAL DRAFT VERSION. Please do NOT distribute.}}
  }
  \fi

\ifcameraready
  \newcommand{\ch}[0]{}
  \newcommand{\todo}[1][]{}

  \newcommand{\chx}[0]{}
  \newcommand{\chxi}[0]{}
  \newcommand{\chxii}[0]{}
  
  \newcommand{\sg}[0]{}

\else

  \newcommand{\todo}[1][]{\textbf{\scriptsize \fcolorbox{black}{red}{\color{white}{TODO}}} \underline{$\overline{\hbox{\emph{#1}}}$}}

  \newcommand{\chx}[0]{}
  \newcommand{\chxi}[0]{}
  \newcommand{\chxii}[1]{\textcolor{Blue}{#1}}
  \newcommand{\ch}[0]{}
  \newcommand{\sg}[0]{}

\fi

\begin{document}

\begin{frontmatter}



\title{\huge \textbf{Energy-Efficient Deflection-based On-chip Networks: \\
Topology, Routing, Flow Control}}


\author[1]{Rachata Ausavarungnirun}
\author[2]{Onur Mutlu}

\address{SAFARI Research Group}
\address[2]{ETH Z\"urich}
\address[1]{King Mongkut's University of Technology North Bangkok\vspace{-10pt}}

\begin{abstract}

As the number of cores scales to tens and \chx{hundreds,} the energy consumption of routers across various types of on-chip networks in chip muiltiprocessors (CMPs) increases significantly. A major source of this energy consumption comes from the input buffers inside Network-on-Chip (NoC) routers, which are traditionally designed to \chx{maximize} performance. To mitigate this high energy cost, many works propose bufferless router designs that utilize deflection routing to resolve port contention. While this approach is able to maintain high performance relative to its buffered counterparts at low network traffic, the bufferless router design suffers performance \chx{degradation} under high network load. 

In order to maintain high performance and energy efficiency under \emph{both} low and high network loads, this chapter discusses critical drawbacks of traditional bufferless designs and describes recent research works focusing on two major modifications to improve the overall performance of the traditional bufferless network-on-chip design. \ch{The first modification is} a \ch{minimally-buffered} design \chx{that introduces limited buffering} inside critical parts of the on-chip network in order to reduce the number of deflections. \ch{The} second modification \ch{is} a hierarchical bufferless interconnect design that aims to further improve performance by limiting the number of hops each packet needs to travel \ch{while in the network}. In both approaches, we discuss design tradeoffs and provide evaluation results based on common CMP configurations with various network topologies to show the effectiveness of each proposal.


\end{abstract}

\end{frontmatter}

\ifcameraready
\else
  \thispagestyle{firststyle}
  \pagestyle{allpages}
\fi

\clearpage

\renewcommand{\baselinestretch}{1.0}\normalsize



\section{Introduction}
\label{sec:introduction}

As commercial processors \ch{incorporate more} cores, scalability and energy efficiency demand better interconnection substrates. Different interconnect designs such as a \chx{2D mesh~\cite{principles,casebufferless,minbd,chipper,hat-sbac-pad,carpool,xiyue-iccd16,hotnets2010,sigcomm12,grot09,pvc,grot2010,kilonocs,Nicopoulos06,stc,aergia,a2c}} or a flattened \ch{butterfly~\cite{flattened_bfly} have} become increasingly popular as scalable, high-performance on-chip networks for chip multiprocessors (CMPs) as the number of core grows to hundreds. Unfortunately, these high-performance Network-on-Chip (NoC) designs are projected to consume significant amount of power. For example, 28\% of the chip power is consumed by the NoC in the Intel Terascale 80-core chip~\cite{tera}, 36\% in MIT RAW~\cite{taylor02}, and 36\% in Intel 48-core SCC~\cite{intel-scc}. To reduce the overall power consumption of modern processors, NoC energy efficiency is \ch{a critically} important design goal~\cite{casebufferless,chipper,slimnoc,hird,minbd,hird-parco,fattah-maze}.

Previous on-chip interconnection network designs commonly assume that each router in the network needs to contain buffers to buffer the packets (or flits) transmitted within the network. Indeed, buffering within each router improves the bandwidth efficiency in the network because buffering reduces the number of dropped or ``misrouted'' packets~\cite{principles}, i.e. packets that are sent to a less desirable destination port. On the other hand, buffering has several disadvantages. First, buffers consume significant energy/power: dynamic energy when read/written and static energy even when they are not occupied. Second, having buffers increases the complexity of the network design because logic needs to be implemented to place packets into and out of buffers. Third, buffers can consume significant chip area: even with a small number (16) of total buffer entries per node where each entry can store 64 bytes of data, a network with 64 nodes requires 64KB of buffer storage. In fact, in the TRIPS prototype chip, input buffers of the routers were shown to occupy 75\% of the total on-chip network area~\cite{gratz-iccd06}. Energy consumption and hardware storage cost of buffers increases as future many-core chips incorporate more network nodes.

Mechanisms have been proposed to make conventional input-buffered NoC routers more energy-efficient (i.e., use less energy per unit of performance). For example, bypassing empty input buffers~\cite{michelog10,wang03} reduces some dynamic buffer power, but static power remains.
Such bypassing is also less effective when buffers are not frequently empty. Bufferless deflection routers~\cite{casebufferless} remove router input buffers completely (\ch{thereby} eliminating their static and dynamic power) to reduce router power. \ch{In a conventional bufferless deflection network, flits (several of which make up one packet) are independently routed, unlike most buffered networks, where a packet is the smallest independently-routed unit of traffic.} When two flits contend for a single router output, one must be deflected~\cite{hotpotato} to another output. Thus, a flit never requires a buffer in a router. By controlling which flits are deflected, a bufferless deflection router can ensure that all traffic is eventually delivered. Removing buffers yields simpler and more energy-efficient NoC designs: \chx{for example}, CHIPPER~\cite{chipper} reduces average network power by 54.9\% in a 64-node system compared to a conventional buffered router.

Unfortunately, at high network load, deflection routing reduces performance and efficiency. This is because deflections occur more frequently when many flits contend in the network. Each deflection sends a flit further from its destination, causing unnecessary link and router traversals. Relative to a buffered network, a bufferless network with a high deflection rate wastes energy, and suffers \ch{from} worse congestion, because of these unproductive network hops. In contrast, a buffered router is able to hold flits (or packets) in its input buffers until the required output port is available, incurring no unnecessary hops. Thus, a buffered network can sustain higher performance at peak load~\cite{casebufferless}, but at the cost of large buffers, which can consume significant power and die area.

The best interconnect design would obtain the energy efficiency of the bufferless approach with the high performance of the buffered approach. To \chx{obtain} the best of both worlds, a router would contain only a small amount of buffering for flits that actually require it, and the \ch{network topology would allow} flits to reach their destination \ch{using} the fewest number of hops.

On top of the bufferless approach, to achieve scalability to many (tens, hundreds, or thousands) of cores, \emph{mesh} or other higher-radix topologies are used. Both the Intel SCC~\cite{borkar12} and Terascale~\cite{mesh5ghz} CMPs, and several other many-core CMPs (e.g., the MIT RAW~\cite{taylor02} prototype, the UT-Austin TRIPS chip~\cite{gratz06}, several Tilera products~\cite{tile64,tile100}), and more recently, the Intel Skylake~\cite{intel-skylake}, Intel Cascade Lake~\cite{intel-cascade-lake}, and Intel Ice Lake~\cite{intel-ice-lake} server processors exchange packets on a mesh. A mesh topology has good scalability because there is no central structure which needs to scale (unlike a central bus or crossbar-based design), and bisection bandwidth increases as the network grows. However, routers in a 2D mesh can consume significant energy and die area \chx{due to} overheads in buffering, in routing and flow control, and in the switching elements (crossbars) \ch{that} connect multiple inputs with multiple outputs.

Mainstream commercial CMPs today most commonly use \emph{ring}-based interconnects.  Rings are a well-known network topology~\cite{principles}, and the idea behind a ring topology is very simple: all routers (also called ``ring stops'') are connected by a loop that carries network traffic. At each router, new traffic can be injected into the ring, and traffic in the ring can be removed from the ring when it reaches its destination. When traffic is traveling on the ring, it continues uninterrupted until it reaches its destination. A ring router thus \emph{needs no in-ring buffering or flow control} because it prioritizes on-ring traffic. In addition, the router's datapath is very simple compared to a mesh router, because the router has fewer inputs and requires no large, power-inefficient crossbars; typically it consists only of several MUXes to allow traffic to enter and leave, and one pipeline register. Its latency is typically only one \chx{clock} cycle, because no routing decisions or output port allocations are necessary (other than removing traffic from the ring when it arrives). Because of these advantages, several prototype and commercial multicore processors have utilized ring interconnects. \chxi{Examples} of these processors include the Intel Larrabee~\cite{larrabee}, IBM Cell~\cite{cell}, Intel Sandy~Bridge~\cite{sandybridge}, Intel Skylake~\cite{intel-skylake}, Intel Coffee Lake~\cite{intel-coffee-lake} and more recently, the Intel Ice Lake~\cite{intel-ice-lake} and the three interwoven \chxi{rings} in the NVIDIA DGX-1 NVLink~\cite{dgx1}.

Past work has shown that rings are competitive with meshes up to tens of nodes~\cite{kim09nocarc,farkas-sc92,hector}.  Unfortunately, rings suffer from a fundamental scaling problem because a ring's bisection bandwidth does not scale with the number of nodes in the network. Building more rings, or a wider ring, serves as a stopgap measure but increases the cost of every router on the ring in proportion to the bandwidth increase. As commercial CMPs \ch{incorporate more cores}, a new network design will be needed that balances the simplicity and low overhead of rings with the scalability of more complex topologies.

\begin{figure}[h!]
\centering
\includegraphics[width=2in]{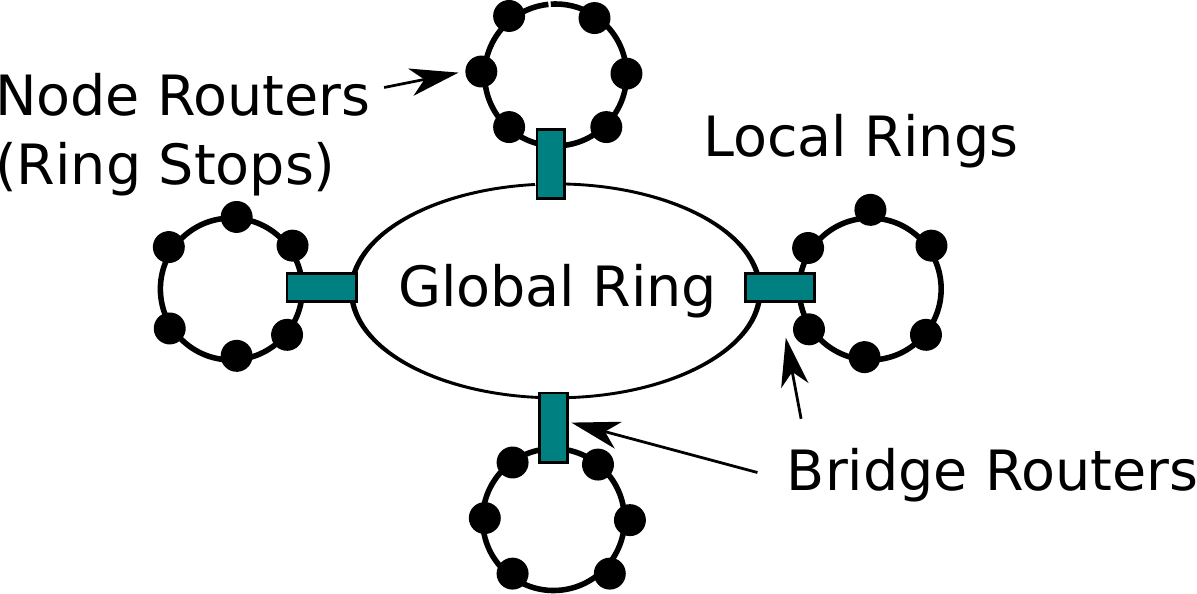}
\caption{A traditional hierarchical ring
  design~\protect\cite{ravindran97,xiangdong95,hr-model,ravindran98,numachine}
  allows ``local rings'' with simple node routers to scale by
  connecting to a ``global ring'' via bridge routers. \ch{Reproduced from~\cite{hird-safari-tr}.}}
  \label{fig:hring}
\end{figure}

A hybrid design is possible. In this chapter, we introduce an approach to construct rings in a \emph{hierarchy} such that groups of nodes share a simple ring interconnect, and these ``local'' rings are joined by one or more ``global'' rings. Figure~\ref{fig:hring} shows an example of such a \emph{hierarchical ring} design~\cite{ravindran97,xiangdong95,hr-model,ravindran98,numachine,hird-parco,hird,hector,ravindran98,farkas-sc92,holliday94}. Past works~\cite{ravindran97,xiangdong95,hr-model,ravindran98,numachine,hector,ravindran98,farkas-sc92,holliday94} \chx{propose} hierarchical rings as a
scalable alternative to single ring and mesh networks. These proposals join rings with \emph{bridge routers}, which reside on multiple rings and transfer traffic between rings. This design was shown to yield good performance and scalability~\cite{ravindran97}. \ch{A} state-of-the-art design~\cite{ravindran97} requires \emph{flow control and buffering} at every node router (ring stop), because a ring transfer can \chx{cause one ring to} back up and stall when another ring is congested. While this previously proposed hierarchical ring is much more scalable than a single ring~\cite{ravindran97}, the reintroduction of in-ring buffering and flow control nullifies one of the primary advantages of using ring networks in the first place (i.e., the lack of buffering and buffered flow control within each ring).

To combine the concept \ch{of a minimally-buffered router with a} hierarchical ring, previous work allows a bridge router with a full buffer to \emph{deflect} packets, called \emph{HiRD} (i.e., Hierarchical Rings with Deflection~\cite{hird-parco,hird-safari-tr,hird}). Rather than requiring buffering and flow control in the ring, packets simply cycle through the network and try again.  While deflection-based, bufferless networks have been proposed and evaluated in the \chx{past~\cite{cm,hep,tera,casebufferless,scarab,gomez08,glsvlsi,deflection_routing,chipper,minbd,hird,hird-parco, cai-isqed,chich01,greenberg92,barnoy93,busch07,michelog10,cfc-cal,carpool,bless_switching},} this minimally-buffered hierarchical ring approach is effectively an elegant hybridization of bufferless (rings) and buffered (bridge routers) styles. To prevent packets from potentially deflecting around a ring arbitrarily many times (i.e., to prevent livelock), we introduce two new mechanisms, the \emph{injection guarantee} and the \emph{transfer guarantee}, that ensure packet delivery even for adversarial/pathological conditions (as \ch{discussed} in ~\cite{hird,hird-parco,hird-safari-tr} and \ch{evaluated} with worst-case traffic in Section~\ref{sec:eval-guarantees}). This simple hierarchical ring design provides a more scalable network architecture while retaining the key simplicities of ring networks (no buffering or flow control within each ring). Section~\ref{sec:eval-guarantees} show in our evaluations that HiRD provides better performance, lower power, and better energy efficiency with respect to \ch{various} buffered hierarchical ring \ch{designs~\cite{ravindran97} as well as other NoC designs.}

\section{Bufferless Routing}
\label{sec:background}

\subsection{Why Bufferless? (and When?)}
Bufferless\footnote{More precisely, a “bufferless” NoC has no in-router (e.g., virtual channel) buffers, only pipeline latches. Baseline bufferless designs, such as BLESS~\cite{casebufferless}, still require reassembly buffers and injection queues. Another subsequent work CHIPPER~\cite{chipper} eliminates these buffers as well.} NoC design has recently been evaluated as an alternative to traditional virtual-channel \chx{buffered routers~\cite{cm,hep,tera,casebufferless,scarab,gomez08,glsvlsi,deflection_routing,chipper,minbd,hird,hird-parco, cai-isqed,chich01,greenberg92,barnoy93,busch07,michelog10,cfc-cal,carpool,bless_switching}.} It is appealing mainly for two reasons: reduced \chx{power consumption}, and simplicity in design. As core count in modern CMPs continues to increase, the interconnect becomes a more significant component of system \chx{power consumption}. Several prototype manycore systems point toward this trend: in MIT RAW, interconnect consumes ~40\% of system power; in the Intel Terascale chip, 30\%. Buffers consume a significant portion of this power. A recent work~\cite{casebufferless}, described in Section~\ref{ref:bless} \chx{reduces} network energy by 40\% by eliminating buffers. Furthermore, the complexity reduction of the design at the high level could be substantial: a bufferless router requires only pipeline registers, a crossbar, and arbitration logic. This can translate into reduced system design and verification cost. 

Bufferless NoCs present a tradeoff: by eliminating buffers, the peak network throughput is reduced, potentially degrading performance. However, network power is often significantly reduced. For this tradeoff to be effective, the power reduction must outweigh the slowdown’s effect on total energy. Moscibroda and Mutlu~\cite{casebufferless} reported minimal performance reduction with bufferless when NoC is lightly loaded, which constitutes many of the applications they evaluated. Bufferless NoC design thus represents a compelling design point for many systems with low-to-medium network load, eliminating unnecessary capacity for significant savings. 

\subsection{BLESS: Baseline Bufferless Deflection Routing}
\label{ref:bless}
Here we briefly introduce bufferless deflection routing in the context of BLESS~\cite{casebufferless}. BLESS routes flits, the minimal routable units of packets, between nodes in a mesh interconnect. Each flit in a packet contains header bits and can travel independently, although in the best case, all of a packet’s flits remain contiguous in the network. Each node contains an injection buffer and a reassembly buffer; there are no buffers within the network, aside from the router pipeline itself. Every cycle, flits that arrive at the input ports contend for the output ports. When two flits contend for one output port, BLESS avoids the need to buffer by misrouting one flit to another port. The flits continue through the network until ejected at their destinations, possibly out of order, where they are reassembled into packets and delivered. Deflection routing is not new: it was first proposed in~\cite{hotpotato}, and is used in optical networks because of the cost of optical buffering~\cite{chich01}. It works because a router has as many output links as input links (in a 2D mesh, 4 for neighbors and 1 for local access). Thus, the flits that arrive in a given cycle can always leave exactly $N$ cycles later, for an $N$-stage router pipeline. If all flits request unique output links, then a deflection router can grant every flit’s requested output. However, if more than one flit contends for the same output, all but one must be deflected to another output that is free.

\subsubsection{Livelock Freedom}
Whenever a flit is deflected, it moves further from its destination. If a flit is deflected continually, it may never reach its destination. Thus, a routing algorithm must explicitly avoid livelock. It is possible to probabilistically bound network latency in a deflection network~\cite{busch00,chaosrouter}. However, a deterministic bound is more desirable. BLESS~\cite{casebufferless} uses an Oldest-First prioritization rule to give a deterministic bound on network latency. Flits arbitrate based on packet timestamps. Prioritizing the oldest traffic creates a consistent total order and allows this traffic to make forward progress. Once the oldest packet arrives, another packet becomes oldest. Thus livelock freedom is guaranteed inductively. However, this age-based priority mechanism is expensive~\cite{scarab,michelog10} both in header information and in arbitration critical path. Alternatively, some bufferless routing proposals do not provide or explicitly show deterministic livelock-freedom guarantees~\cite{scarab,gomez08,glsvlsi}. This can lead to faster arbitration if it allows for simpler priority schemes. However, a provable guarantee of livelock freedom is necessary to show system correctness in all cases. \chx{CHIPPER~\cite{chipper}, described in Section~\ref{sec:chipper}, provides strong livelock freedom guarantees in a bufferless design.}

\begin{figure*}[h!]
\centering
\includegraphics[width=6.5in]{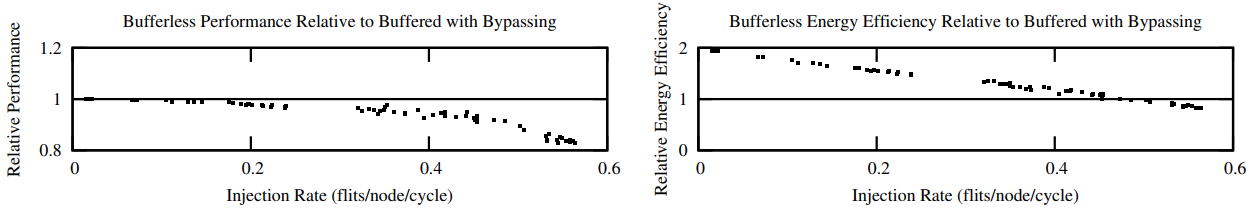}
\caption{System performance and energy efficiency (performance per watt) of bufferless deflection routing, relative to conventional input-buffered routing (4 VCs, 4 flits/VC) that employs buffer bypassing, in a 4x4 2D mesh. Injection rate (X axis) for each workload is measured in the baseline buffered network. \ch{Reproduced from~\cite{minbd}.}}
  \label{fig:chipper-perf}
\end{figure*}

\subsubsection{Injection} 
BLESS guarantees that all flits entering a router can leave it, because there are at least as many output links as input links. However, this does not guarantee that new traffic from the local node (e.g., core or shared cache) can always enter the network. A BLESS router can inject a flit whenever an input link is empty in a given cycle. In other words, BLESS requires a “free slot” in the network in order to insert new traffic. When a flit is injected, it contends for output ports with the other flits in that cycle. Note that the injection decision is purely local: that is, a router can decide whether to inject without coordinating with other routers.

\subsubsection{Ejection and Packet Reassembly} 
A BLESS router can eject one flit per cycle when that flit reaches its destination. In any bufferless deflection network, flits can arrive in random order; therefore, a packet reassembly buffer is necessary. Once all flits in a packet arrive, the packet can be delivered to the local node. Importantly, this buffer must be managed so that it does not overflow, and in such a way that maintains correctness. BLESS~\cite{casebufferless} does not consider this problem in detail. Instead, it assumes an infinite reassembly buffer, and reports maximum occupancies for the evaluated workloads. 

\subsection{Deflection Routing Complexities}
\label{sec:chipper}

While bufferless deflection routing is conceptually and algorithmically simple, a straightforward hardware implementation leads to numerous complexities. In particular, two types of problem plague baseline bufferless deflection routing: high hardware cost, and unaddressed correctness issues. The hardware cost of a direct implementation of bufferless deflection routing is nontrivial, due to expensive control logic. Just as importantly, correctness issues arise in the reassembly buffers when they have practical (non-worst-case) sizes, and this fundamental problem is unaddressed by current work. Here, we describe the major difficulties: output port allocation, expensive priority arbitration, and reassembly buffer cost and correctness. Prior work cites these weaknesses~\cite{casebufferless}.

To address these drawbacks, CHIPPER~\cite{chipper} proposes a new bufferless router architecture, CHIPPER, that solves these problems through three key insights. First, \chx{CHIPPER eliminates} the expensive port allocator and the crossbar, and replace both with a permutation network; \chx{deflection routing} maps naturally to this arrangement, reducing critical path length and power/area cost. Second, \chx{CHIPPER provides} a strong livelock guarantee through an implicit token passing scheme, eliminating the cost of a traditional priority scheme. Finally, \chx{CHIPPER proposes} a simple flow control mechanism for correctness with reasonable reassembly buffer sizes, and propose using cache miss buffers (MSHRs~\cite{mshr-isca81,mshr-isca94,mshr-micro06}) as reassembly buffers. \chx{CHIPPER~\cite{chipper} shows that at low-to-medium load, the reduced-complexity CHIPPER} design performs competitively to a traditional buffered router (in terms of application performance and operational frequency) with significantly reduced network power, and very close to baseline bufferless (BLESS~\cite{casebufferless}) with a reduced critical path.

For low-to-medium network load, CHIPPER delivers performance close to a conventional buffered network as shown in Figure~\ref{fig:chipper-perf}, because the deflection rate is low: thus, most flits take productive network hops \chx{in} every cycle, just as in the buffered network. In addition, the bufferless router has significantly reduced power (hence improved energy efficiency), because the buffers in a conventional router consume significant power. However, as network load increases, the deflection rate in a bufferless deflection network also rises, because flits contend with each other more frequently. With a higher deflection rate, the dynamic power of a bufferless deflection network rises more quickly with load than dynamic power in an equivalent buffered network, because each deflection incurs some extra work. Hence, bufferless deflection networks lose their energy efficiency advantage at high load. Just as important, the high deflection rate causes each flit to take a longer path to its destination, and this increased latency reduces the network throughput and system performance. 

Overall, neither design obtains both good performance and good energy efficiency at all loads. If the system usually experiences low-to-medium network load, then the bufferless design provides adequate performance with low power (hence high energy efficiency). But, if we use a conventional buffered design to obtain high performance, then energy efficiency is poor in the low-load case, and even buffer bypassing does not remove this overhead because buffers consume static power regardless of use. Finally, simply switching between these two extremes at a per-router granularity, as previously proposed~\cite{afc}, does not address the fundamental inefficiencies in the bufferless routing mode, but rather, uses input buffers for all incoming flits at a router when load is too high for the bufferless mode (hence retains the energy-inefficiency of buffered operation at high load). 

\chxi{This chapter (in Section~\ref{sec:minBD}) describes MinBD~\cite{minbd,minbd-tr,minbd-book}, a minimally-buffered deflection router that} combines bufferless and buffered routing in a new way to reduce this overhead.

\section{Scalability in Mesh-Based Interconnects}

Despite the simplicity advantage of a ring-based network, rings have a fundamental \emph{scalability} limit: as compared to a mesh, a ring stops scaling at fewer nodes because its bisection bandwidth is \emph{constant} (proportional only to link width) and the average hop count (which translates to latency for a packet) increases linearly with the number of nodes. (Intuitively, a packet visits half the network on its way to the destination, in the worst case, and a quarter in the average case, for a bidirectional ring.) In contrast, a mesh has bisection bandwidth proportional to one dimension of the layout (e.g., the square-root of the node count for a 2D mesh) and also has an average hop count that scales only with one dimension. The higher radix, and thus higher connectivity, of the mesh allows for more path diversity and lower hop counts which increases performance.

\begin{figure}[h!]
\centering
\includegraphics[width=2in]{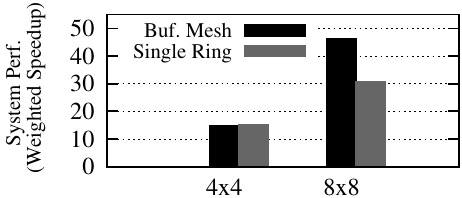}
\caption{Performance as mesh and ring networks to 64 nodes.}
\label{fig:motivation-scale-perf}
\end{figure}

To demonstrate this problem quantitatively, Figure~\ref{fig:motivation-scale-perf} shows application performance averaged over a representative set of network-intensive workloads on (i) a single ring network and (ii) a conventional 2D-mesh buffered interconnect. As the plot shows, although the single ring is able to match the mesh's performance at the 16-node design point, it degrades when node count increases to 64. Note that in this case, the ring's bisection bandwidth is kept equivalent to the mesh, so the performance degradation is solely due to higher hop count; considerations of practical ring width might reduce performance further.

\begin{figure*}[h!]
\centering
\includegraphics[width=3.4in]{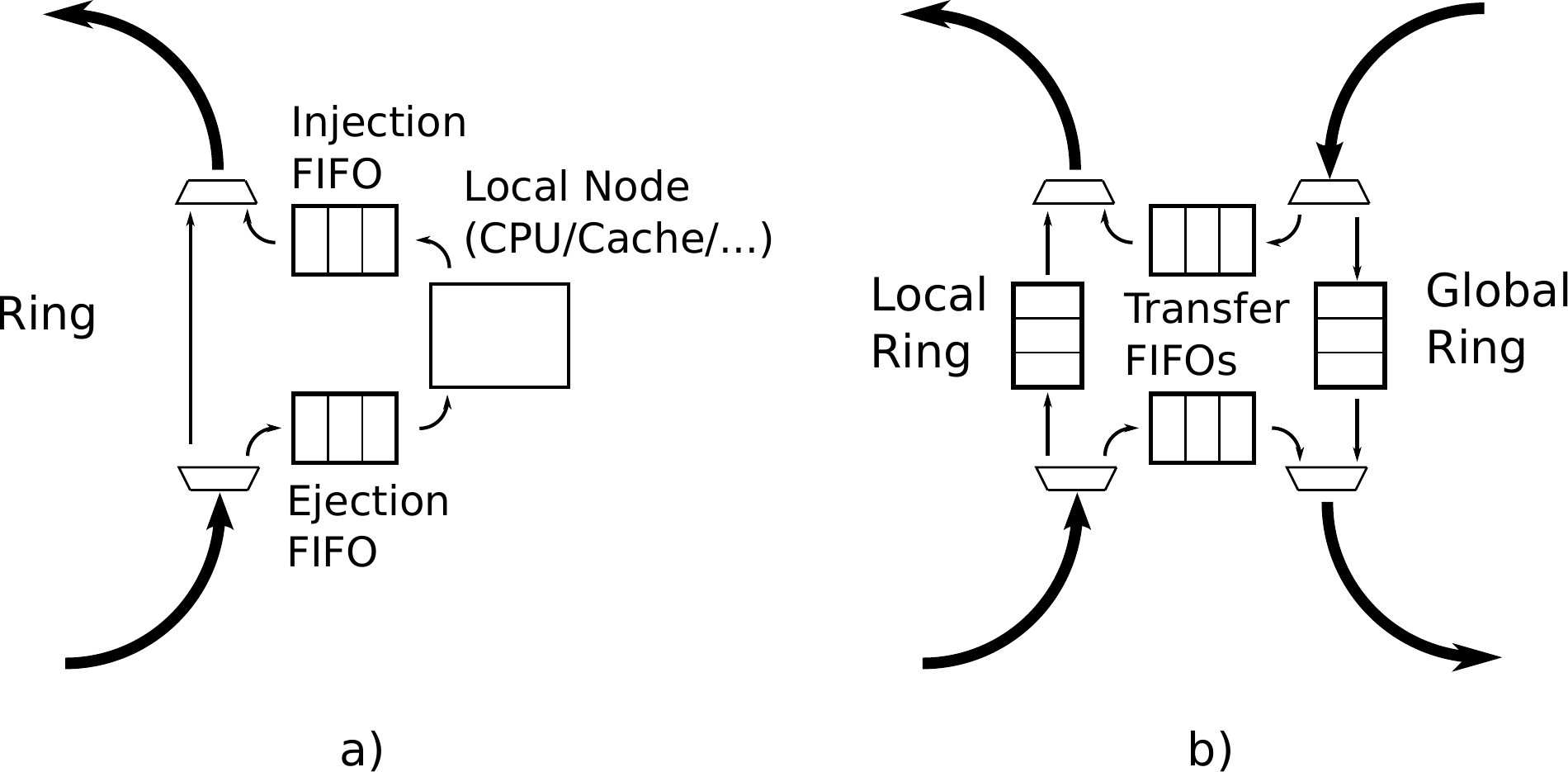}
\caption{ a) A ring router. b) Buffered hierarchical ring detail, as proposed in prior
  work~\protect\cite{ravindran97}: one \emph{bridge router} which connects two
  rings, as proposed in HiRD~\cite{hird}. \ch{Reproduced from~\cite{hird-safari-tr}.}}
\vspace{-2em}
\label{fig:ring-router}
\end{figure*}

\subsection{Simplicity in Ring-Based Interconnects}

Ring interconnects are attractive in current small-to-medium-scale commercial CMPs because ring routers (ring stops) are simple, which leads to smaller die area and energy overhead. In its simplest form, a ring router needs to perform only two functions: injecting new traffic into the ring, and removing traffic from the ring when it has arrived at its destination. Figure~\ref{fig:ring-router} a) depicts the router datapath and control logic (at a high level) necessary to implement this functionality. For every incoming \emph{flit} (unit of data transfer as wide as one link), the router only needs to determine whether the flit stays on the ring or exits at this node. Then, if a flit is waiting to inject, the router checks whether a flit is already present on the ring in this timeslot, and if not, injects the new flit using the in-ring MUX.

Rings achieve very good energy efficiency at low-to-medium core counts~\cite{kim09nocarc}; However, the simplicity advantage of a ring-based network, rings have a fundamental \emph{scalability} limit: a ring stops scaling at fewer nodes because its bisection bandwidth is \emph{constant} (proportional only to link width) and the average hop count (which translates to latency for a packet) increases linearly with the number of nodes. (In the worst case for a bidirectional ring, a packet visits half the network on its way to its destination, and a quarter on average.)

\subsection{Hierarchical Rings for Scalability}


Fortunately, past work has observed that \emph{hierarchy} allows for additional scalability in many interconnects: in ring-based designs~\cite{ravindran97,xiangdong95,hr-model,ravindran98,numachine}, with hierarchical buses~\cite{udipi10}, and with hierarchical meshes~\cite{das09}.  The state-of-the-art hierarchical ring design~\cite{ravindran97} in particular reports promising results by combining local rings with one or more levels of higher-level rings, which we refer to as global rings, that connect lower level rings together.  Rings of different levels are joined by Inter-Ring Interfaces (IRIs), which we call ``bridge routers'' in this work. Figure~\ref{fig:ring-router} b) graphically depicts the details of one bridge router in the previously proposed buffered hierarchical ring network.


Unfortunately, connecting multiple rings via bridge routers introduces a new problem.  Recall that injecting new traffic into a ring requires an open slot in that ring. If the ring is completely full with flits (i.e., a router attempting to inject a new flit must instead pass through a flit already on the ring every cycle), then no new traffic will be injected. But, a bridge router must inject transferring flits into those flits' destination ring in exactly the same manner as if they were newly entering the network. If the ring on one end of the bridge router is completely full (cannot accept any new flits), and the transfer FIFO into that ring is also full, then any other flit requiring a transfer must \emph{block} its current ring. In other words, ring transfers create new dependences between adjacent rings, which creates the need for end-to-end flow control. This flow control forces every node router (ring stop) to have an in-ring FIFO and flow control logic, which increases energy and die area overhead and significantly reduces the appeal of a simple ring-based design. Table~\ref{table:power_area} compares the power consumption for a previously proposed hierarchical ring design (Buffered HRing) and a bufferless hierarchical ring design on a system with 16 nodes using DSENT 0.91~\cite{dsent} and a 45nm technology commercial design library. In the examined idealized bufferless design, each ring has no in-ring buffers, but there are buffers between the hierarchy levels. When a packet needs a buffer that is full, it gets deflected and circles its local ring to try again. Clearly, eliminating in-ring buffers in a hierarchical ring network can reduce power and area significantly.

\begin{table}[h!]
\centering
\footnotesize{
\vspace{-0.1in}
\begin{tabular}{|p{2.2cm}||p{2.2cm}|p{2.2cm}|}
\hline
Metric & Buffered HRing & Bufferless HRing \\
\hline
\hline
Normalized power & 1 & 0.535 \\
\hline
Normalized area & 1 & 0.495 \\
\hline
\end{tabular}
}
\caption{Power and area comparison (16-node network). \ch{Reproduced from~\cite{hird-safari-tr}.}}
\label{table:power_area}
\vspace{-0.15in}
\end{table}

However, simply removing in-ring buffers can introduce livelock, as a deflected packet may circle its local ring indefinitely. Our goal in this work is to introduce a hierarchical ring design that maintains simplicity and low cost, while ensuring livelock and deadlock freedom for packets.

\section{MinBD: Minimally-Buffered Deflection Router Design}
\label{sec:minBD}

\begin{figure*}[h]
\centering
\includegraphics[width=6.3in]{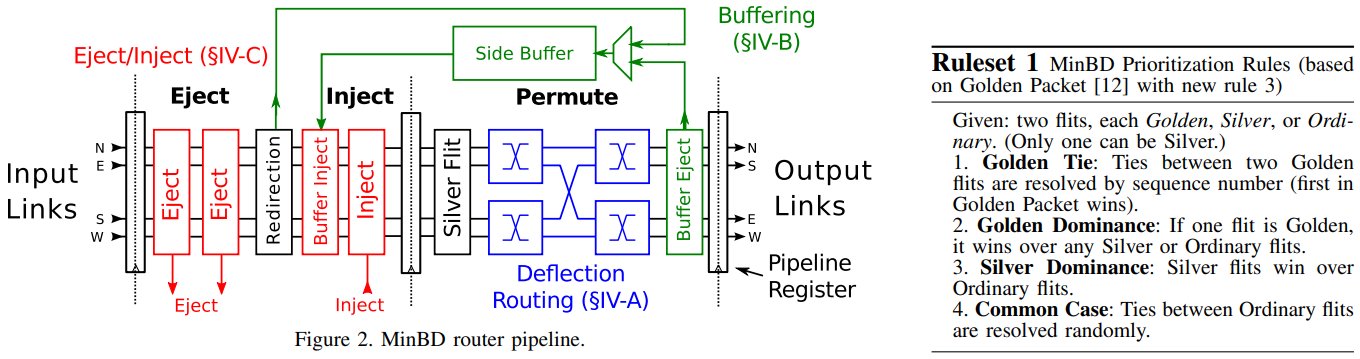}
\caption{MinBD router pipeline. \ch{Reproduced from~\cite{minbd}.}}
\label{fig:minbd}
\end{figure*}

The MinBD (minimally-buffered deflection) router~\cite{minbd,minbd-tr,minbd-book} is a new router design that combines bufferless deflection routing with
a small buffer, which we call the \emph{side buffer}. We start by
outlining the key principles we follow to reduce deflection caused inefficiency by using buffering:
\begin{enumerate}
\item When a flit would be deflected by a router, it is often
better to buffer the flit and arbitrate again in a later cycle.
Some buffering can avoid many deflections.
\item However, buffering every flit leads to unnecessary power
overhead and buffer requirements, because many flits will
be routed productively on the first try. The router should
buffer a flit only if necessary.
\item Finally, when a flit arrives at its destination, it should be
removed from the network (ejected) quickly, so that it
does not continue to contend with other flits.
\end{enumerate} 

\noindent\textbf{Basic High-Level Operation.} The MinBD router does not use
input buffers, unlike conventional buffered routers. Instead, a
flit that arrives at the router proceeds directly to the routing
and arbitration logic. This logic performs deflection routing,
so that when two flits contend for an output port, one of
the flits is sent to another output instead. However, unlike a
bufferless deflection router, the MinBD router can also buffer
up to one flit per cycle in a single FIFO-queue side buffer.
The router examines all flits at the output of the deflection
routing logic, and if any are deflected, one of the deflected
flits is removed from the router pipeline and buffered (as long
as the buffer is not full). From the side buffer, flits are reinjected into the network by the router, in the same way that
new traffic is injected. Thus, some flits that would have been
deflected in a bufferless deflection router are removed from
the network temporarily into this side buffer, and given a
second chance to arbitrate for a productive router output when
re-injected. This reduces the network’s deflection rate (hence
improves performance and energy efficiency) while buffering
only a fraction of traffic.

We will describe the operation of the MinBD router in stages.
First, Section~\ref{sec:minbd-deflection} describes the deflection routing logic that computes
an initial routing decision for the flits that arrive in every
cycle. Then, Section~\ref{sec:minbd-buffer} describes how the router chooses to buffer
some (but not all) flits in the side buffer. Section~\ref{sec:minbd-ejection} describes how
buffered flits and newly-generated flits are injected into the
network, and how a flit that arrives at its destination is ejected.
Finally, Section~\ref{sec:minbd-correctness} discusses correctness issues, and describes how
MinBD ensures that all flits are eventually delivered.

\subsection{Deflection Routing}
\label{sec:minbd-deflection}
The MinBD router pipeline is shown in Figure~\ref{fig:minbd} Flits travel through the pipeline from the inputs (on the left) to outputs (on the right). We first discuss the deflection routing logic, located in the Permute stage on the right. This logic implements deflection routing: it sends each input flit to its preferred output when possible, deflecting to another output otherwise. MinBD uses the deflection logic organization first proposed in CHIPPER~\cite{chipper}. The permutation network in the Permute stage consists of two-input blocks arranged into two stages of two blocks each. This arrangement can send a flit on any input to any output. (Note that it cannot perform all possible permutations of inputs to outputs, but as we will see, it is sufficient for correct operation that at least one flit obtains its preferred output.) In each two-input block, arbitration logic determines which flit has a higher priority, and sends that flit in the direction of its preferred output. The other flit at the two-input block, if any, must take the block’s other output. By combining two stages of this arbitration and routing, deflection arises as a distributed decision: a flit might be deflected in the first stage, or the second stage. Restricting the arbitration and routing to two-flit subproblems reduces complexity and allows for a shorter critical path, as demonstrated in CHIPPER~\cite{chipper}. In order to ensure correct operation, the router must arbitrate between flits so that every flit is eventually delivered, despite deflections. We adapt a modified version of the Golden Packet priority scheme~\cite{chipper}, which solves this livelock-freedom problem. This priority scheme is summarized in Ruleset 1. The basic idea of the Golden Packet priority scheme is that at any given time, at most one packet in the system is golden. The flits of this golden packet, called “golden flits,” are prioritized above all other flits in the system (and contention between golden flits is resolved by the flit sequence number). While prioritized, golden flits are never deflected by non-golden flits. The packet is prioritized for a period long enough to guarantee its delivery. Finally, this “golden” status is assigned to one globally-unique packet ID (e.g., source node address concatenated with a request ID), and this assignment rotates through all possible packet IDs such that any packet that is “stuck” will eventually become golden. In this way, all packets will eventually be delivered, and the network is livelock-free. (See CHIPPER~\cite{chipper} for the precise way in which the Golden Packet is determined; we use the same rotation schedule.) 

However, although Golden Packet arbitration provides correctness, a performance issue occurs with this priority scheme. Consider that most flits are not golden: the elevated priority status provides worst-case correctness, but does not impact common-case performance (prior work reported over 99\% of
flits are delivered without becoming golden~\cite{chipper}). However, when no flits are golden and ties are broken randomly, the arbitration decisions in the two permutation network stages are not coordinated. Hence, a flit might win arbitration in the first stage, and cause another flit to be deflected, but then lose arbitration in the second stage, and also be deflected. Thus, unnecessary deflections occur when the two permutation network stages are uncoordinated.

In order to resolve this performance issue, we observe that it is enough to ensure that in every router, at least one flit is prioritized above the others in every cycle. In this way, at least one flit will certainly not be deflected. To ensure this when no golden flits are present, we add a “silver” priority level, which wins arbitration over common-case flits but loses to the golden flits. One silver flit is designated randomly among the set of flits that enter a router at every cycle (this designation is local to the router, and not propagated to other routers). This modification helps to reduce deflection rate. Prioritizing a silver flit at every router does not impact correctness, because it does not deflect a golden flit if one is present, but it ensures that at least one flit will consistently win arbitration at both stages. Hence, deflection rate is reduced, improving performance.

\subsection{Using a Small Buffer to Reduce Deflections}
\label{sec:minbd-buffer}

The key problem addressed by MinBD is deflection inefficiency at high load. In other words, when the network is highly utilized, contention between flits occurs often. This results in many deflected flits. We observe that adding a small buffer to a deflection router can reduce deflection rate, because the router can choose to buffer rather than deflect a flit when its output port is taken by another flit. Then, at a later time when output ports may be available, the buffered flit can re-try arbitration. 

Thus, to reduce deflection rate, MinBD adds a “side buffer” that buffers only some flits that otherwise would be deflected. This buffer is shown in Figure~\ref{fig:minbd} above the permutation network. In order to make use of this buffer, a “buffer eject” block is placed in the pipeline after the permutation network. At this point, the arbitration and routing logic has determined which flits to deflect. The buffer eject block recognizes flits that have been deflected, and picks up to one such deflected flit per cycle. It removes a deflected flit from the router pipeline, and places this flit in the side buffer, as long as the side buffer is not full. (If the side buffer is full, no flits are removed from the pipeline into the buffer until space is created.) This flit is chosen randomly among deflected flits (except that a golden flit is never chosen: see Section~\ref{sec:minbd-correctness}). In this way, some deflections are avoided. The flits placed in the buffer will later be re-injected into the pipeline, and will re-try arbitration at that time. This re-injection occurs in the same way that new traffic is injected into the network, which we discuss below.

\subsection{Injection and Ejection}
\label{sec:minbd-ejection}

So far, we have considered the flow of flits from router input ports (i.e., arriving from neighbor routers) to router output ports (i.e., to other neighbor routers). A flit must enter and leave the network at some point. To allow traffic to enter (inject) and leave (eject), the MinBD router contains inject and eject blocks in its first pipeline stage (see Figure~\ref{fig:minbd}). When a set of flits arrive on router inputs, these flits first pass through the ejection logic. This logic examines the destination of each flit, and if a flit is addressed to the local router, it is removed from the router pipeline and sent to the local network node.\footnote{Note that flits are reassembled into packets after ejection. To implement this reassembly, we use the Retransmit-Once scheme, as used by CHIPPER~\cite{chipper}, which uses MSHRs (Miss-Status Handling Registers~\cite{kroft81}, or existing buffers
in the cache system) to reassemble packets in place.} If more than one locally-addressed flit is present, the ejector picks one, according to the same priority scheme used by routing arbitration. 

However, ejecting a single flit per cycle can produce a bottleneck and cause unnecessary deflections for flits that could not be ejected. In the workloads we evaluate, at least one flit is eligible to eject 42.8\% of the time. Of those cycles, 20.4\% of the time, at least two flits are eligible to eject. Hence, in ~8.5\% of all cycles, a locally-addressed flit would be deflected rather than ejected if only one flit could be ejected per cycle. To avoid this significant deflection-rate penalty, we double the ejection bandwidth. To implement this, a MinBD~\cite{minbd-book,minbd,minbd-tr} router contains two ejector blocks. Each of these blocks is identical, and can eject up to one flit per cycle. Duplicating the ejection logic allows two flits to leave the network per cycle at every node. After locally-addressed flits are removed from the pipeline, new flits are allowed to enter. There are two injector blocks in the router pipeline shown in Figure~\ref{fig:minbd}: (i) re-injection of flits from the side buffer, and (ii) injection of new flits from the local node. (The “Redirection” block prior to the injector blocks will be discussed in the next section.) Each block operates in the same way. A flit can be injected into the router pipeline whenever one of the four inputs does not have a flit present in a given cycle, i.e., whenever there is an “empty slot” in the network. Each injection block pulls up to one flit per cycle from an injection queue (the side buffer, or the local node’s injection queue), and places a new flit in the pipeline when a slot is available. Flits from the side buffer are re-injected before new traffic is injected into the network. However, note that there is no guarantee that a free slot will be available for an injection in any given cycle. We now address this starvation problem for side buffer re-injection.

\subsection{Ensuring Side Buffered Flits Make Progress}
\label{sec:minbd-correctness}
When a flit enters the side buffer, it leaves the router pipeline, and must later be re-injected. As we described above, flit reinjection must wait for an empty slot on an input link. It is possible that such a slot will not appear for a long time. In this case, the flits in the side buffer are delayed unfairly while other flits make forward progress.

To avoid this situation, we implement buffer redirection. The key idea of buffer redirection is that when this side buffer starvation problem is detected, one flit from a randomly-chosen router input is forced to enter the side buffer. Simultaneously, the flit at the head of the side buffer is allowed to inject into the slot created by the forced flit buffering. In other words, one router input is “redirected” into the FIFO buffer for one cycle, in order to allow the buffer to make forward progress. This redirection is enabled for one cycle whenever the side buffer injection is starved (i.e., has a flit to inject, but no free slot allows the injection) for more than some threshold $C_{threshold}$
cycles (in our evaluations, $C_{threshold} = 2$). Finally, note that if a golden flit is present, it is never redirected to the buffer, because this would break the delivery guarantee.

\subsection{Livelock and Deadlock-free Operation}

MinBD provides livelock-free delivery of flits using Golden Packet and buffer redirection. If no flit is ever buffered, then Golden Packet~\cite{chipper} ensures livelock freedom (the “silver flit” priority never deflects any golden flit, hence does not break the guarantee). Now, we argue that adding side buffers does not cause livelock. First, the buffering logic never places a golden flit in the side buffer. However, a flit could enter a buffer and then become golden while waiting. Redirection ensures correctness in this case: it provides an upper bound on residence time in a buffer (because the flit at the head of the buffer will leave after a certain threshold time in the worst case). If a flit in a buffer becomes golden, it only needs to remain golden long enough to leave the buffer in the worst case, then progress to its destination. We choose the threshold parameter ($C_{threshold}$) and golden epoch length so that this is always possible. More details can be found in our extended technical report~\cite{minbd-tr}.

MinBD achieves deadlock-free operation by using Retransmit-Once~\cite{chipper}, which ensures that every node always consumes flits delivered to it by dropping flits when no reassembly/request buffer is available. This avoids packet reassembly deadlock (as described in~\cite{chipper}), as well as protocol level deadlock, because message-class dependencies~\cite{hansson07} no longer exist.

\section{HiRD: Simple Hierarchical Rings with Deflection}
\label{sec:mech}

With the design of a minimally-buffered deflection router, minBD~\cite{minbd-book,minbd-tr,minbd}, we now describe how a similar concept can be integrated to a deflection-based hirarchical ring interconnect called HiRD~\cite{hird,hird-parco} in order to further improve performance, energy efficiency and scalability of NoCs. HiRD is built on several basic operation principles:

\begin{enumerate}

\item Every node (e.g., CPU, cache slice, or memory controller)
resides on one \emph{local ring}, and connects to one \emph{node
router} on that ring.

\item Node routers operate exactly like routers (ring stops) in a
single-ring interconnect: locally-destined flits are removed from
the ring, other flits are passed through, and new flits can inject
whenever there is a free slot (no flit present in a given
cycle). There is no buffering or flow control within any local ring;
flits are buffered only in ring pipeline registers. Node routers
have a single-cycle latency.

\item Local rings are connected to one or more levels of \emph{global
rings} to form a tree hierarchy.

\item Rings are joined via \emph{bridge routers}. A bridge router has
a node-router-like interface on each of the two rings it
connects, and has a set of transfer FIFOs (one in each direction)
between the rings.

\item Bridge routers consume flits that require a transfer whenever
the respective transfer FIFO has available space. The head flit in a
transfer FIFO can inject into its new ring whenever there is a free
slot (exactly as with new flit injections). When a flit requires a
transfer but the respective transfer FIFO is full, the flit remains
in its current ring. It will circle the ring and try again next time
it encounters the correct bridge router (this is a
\emph{deflection}).

\end{enumerate}

By using \emph{deflections} rather than buffering and blocking flow control to manage ring transfers, HiRD retains node router simplicity, unlike past hierarchical ring network designs. This change comes at the cost of potential livelock (if flits are forced to deflect forever). We introduce two mechanisms to provide a deterministic guarantee of livelock-free operation in~\cite{hird,hird-parco}.

While deflection-based bufferless routing has been previously proposed and evaluated for a variety of off-chip and on-chip interconnection networks (e.g.,~\cite{hotpotato,casebufferless,bless_switching,chipper,minbd,hotnets2010,sigcomm12}), deflections are trivially implementable in a ring: if deflection occurs, the flit\footnote{All operations in the network happen in a flit level similar to previous works~\cite{casebufferless,bless_switching,chipper,minbd,hotnets2010,sigcomm12,principles,hat-sbac-pad,carpool,xiyue-iccd16,grot09,pvc,grot2010,kilonocs,Nicopoulos06,das09,stc,aergia,a2c}.} continues circulating in the ring.  Contrast this to past deflection-based schemes that operated on mesh networks where multiple incoming flits may need to be deflected among a multitude of possible out-bound ports, leading to much more circuit complexity in the router microarchitecture, as shown by~\cite{chipper,scarab,michelog10}.  Our application of deflection to rings leads to a simple and elegant embodiment of bufferless routing.

\subsection{Node Router Operation}
\label{sec:mech-node-router}

\begin{figure}[ht]
\centering
\includegraphics[width=2.5in]{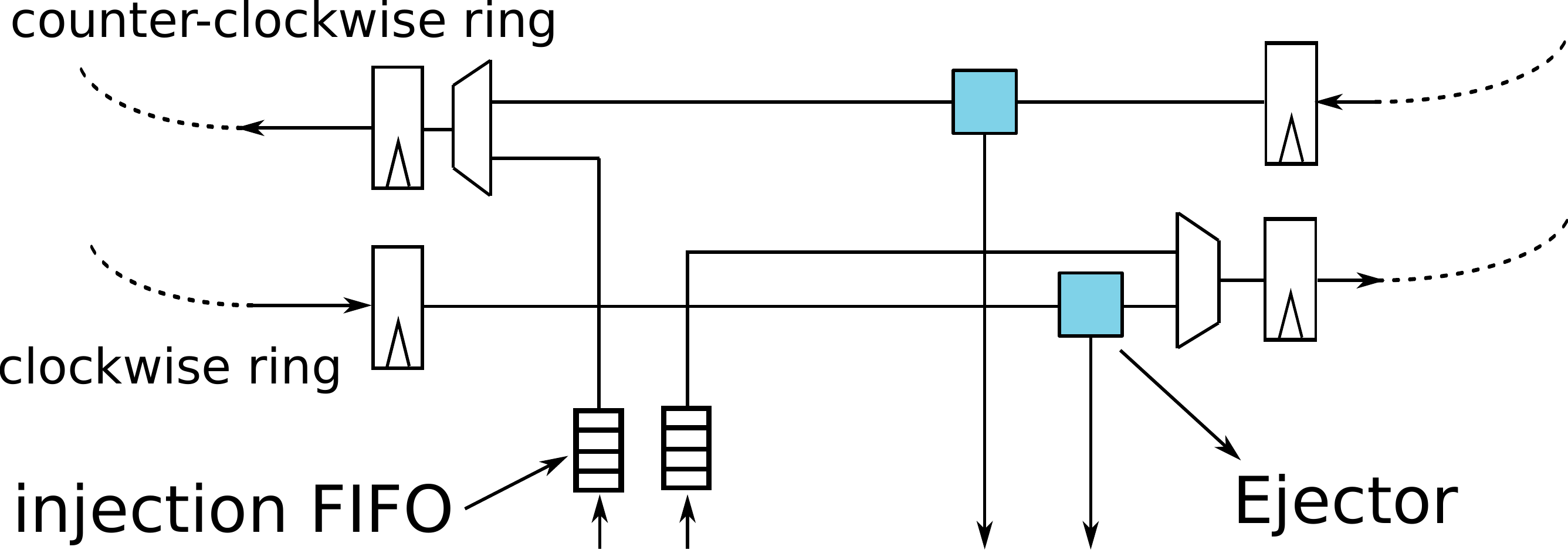}
\caption{Node router. \ch{Reproduced from~\cite{hird-safari-tr}.}}
\label{fig:node-router}
\end{figure}

At each node on a local ring, we place a single node router, shown in Figure~\ref{fig:node-router}. A node router is very simple: it passes through circulating traffic, allows new traffic to enter the ring through a MUX, and allows traffic to leave the ring when it arrives at its destination. Each router contains one pipeline register for the router stage, and one pipeline register for link traversal, so the router latency is exactly one cycle and the per-hop latency is two cycles. Such a design is very common in ring-based and ring-like designs (e.g.,~\cite{kim09}).

As flits enter the router on the ring, they first travel to the ejector. Because we use bidirectional rings, each node router has two ejectors, one per direction.\footnote{For simplicity, we assume that up   to two ejected flits can be accepted by the processor or reassembly buffers in a single cycle. For a fair comparison, we also implement two-flit-per-cycle ejection in our baselines.} Note that the flits constituting a packet may arrive out-of-order and at widely separated times. Re-assembly into packets is thus necessary. Packets are re-assembled and reassembly buffers are managed using the Retransmit-Once scheme, borrowed from the CHIPPER bufferless router design~\cite{chipper}. With this scheme, receivers reassemble packets in-place in MSHRs (Miss-Status Handling Registers~\cite{kroft81}), eliminating the need for separate reassembly buffers. The key idea in Retransmit-Once is to avoid ejection backpressure-induced deadlocks by ensuring that all arriving flits are consumed immediately at their receiver nodes. When a flit from a new packet arrives, it allocates a new reassembly buffer slot if available. If no slot is available, the receiver drops the flit and sets a bit in a retransmit queue which corresponds to the sender and transaction ID of the dropped flit. Eventually, when a buffer slot becomes available at the receiver, the receiver reserves the slot for a sender/transaction ID in its retransmit queue and requests a retransmit from the sender. Thus, all traffic arriving at a node is consumed (or dropped) immediately, so ejection never places backpressure on the ring. Retransmit-Once hence avoids protocol-level deadlock~\cite{chipper}. Furthermore, it ensures that a ring full of flits always drains, thus ensuring forward progress (as we will describe more fully in~\cite{hird,hird-parco}).

After locally-destined traffic is removed from the ring, the remaining traffic travels to the injection stage. At this stage, the router looks for ``empty slots,'' or cycles where no flit is present on the ring, and injects new flits into the ring whenever they are queued for injection. The injector is even simpler than the ejector, because it only needs to find cycles where no flit is present and insert new flits in these slots. Note that we implement two separate injection buffers (FIFOs), one per ring direction; thus, two flits can be injected into the network in a single cycle. A flit enqueues for injection in the direction that yields a shorter traversal toward its destination.

\subsection{Bridge Routers}
\label{sec:bridge_router}

\begin{figure}[h]
\centering
\includegraphics[width=2.8in]{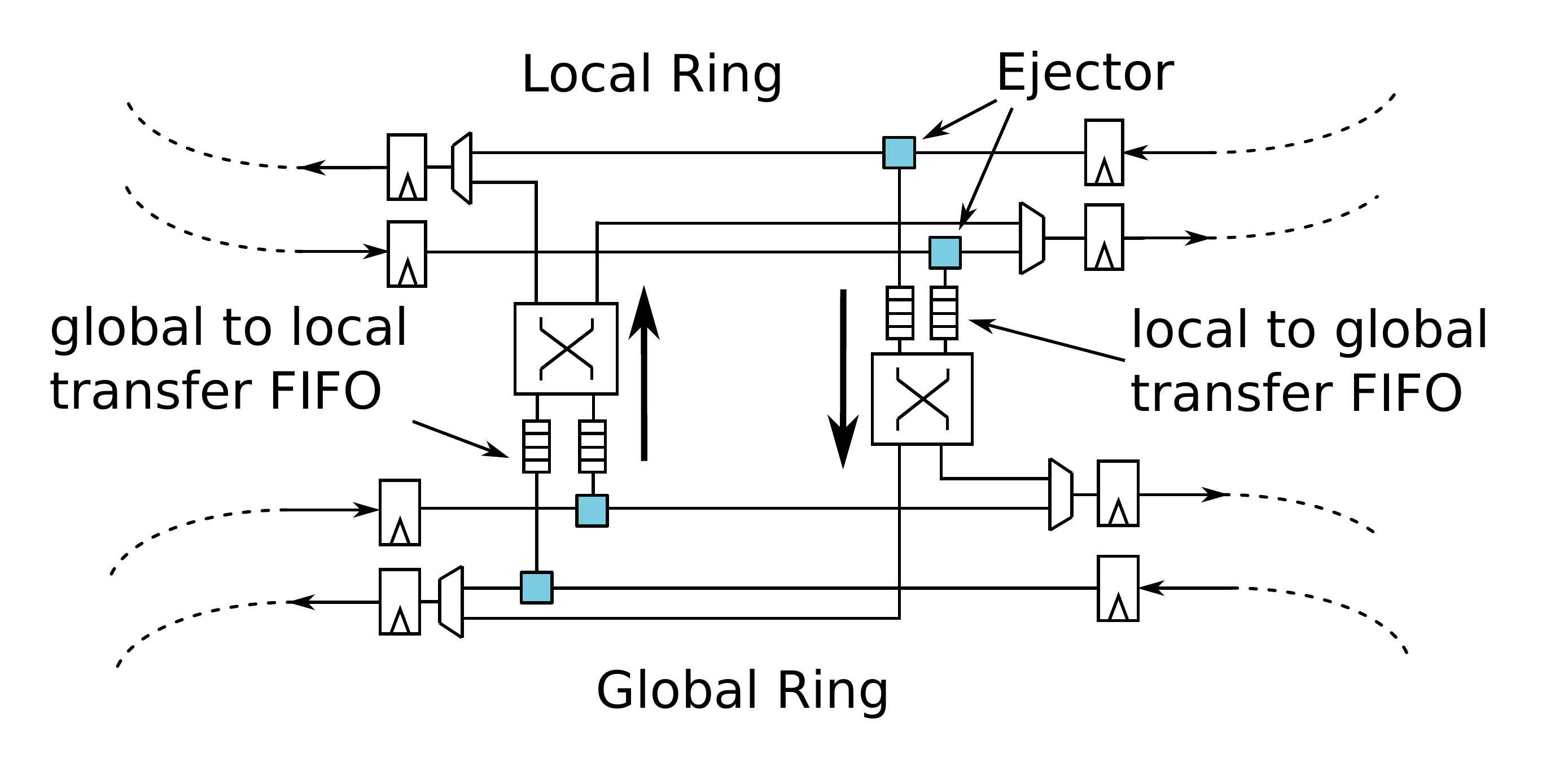}
\caption{Bridge router. \ch{Reproduced from~\cite{hird-safari-tr}.}}
\label{fig:bridge-router}
\end{figure}

The \emph{bridge routers} connect a local ring and a global ring, or a global ring with a higher-level global ring (if there are more than two levels of hierarchy).  A high-level block diagram of a bridge router is shown in Figure~\ref{fig:bridge-router}.  A bridge router resembles two node routers, one on each of two rings, connected by FIFO buffers in both directions. When a flit arrives on one ring that requires a transfer to the other ring (according to the routing function described below in \S\ref{sec:routing}), it can leave its current ring and wait in a FIFO as long as there is space available. These \emph{transfer FIFOs} exist so that a transferring flit's arrival need not be perfectly aligned with a free slot on the destination ring.  However, this transfer FIFO will sometimes fill. In that case, if any flit arrives that requires a transfer, the bridge router simply does not remove the flit from its current ring; the flit will continue to travel around the ring, and will eventually come back to the bridge router, at which point there may be an open slot available in the transfer FIFO. This is analogous to a \emph{deflection} in hot-potato routing~\cite{hotpotato}, also known as deflection routing, and has been used in recent on-chip mesh interconnect designs to resolve contention~\cite{casebufferless,chipper,minbd,glsvlsi,hotnets2010,sigcomm12}.  Note that to ensure that flits are \emph{eventually} delivered, despite any deflections that may occur, we introduce two \emph{guarantee mechanisms} in~\cite{hird,hird-parco}.  Finally, note that deflections may cause flits to arrive out-of-order (this is fundamental to any non-minimal adaptively-routed network). Because we use Retransmit-Once~\cite{chipper}, packet reassembly works despite out-of-order arrival.

\begin{figure*}[h!]
\centering
\vspace{-0.4in}
\subfloat[4-, 8-, and 16-bridge hierarchical ring designs.]{
\includegraphics[width=4in]{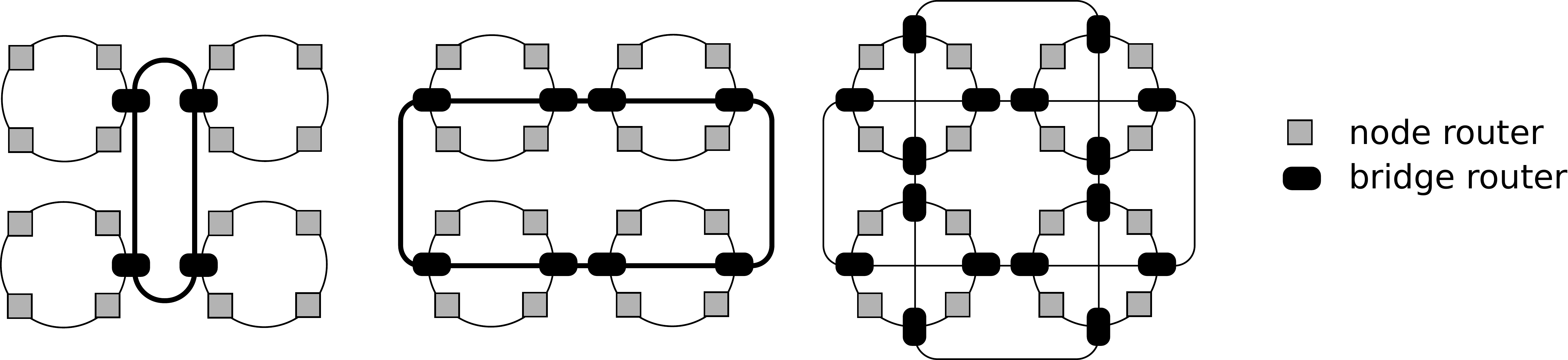}
\label{fig:topology}
}
\subfloat[Three-level hierarchy (8x8).]{
\includegraphics[width=2.3in]{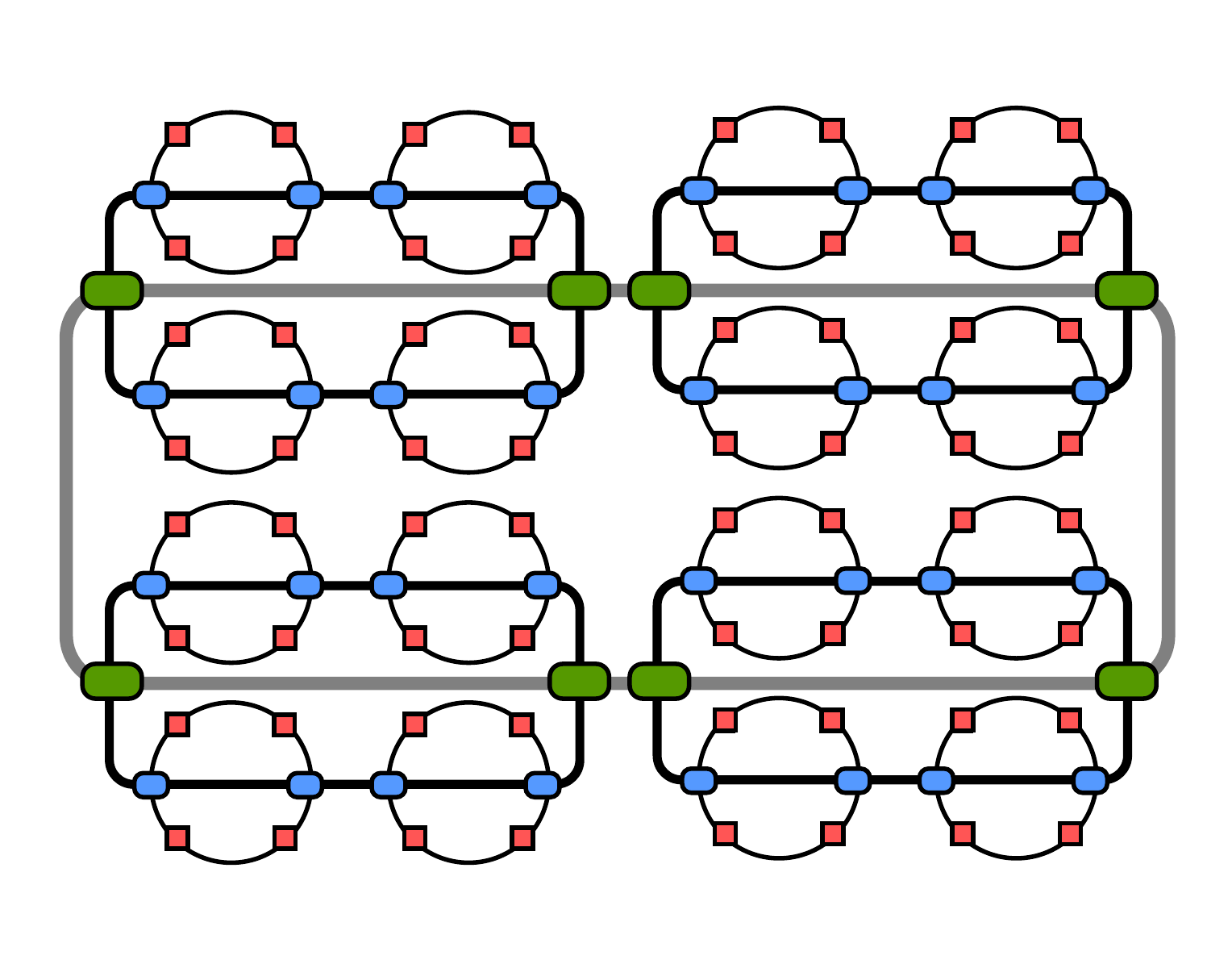}
\label{fig:scale}
}
\caption{Hierarchical ring design of HiRD. \ch{Reproduced from~\cite{hird-safari-tr}.}}
\end{figure*}

The bridge router uses \emph{crossbars} to allow a flit ejecting from either ring direction in a bidirectional ring to enqueue for injection in either direction in the adjoining ring. When a flit transfers, it picks the ring direction that gives a shorter distance, as in a node router. However, these crossbars actually allow for a more general case: the bridge router can actually join several rings together by using larger crossbars. For our network topology, we use hierarchical rings. We use wider global rings than local rings (analogous to a \emph{fat tree}~\cite{cm}) for performance reasons. These wider rings perform logically as separate rings as wide as one flit. Although not shown in the figure for simplicity, the bridge router in such a case uses a larger crossbar and has one ring interface (including transfer FIFO) per ring-lane in the wide global ring. The bridge router then load-balances flits between rings when multiple lanes are available. (The crossbar and transfer FIFOs are fully modeled in our evaluations.)

When building a two-level design, there are many different arrangements of global rings and bridge routers that can efficiently link the local rings together. Figure~\ref{fig:topology} shows three designs denoted by the number of bridge routers in total: 4-bridge, 8-bridge, and 16-bridge. We assume an 8-bridge design for the remainder of this paper. Also, note that the hierarchical structure that we propose can be extended to more than two levels. We use a 3-level hierarchy, illustrated in Figure~\ref{fig:scale}, to build a 64-node network.

Finally, in order to address a potential deadlock case (which will be explained more in~\cite{hird,hird-parco,hird-safari-tr}), bridge routers implement a special \emph{Swap Rule}. The Swap Rule states that when the flit that just arrived on each ring requires a transfer to the other ring, the flits can be \emph{swapped}, bypassing the transfer FIFOs altogether. This requires a bypass datapath (which is fully modeled in our hardware evaluations). It ensures correct operation in the case when transfer FIFOs in both directions are full. Only one swap needs to occur in any given cycle, even when the bridge router connects to a wide global ring. Note that because the swap rule requires this bypass path, the behavior is always active (it would be more difficult to definitively identify a deadlock and enable the behavior only in that special case). The Swap Rule may cause flits to arrive out-of-order when some are bypassed in this way, but the network already delivers flits out-of-order, so correctness is not compromised.

\subsection{Routing}
\label{sec:routing}

Finally, we briefly address routing. Because a hierarchical ring design is
fundamentally a \emph{tree}, routing is very simple: when a flit is destined
for a node in another part of the hierarchy, it first travels \emph{up} the
tree (to more global levels) until it reaches a common ancestor of its source
and its destination, and then it travels \emph{down} the tree to its
destination. Concretely, each node's address can be written as a series of
parts, or digits, corresponding to each level of the hierarchy (these trivially
could be bitfields in a node ID). A ring can be identified by the common prefix
of all routers on that ring; the root global ring has a null (empty) prefix,
and local rings have prefixes consisting of all digits but the last one. If a
flit's destination does not match the prefix of the ring it is on, it takes any
bridge router to a more global ring. If a flit's destination does match the
prefix of the ring it is on (meaning that it is traveling down to more
local levels), it takes any bridge router which connects to the next level, until
it finally reaches the local ring of its destination and ejects at the node
with a full address match.


\subsection{Guaranteed Delivery: Correctness in Hierarchical Ring Interconnects}
\label{sec:livelock}
\label{sec:guarantees}

In order for the system to operate correctly, the interconnect must guarantee
that every flit is eventually delivered to its destination. HiRD ensures
correct operation through two mechanisms that provide two guarantees: the
\emph{injection guarantee} and the \emph{transfer guarantee}. The injection
guarantee ensures that any flit waiting to inject into a ring will eventually
be able to enter that ring. The transfer guarantee ensures that any flit
waiting to enter a bridge router's transfer queue will eventually be granted a
slot in that queue.

\begin{figure}[h!]
\centering
\vspace{-0.2in}
\includegraphics[width=2.8in]{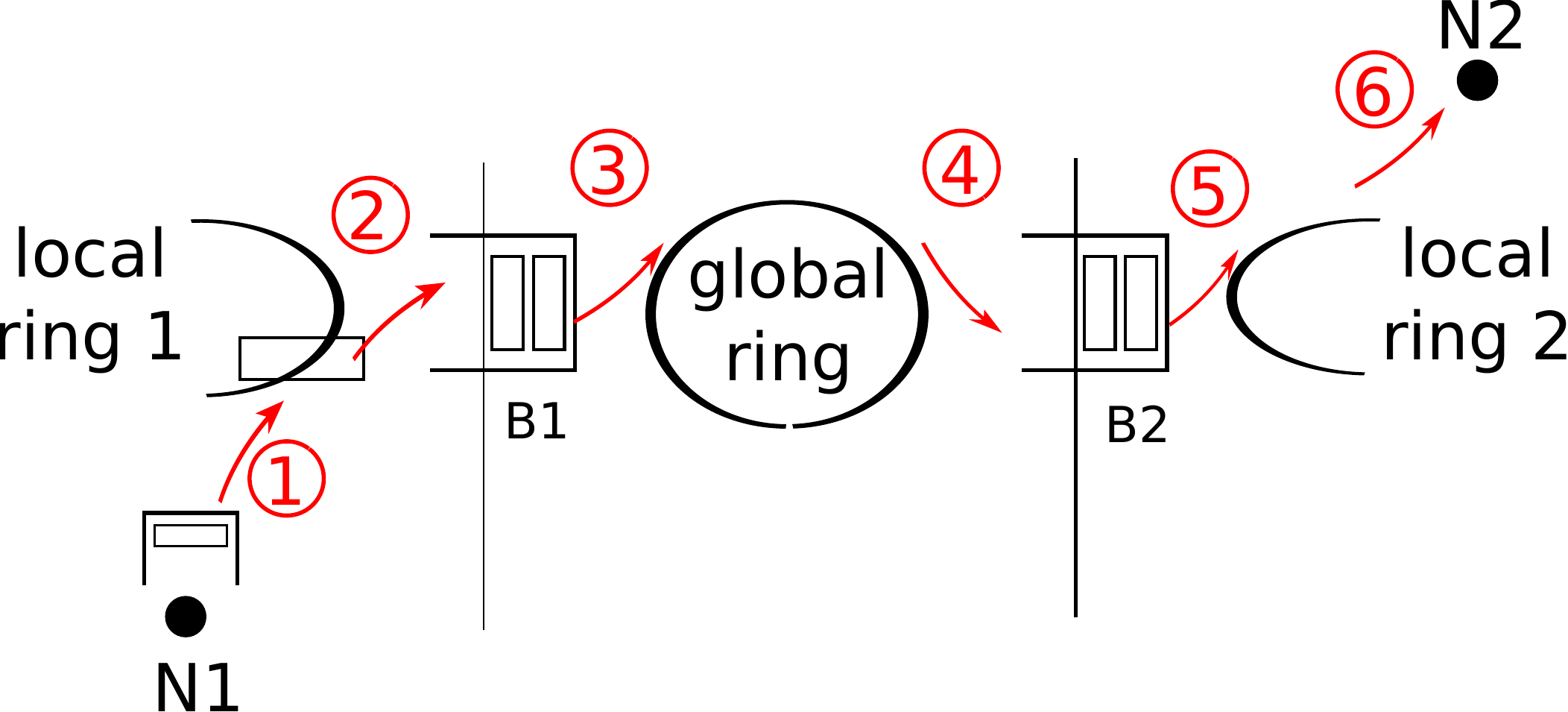}
\caption{The need for the injection and transfer guarantees:
  contention experienced by a flit during its journey. \ch{Reproduced from~\cite{hird-safari-tr}.}}
\label{fig:guarantee-motivation}
\vspace{-0.15in}
\end{figure}

To understand the need for each guarantee, let us consider an example, shown in
Figure~\ref{fig:guarantee-motivation}. A flit is enqueued for network injection
at node N1 on the leftmost local ring. This flit is destined for node N2 on the
rightmost local ring; hence, it must traverse the leftmost local ring, then the
global ring in the center of the figure, followed by the rightmost local ring.
The flit transfers rings twice, at the two bridge routers B1 and B2 shown in
the figure. The figure also indicates the six points (labeled as \mycirc{1} to
\mycirc{6}) at which the flit moves from a queue to a ring or vice-versa: the
flit first enters N1's injection queue, transfers to the leftmost local ring
\mycirc{1}, the bridge router B1 \mycirc{2}, the global ring \mycirc{3}, the
bridge router B2 \mycirc{4}, the rightmost local ring \mycirc{5}, and finally
the destination node N2 \mycirc{6}.

In the worst case, when the network is heavily contended, the flit
could wait for an unbounded amount of time at \mycirc{1} to
\mycirc{5}. First, recall that to enter any ring, a flit must wait for an empty
slot on that ring (because the traffic on the ring continues along the
ring once it has entered, and thus has higher priority than any new
traffic). Because of this, the flit traveling from node N1 to N2 could
wait for an arbitrarily long time at \mycirc{1}, \mycirc{3}, and \mycirc{5}  if no other
mechanism intercedes. This first problem is one of \emph{injection
  starvation}, and we address it with the \emph{injection guarantee}
mechanism described below. Second, recall that a flit that needs to
transfer from one ring to another via a bridge router enters that
bridge router's queue, but if the bridge router's queue is full, then
the transferring flit must make another trip around its current ring
and try again when it next encounters a bridge router. Because of this
rule, the flit traveling from N1 to N2 could be \emph{deflected} an
arbitrarily large number of times at \mycirc{2} and \mycirc{4} (at entry to
bridge routers B1 and B2) if no other mechanism intercedes. This
second problem is one of \emph{transfer starvation}, and we address it
with the \emph{transfer guarantee} mechanism described below.

\emph{Our goal} in this section is to demonstrate that HiRD provides both the
injection guarantee (\S\ref{sec:guarantee-injection}) and the transfer guarantee
(\S\ref{sec:guarantee-transfer}) mechanisms. We show correctness
in \S\ref{sec:guarantee-correctness}, and quantitatively evaluate both
mechanisms in \S\ref{sec:eval-guarantees} and in~\cite{hird-safari-tr}.

\subsection{Preventing Injection Starvation: Injection Guarantee}
\label{sec:guarantee-injection}

The \emph{injection guarantee} ensures that every router on a ring can
eventually inject a flit. This guarantee is provided by a very simple
throttling-based mechanism: when any node is starved (cannot inject a flit)
past a threshold number of cycles, it asserts a signal to a global controller,
which then throttles injection from every other node. No new traffic will enter
the network while this throttling state is active. All existing flits in the
network will eventually drain, and the starved node will be able to finally
inject its new flit. At that time, the starved node de-asserts its throttling
request signal to the global controller, and the global controller subsequently
allows all other nodes to resume normal operation. 

Note that this injection guarantee can be implemented in a hierarchical manner
to improve scalability. In the hierarchical implementation, each individual
local ring in the network monitors only its own injection and throttles
injection locally if any node in it is starved. After a threshold number of
cycles.\footnote{In our evaluation, we set this threshold to be 100 cycles.} if
any node in the ring still cannot inject, the bridge routers connected to that
ring start sending throttling signals to any other ring in the next level of
the ring hierarchy they are connected to. In the worst case, every local ring
stops accepting flits and all the flits in the network drain and eliminate any
potential livelock or deadlock. Designing the delivery guarantee this way 
requires two wires in each ring and small design overhead at
the bridge router to propagate the throttling signal across hierarchy levels. In our
evaluation, we faithfully model this hierarchical design.

\subsubsection{Ensuring Ring Transfers: Transfer Guarantee}
\label{sec:guarantee-transfer}

The \emph{transfer guarantee} ensures that any flit waiting to transfer from
its current ring to another ring via a bridge router will eventually be able to
enter that bridge router's queue. Such a guarantee is non-trivial because the
bridge router's queue is finite, and when the destination ring is congested, a
slot may become available in the queue only infrequently. In the worst case, a
flit in one ring may circulate indefinitely, finding a bridge router to its
destination ring with a completely full queue each time it arrives at the
bridge router. The transfer guarantee ensures that any such circulating flit
will eventually be granted an open slot in the bridge router's transfer queue.
Note in particular that this guarantee is \emph{separate from} the injection
guarantee: while the injection guarantee ensures that the bridge router will be
able to inject flits from its transfer queue into the destination ring (and
hence, have open slots in its transfer queue eventually), these open transfer
slots may not be distributed \emph{fairly} to flits circulating on a ring
waiting to transfer through the bridge router. In other words, some flit may
always be ``unlucky'' and never enter the bridge router if slots open at the
wrong time. The transfer guarantee addresses this problem.

In order to ensure that any flit waiting to transfer out of a ring
eventually enters its required bridge router, each bridge router
\emph{observes a particular slot on its source ring} and monitors for
flits that are ``stuck'' for more than a threshold number of
retries. (To observe one ``slot,'' the bridge router simply examines
the flit in its ring pipeline register once every N cycles, where N is
the latency for a flit to travel around the ring once.) If any flit
circulates in its ring more than this threshold number of times, the
bridge router reserves the next open available entry in its transfer queue for
this flit (in other words, it will refuse to accept other flits for
transfer until the ``stuck'' flit enters the queue). Because of the
injection guarantee, the head of the transfer queue must inject into
the destination ring eventually, hence an entry must become available eventually,
and the stuck flit will then take the entry in the transfer queue the
next time it arrives at the bridge router. Finally, the slot which the
bridge router observes rotates around its source ring: whenever the
bridge router observes a slot the second time, if the flit that
occupied the slot on first observation is no longer present (i.e.,
successfully transferred out of the ring or ejected at its
destination), then the bridge router begins to observe the \emph{next}
slot (the slot that arrives in the next cycle). In this way, every
slot in the ring is observed eventually, and any stuck flit will thus
eventually be granted a transfer.

\subsubsection{Putting it Together: Guaranteed Delivery}
\label{sec:guarantee-correctness}

Before we prove the correctness of these mechanisms in detail, it is helpful to
summarize the basic operation of the network once more. A flit can inject into
a ring whenever a free slot is present in the ring at the injecting router
(except when the injecting router is throttled by the injection guarantee
mechanism). A flit can eject at its destination whenever it arrives, and
destinations always consume flits as soon as they arrive (which is ensured
despite finite reassembly buffers using the Retransmit-Once mechanism~\cite{chipper}, as
already described in \S\ref{sec:mech-node-router}). A flit transfers between
rings via a transfer queue in a bridge router, first leaving its source ring to
wait in the queue and then injecting into its destination ring when at the head
of the queue, and can enter a transfer queue whenever there is a free entry in
that transfer queue (except when the entry is reserved for another flit by the
transfer guarantee mechanism). Finally, when two flits at opposite ends of a
bridge router each desire to transfer through the bridge router, the
\emph{Swap Rule} allows these flits to exchange places directly, bypassing the
queues (and ensuring forward progress).

Our proof is structured as follows: we first argue that if no new flits enter
the network, then the network will drain in finite time. The injection
guarantee ensures that any flit can enter the network. Then, using the
injection guarantee, transfer guarantee, the swap rule, and the fact that the
network is hierarchical, any flit in the network can eventually reach any ring
in the network (and hence, its final destination ring). Because all flits in a
ring continue to circulate that ring, and any node on a ring must consume any
flits that are destined for that node, final delivery is ensured once a flit
reaches its final destination ring.

\noindent\textbf{Network drains in finite time.} Assume no new flits enter the
network (for now). A flit could only be stuck in the network indefinitely if
transferring flits create a cyclic dependence between completely full rings.
Otherwise, if there are no dependence cycles, then if one ring is full and
cannot accept new flits because other rings will not accept \emph{its} flits,
then eventually there must be some ring which depends on no other ring (e.g., a
local ring with all locally-destined flits), and this ring will drain first,
followed by the others feeding into it. However, because the network is
hierarchical (i.e., a tree), the only cyclic dependences possible are between
rings that are immediate parent and child (e.g., global ring and local ring, in
a two-level hierarchy). The \emph{Swap Rule} ensures that when a parent and
child ring are each full of flits that require transfer to the other ring, then
transfer is always possible, and forward progress will be ensured. Note in
particular that we do not require the injection or transfer guarantee for the
network to drain. Only the \emph{Swap Rule} is necessary to ensure that no deadlock
will occur.

\linespread{1}
\begin{table*}[h]
\centering
\footnotesize{
\begin{tabular}{|l||p{9.3cm}|}
\hline
\textbf{Parameter} & \textbf{Setting} \\
\hline
\hline
System topology & CPU core and shared cache slice at every node \\
\hline
Core model & Out-of-order, 128-entry ROB, 16 MSHRs (maximum simultaneous outstanding requests) \\
\hline
Private L1 cache & 64 KB, 4-way associative, 32-byte block size \\
\hline
Shared L2 cache & Perfect (always hits) to stress the network and
penalize our reduced-capacity deflection-based design; cache-block-interleaved
 \\
\hline
Cache coherence & Directory-based protocol (based on SGI
Origin~\cite{laudon97}), directory entries co-located with
shared cache blocks \\
\hline
Simulation length & 5M-instruction warm-up, 25M-instruction active execution per node
~\cite{casebufferless,chipper,hat-sbac-pad,minbd} \\
\hline
\end{tabular}
}
\caption{Simulation and system configuration parameters. Reproduced from~\cite{hird-safari-tr}.}
\label{table:sysparams}
\end{table*}

\begin{table*}
\centering
\footnotesize{
\begin{tabular}{|l|l|p{8.5cm}|}
\hline
Parameter & Network & Setting \\
\hline
\hline
\multirow{4}{*}{Interconnect Links} & Single Ring & Bidirectional, \textbf{4x4:} 64-bit and 128-bit width, \textbf{8x8:} 128-bit and 256-bit width \\
\cline{2-3}
& Buffered HRing & Bidirectional, \textbf{4x4:} 3-cycle per-hop latency (link+router); 64-bit local and 128-bit global rings, \textbf{8x8:} three-level hierarchy, 4x4 parameters, with second-level rings connected by a 256-bit third-level ring \\
\cline{2-3}
& HiRD & \textbf{4x4:} 2-cycle (local), 3-cycle (global) per-hop
latency (link+router); 64-bit local ring, 128-bit global ring; \textbf{8x8:} 4x4
parameters, with second-level rings connected by a 256-bit third-level
ring \\
\hline 
\multirow{6}{*}{Router} & Single Ring & 1-cycle per-hop latency (as in~\cite{kim09nocarc}) \\
\cline{2-3}
& Buffered HRing & Node (NIC) and bridge (IRI) routers based
on~\cite{ravindran97}; 4-flit in-ring and transfer
FIFOs. Bidirectional links of dual-flit width (for fair comparison
with our design). Bubble flow control~\cite{bubbleflow} for deadlock
freedom. \\
\cline{2-3}
& HiRD & Local-to-global buffer depth of 1, global-to-local buffer depth of 4 \\
\hline
\end{tabular}
}
\caption{Network parameters for HiRD evaluation. Reproduced from~\cite{hird-parco}.}
\vspace{-0.25in}
\label{table:topoparams}
\end{table*}

\noindent\textbf{Any node can inject.} Now that we have shown that the network
will drain if no new flits are injected, it is easy to see that the injection
guarantee ensures that any node can eventually inject a flit: if any node is
starved, then all nodes are throttled, no new flit enters the network, and the
network must eventually drain (as we just showed), at which point the starved
node will encounter a completely empty network into which to inject its flit.
(It likely will be able to inject before the network is completely empty, but
in the worst case, the guarantee is ensured in this way.)

\noindent\textbf{All flits can transfer rings and reach their destination
rings.} With the injection guarantee in place, the transfer guarantee can be
shown to provide its stated guarantee as follows: because of the injection
guarantee, a transfer queue in a bridge router will always inject its head flit
in finite time, hence will have an open entry to accept a new transferring flit
in finite time. All that is necessary to ensure that \emph{all} transferring
flits eventually succeed in their transfers is that \emph{any} flit stuck for
long enough gets an available entry in the transfer queue. The transfer
guarantee does exactly this by observing ring slots in sequence and reserving a
transfer queue entry when a flit becomes stuck in a ring. Because the mechanism
will eventually observe every slot in the ring, all flits will be allowed to make their
transfers eventually. Hence, all flits can continue to transfer rings until
reaching their destination rings (and thus, their final destinations).

\subsubsection{Hardware Cost}

Our injection and transfer guarantee mechanisms have low hardware overhead. To implement the
injection guarantee, one counter is required for each injection point.
This counter tracks how many cycles have elapsed while injection is starved,
and is reset whenever a flit is successfully injected. Routers communicate with
the throttling arbitration logic with only two wires, one to signal blocked
injection and one control line that throttles the router. The wiring is done
hierarchically instead of globally to minimize the wiring cost (\S\ref{sec:guarantee-injection}). Because the
correctness of the algorithm does not depend on the delay of these wires, and
the injection guarantee mechanism is activated only rarely (in fact, \emph{never} for
our evaluated realistic workloads), the signaling and central coordinator need not be
optimized for speed. To provide the transfer guarantee, each bridge router
implements ``observer'' functionality for each of the two rings it sits on, and
the observer consists only of three small counters (to track the current
timeslot being observed, the current timeslot at the ring pipeline register in
this router, and the number of times the observed flit has circled the ring)
and a small amount of control logic.  Importantly, note that neither mechanism
impacts the router critical path nor affects the router datapath (which
dominates energy and area).


\subsection{HiRD: Evaluation Methodology}
\label{sec:meth}

We perform our evaluations using a cycle-accurate simulator of a CMP system
with 1.6GHz interconnect to provide application-level performance
results~\cite{NOCulator}. Our simulator is publicly available and includes
the source code of all mechanisms we evaluated~\cite{NOCulator}.
Tables~\ref{table:sysparams}~and~\ref{table:topoparams} provide the configuration
parameters of our simulated systems.

Our methodology ensures a rigorous and isolated evaluation
of NoC capacity for especially cache-resident workloads, and has also been used in
other studies~\cite{casebufferless,chipper,hotnets2010,sigcomm12,minbd}. Instruction traces for the
simulator are taken using a Pintool~\cite{pin} on representative
portions of SPEC CPU2006 workloads.

\begin{figure*}[h!]
\centering
\includegraphics[width=4.5in]{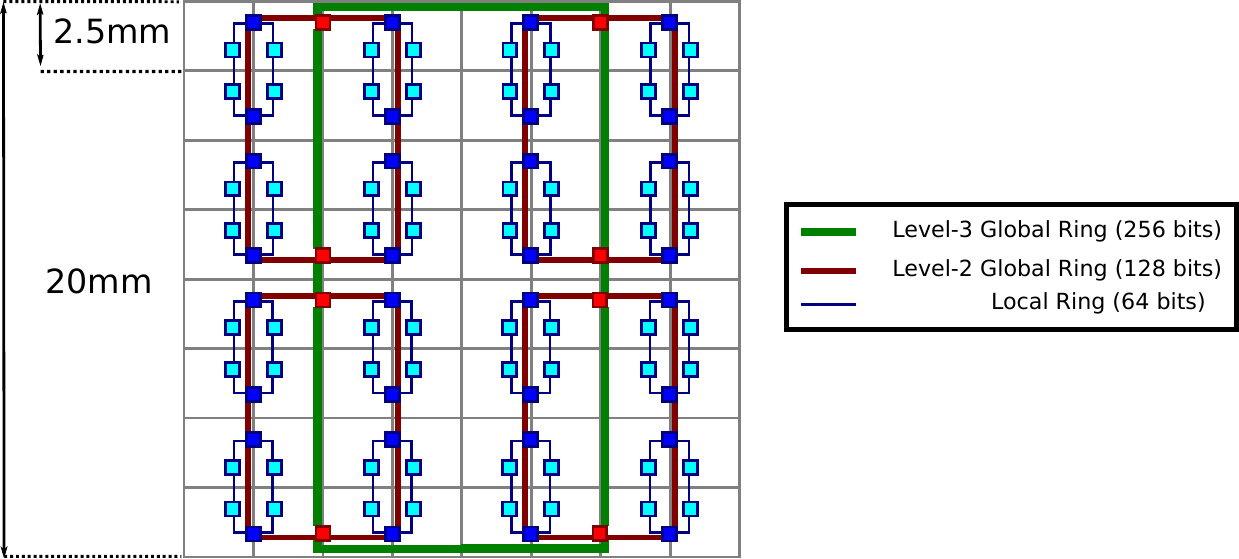}
\caption{Assumed floorplan for HiRD 3-level (64-node)
  network. Two-level (16-node) network consists of one quadrant of
  this floorplan. \ch{Reproduced from~\cite{hird-safari-tr}.}}
\label{fig:hird-floorplan}
\end{figure*}

We mainly compare to a single
bidirectional ring and a state-of-the-art buffered hierarchical
ring~\cite{ravindran97}. 
Also, note that while there are many possible ways to
optimize each baseline (such as congestion control~\cite{hat-sbac-pad,hotnets2010,sigcomm12}, adaptive routing
schemes, and careful parameter tuning), we assume a fairly typical
aggressive configuration for each.

\noindent\textbf{Data Mapping.} We map data in a cache-block-interleaved way to
different shared L2 cache slices. This mapping is agnostic to the underlying locality.
As a result, it does not exploit the low-latency data access in the local ring.
One can design systematically better mapping in order to keep frequently used
data in the local ring as in~\cite{CloudCache,ccraik-safari-tr}. However, such a mapping mechanism is orthogonal
to our proposal and can be applied in all ring-based network designs.

\label{sec:meth-workloads}
\noindent\textbf{Application \& Synthetic Workloads.} The system is run with a
set of 60 multiprogrammed workloads. Each workload consists of one
single-threaded instance of a SPEC CPU2006 benchmark on each core, for a total
of either 16 (4x4) or 64 (8x8) benchmark instances per workload.
Multiprogrammed workloads such as these are representative of many common
workloads for large CMPs.  Workloads are constructed at varying network
intensities as follows: first, benchmarks are split into three classes (Low,
Medium and High) by L1 cache miss intensity (which correlates directly with
network injection rate), such that benchmarks with less than 5 misses per
thousand instructions (MPKI) are ``Low-intensity,'' between 5 and 50 are
``Medium-intensity,'' and above 50 MPKI are ``High-intensity.''  Workloads are
then constructed by randomly selecting a certain number of benchmarks from each
category. We form workload sets with four intensity mixes: High (H), Medium (M),
Medium-Low (ML), and Low (L), with 15 workloads in each (the average network injection
rates for each category are 0.47, 0.32, 0.18, and 0.03 flits/node/cycle,
respectively). 

\begin{figure*}[h!]
\vspace{-0.1in}
\centering
\includegraphics[width=6.5in]{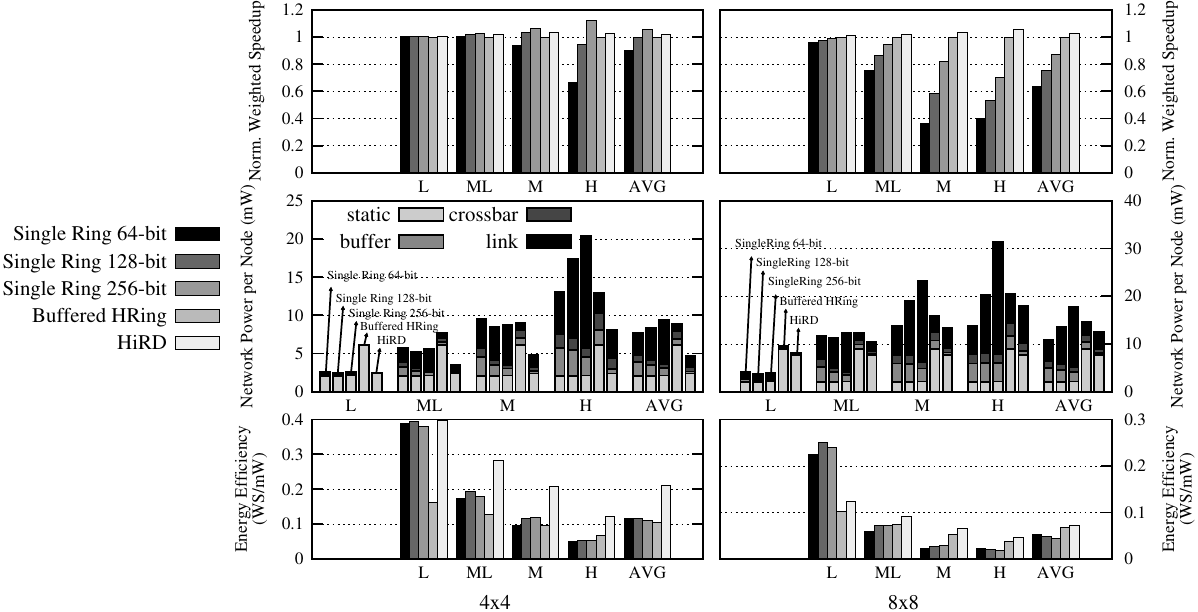}
\vspace{-0.05in}
\caption{HiRD as compared to buffered hierarchical rings and a single-ring network. \ch{Reproduced from~\cite{hird-safari-tr}.}}
\label{fig:hird-buf-hring}
\end{figure*}

\noindent\textbf{Multithreaded Workloads.} We use the GraphChi implementation
of the GraphLab framework~\cite{graphchi,graphlab}. The implementation we use
is designed to run efficiently on multi-core systems. The workload consists of
Twitter Community Detection (CD), Twitter Page Rank (PR), Twitter Connected
Components (CC), Twitter Triangle Counting (TC)~\cite{twitter_rv}, and Graph500
Breadth First Search (BFS). We simulated the representative portion of each
workload and each workload has a working set size of greater than 151.3$\,$MB.
On every simulation of these multithreaded workloads, we warm up the cache with
the first 5 million instructions, then we run the remaining code of the
representative portion.

\noindent\textbf{Energy \& Area.} We measure the energy and area of routers and links by
individually modeling the crossbar, pipeline registers, buffers, control logic,
and other datapath components. For links, buffers and datapath elements, we use
DSENT 0.91~\cite{dsent}. Control logic is modeled in Verilog RTL. Both energy and area are calculated based on a 45nm
technology. The link lengths we assume are based on the floorplan of our designs,
which we describe in the next paragraph.

We assume the area of each core to be 2.5 mm $x$ 2.5 mm. We assume a 2.5 mm link length for single-ring designs. For the hierarchical ring design, we assume 1 mm links between local-ring routers, because the four routers on a local ring can be placed at four corners that meet in a tiled design. Global-ring links are assumed to be 5.0 mm (i.e., five times as long as local links), because they span across two tiles on average if local rings are placed in the center of each four-tile quadrant. Third-level global ring links are assumed to be 10mm (i.e., ten times as long as local links) in the 8x8 evaluations. This floorplan is illustrated in more detail in Figure~\ref{fig:hird-floorplan} for the 3-level (64-node) HiRD network. Note that one quadrant of the floorplan of Figure~\ref{fig:hird-floorplan} corresponds to the floorplan of the 2-level (16-node) HiRD network. We faithfully take into account all link lengths in our energy and area estimates for all designs.

\noindent\textbf{Application Evaluation Metrics.} For multiprogrammed workloads, we present application performance results using the commonly-used Weighted Speedup metric~\cite{weighted_speedup,harmonic_speedup}. We use the maximum slowdown metric to measure unfairness~\cite{stc,atlas,tcm,vandierendonck,fst,eiman-isca11,eiman-micro09,mcp,aergia,a2c,asm-micro15,mise-hpca13,dash-taco16,sms,bliss,bliss-tpds}.

\input{eval}

\section{Other Methods to Improve NoC Scalability}

In this Section, we now discuss other approaches that are designed to improve scalability of NoCs. 

\noindent\textbf{Ring-based NoCs.} Hierarchical ring-based interconnect was proposed in a previous line of work~\cite{ravindran97,xiangdong95,hr-model,ravindran98,numachine,kim-hpca14,farkas-sc92,hector,farkas-sc92,holliday94}. We have already extensively compared to past hierarchical ring proposals qualitatively and quantitatively. The major difference between HiRD and these previous \chx{approaches is that HiRD uses} deflection-based bridge routers with minimal buffering, and node routers with no buffering. In contrast, all of these previous works use routers with in-ring buffering, wormhole switching and flow control. 
Kim et al. propose tNoCs, hybrid packet-flit credit-based flow control~\cite{kim-hpca14} and Clumsy Flow Control (CFC)~\cite{cfc-cal}. However, these two designs add additional complexity because tNoCs~\cite{kim-hpca14} requires an additional credit network to guarantee forward progress while CFC requires coordination between cores and memory controllers. Flow control in HiRD \chx{is different from that in these works due to HiRD's} simplicity (with deflection based flow control, the Retransmit-Once mechanism, and simpler local-to-global and global-to-local buffers). Additionally, throttling decisions in HiRD can be made locally in each local ring as opposed to global decisions in CFC~\cite{cfc-cal} and tNoCs~\cite{kim-hpca14}.

\chx{Udipi et al.~propose} a hierarchical topology using global and local
buses~\cite{udipi10}. Using buses limits scalability in favor of simplicity. In contrast, HiRD design has more favorable scaling, in exchange for using more complex flit-switching routers.  Das et al.~\cite{das09} \chx{examine} several hierarchical designs, including a concentrated mesh (one mesh router shared by several nearby nodes).

A previous system, SCI (Scalable Coherent Interface)~\cite{sci},
also uses rings, and can be configured in many topologies (including
hierarchical rings). However, to handle buffer-full conditions, SCI
NACKs and subsequently retransmits packets, whereas HiRD deflects only
single flits (within a ring), and does not require the sender to
retransmit its flits. SCI was designed for off-chip interconnect,
where tradeoffs in power and performance are very different \chx{from those} in
on-chip interconnects. The KSR (Kendall Square Research) machine~\cite{ksr}
uses a hierarchical ring design that resembles HiRD, yet these techniques
are not disclosed in detail and, to our knowledge, have not been 
publicly evaluated in terms of energy efficiency.

\noindent\textbf{Scalable Topology Design.} While low-radix topologies (e.g., ring, tori~\cite{gemini}, meshes~\cite{balfour06}, Express Cubes~\cite{grot09} and Kilo-NoC~\cite{grot2010,kilonocs,kilo-noc-toppick}) offers low area and power consumptoin, high-radix topologies~\cite{flattened_bfly, hyperx, slimnoc} provide scalable alternatives to large scale systems. A flattened butterfly topology provides a scalable design that allows routers to send flits using two hops~\cite{flattened_bfly}. A HyperX network~\cite{hyperx} extends the hypercube~\cite{greenberg92} design to minimize cost and lowering the latency. A SlimNoC~\cite{slimnoc} network provides a low-diameter, high-radix that further reduces the power and area through a SlimFly topology~\cite{slimfly}.

\noindent\textbf{Low Cost Router Designs.} Kim~\cite{kim09} proposes a low-cost router design that is superficially similar to HiRD's node router design where routers convey traffic along rows and columns in a \emph{mesh} without making use of crossbars, only pipeline registers and MUXes. Once traffic enters a row or column, it continues until it reaches its destination, as in a ring. Traffic also transfers from a row to a column analogously to a ring transfer in our design, using a ``turn buffer.'' However, because a turn is possible at any node in a mesh, every router requires such \chx{a buffer~\cite{lionel-isca92-turn-model,lionel-isca98-turn-model};} in contrast, HiRD require similar transfer buffers only at bridge routers, and their cost is paid for by all nodes.
Additionally, this design does not use deflections when there is contention. 

Mullins et al.~\cite{mullins04} propose a buffered mesh router with
single-cycle arbitration. Abad et al.~\cite{rotary-router} propose the Rotary Router that consists of two independent rings that join the router's
ports and perform packet arbitration similar to standalone ring-based networks. Both the Rotary Router and HiRD allow a packet to circle a ring again in a ``deflection'' if an ejection (ring transfer or router exit) is unsuccessful. \chx{Nicopoulos et al.~\cite{Nicopoulos06} propose a buffer structure that allows the network to dynamically \chxi{regulate} the number of virtual channels}. Kodi et al.~\cite{Kodi08} propose an orthogonal mechanism that reduces buffering by using links as buffer space when necessary. Multidrop Express Channels~\cite{grot09} also provides a low cost mechanism to connect multiple nodes using a multidrop bus without expensive router changes.

\section{Conclusion and Future Outlook}
\label{sec:conclusion}

Scalability and energy are \chx{two major concerns as \chxii{core counts} increase in commercial processors.} To provide a design that \chx{is area-efficient and energy efficient} without sacrificing performance, this chapter first presents MinBD~\cite{minbd,minbd-tr,minbd-book}. MinBD is \chx{a minimally-buffered deflection router design. It} combines deflection routing with a small buffer, such that some network traffic that would have been deflected is placed in the buffer instead. By using the buffer for only a fraction of network traffic, MinBD makes more efficient use of a given buffer size than a conventional input-buffered router. Its average network power is also greatly reduced: relative to an input-buffered router, buffer power is much lower, because buffers are smaller. Relative to a bufferless deflection router, dynamic power is lower, because deflection rate is reduced with \chxi{the use of a small energy-conscious} buffer. 

To further improve scalability, this chapter discusses \emph{HiRD}~\cite{hird-parco,hird,hird-safari-tr}, a simple hierarchical ring-based NoC design \chx{that employs deflection routing}. Past work has shown that a hierarchical ring design yields good performance and scalability relative to both a single ring and a mesh. HiRD has two new contributions: (1) a simple router design that enables ring transfers \emph{without in-ring buffering or flow control}, instead using limited \emph{deflections} (retries) when a flit cannot transfer to \chx{another} ring, and (2) two \emph{guarantee mechanisms} that ensure deterministically-guaranteed forward progress despite deflections. The evaluations show that HiRD enables a \chxi{simple and low-cost} implementation of a hierarchical ring network. \chxi{Our HiRD} evaluations also show that HiRD is more energy-efficient than several other topologies while providing competitive performance.

Despite the extensive design space for low-power NoC we considered so far, a number of key challenges remain to enable \chx{truly scalable and energy-efficient interconnection networks} \chxi{for modern systems and workloads.} We believe that low-cost, energy-efficient network-on-chip design is an important challenge in scaling modern architectures beyond traditional CMPs. For example, heterogeneous architectures in modern SoCs can stress the interconnect through their imbalanced loads that \chxi{are a consequence of largely different demands} across many different types of applications and accelerators. Relatively new technology such as chiplets~\cite{chiplet-isca18,centaur} or new types of memory~\cite{network-on-memory, stringfigure, memory-centric-gpu} can create demands for \chxi{efficient} interconnection \chxi{network} designs that connect multiple memory nodes together. \chxii{Especially processing-in-memory systems~\cite{ahn.tesseract.isca15,graphp,mutlu2020modern,micpro2019, zhu2013accelerating,
pugsley2014ndc,zhang.hpdc14,
farmahini2015nda,cali2020genasm,ahn.pei.isca15,
loh2013processing, hsieh.isca16,impica, DBLP:conf/sigmod/BabarinsaI15, 
DBLP:conf/IEEEpact/LeeSK15, 
DBLP:conf/hpca/GaoK16,  
chi2016prime, gu.isca16, 
kim.isca16, asghari-moghaddam.micro16, 
boroumand2016pim, hashemi.isca16, 
gao.pact15, guo2014wondp, 
sura.cf15, morad.taco15, 
hassan.memsys15, li.dac16, 
kang.icassp14, aga.hpca17, 
shafiee2016isaac,  
nai2017graphpim,kim.arxiv17,
kim.bmc18, li.micro17, 
kim.sc17, boroumand.asplos18,
fernandez2020natsa,singh2019napel, rezaei2020nom,sisa,singh2020nero,wang2020figaro,synchron,gomezluna2021benchmarking,boroumand-pact21,damov,fpga-pim-2021,mutlu-date21,fimdram-isscc21,fimdram-isca21,upmem-hotchip,simdram,Seshadri:2015:ANDOR,chang.hpca16,seshadri.micro17} can require well-connected memory arrays via efficient interconnects to tightly couple computation and communication.} 
A fundamentally low-cost and energy-efficient interconnection network can further push the boundaries of computing \chxi{systems,} leading to significant improvements in performance and energy, and potentially enabling new applications and computing platforms.

\section*{\sg{Acknowledgments}}

This chapter incorporates revised material from another earlier article published in Parallel Computing in 2016~\cite{hird-parco}, the proceedings of the International Symposium on Computer Architecture and High Performance Computing in 2014~\cite{hird}, the proceedings of the International Symposium on Networks-on-Chip in 2012~\cite{minbd} and the proceedings of the International Symposium on High Performance Computer Architecture in 2011~\cite{chipper}.

This article is based on research done over the course of the past \chxi{13} years in the SAFARI Research Group on the topic of Network-on-Chips (NoC). We thank all of the members of the SAFARI Research Group, and our collaborators at Carnegie Mellon, ETH Zürich, and other universities, who have contributed to the various works we describe in this paper. Thanks also goes to our research group’s industrial sponsors over the past \chxi{13} years, especially Alibaba, AMD, ASML, Google, Huawei, Intel, Microsoft, NVIDIA, Samsung, Seagate, and VMware. This work was also partially supported by the Intel Science and Technology Center for Cloud Computing, the Semiconductor Research Corporation, the Data Storage Systems Center at Carnegie Mellon University, various NSF and NIH grants, and various awards, including the NSF CAREER Award, the Intel Faculty Honor Program Award, a number of Google and IBM Faculty Research Awards to Onur Mutlu, and the Royal Thai Scholarship to Rachata Ausavarungnirun.




{
\bibliographystyle{elsarticle-num}
\bibliography{paper}
}






\end{document}

%% file: eval.tex
\subsection{Performance, Energy Efficiency and Scalability of HiRD}
\label{sec:evaluation}

\begin{figure*}
\centering
\includegraphics[width=6.3in]{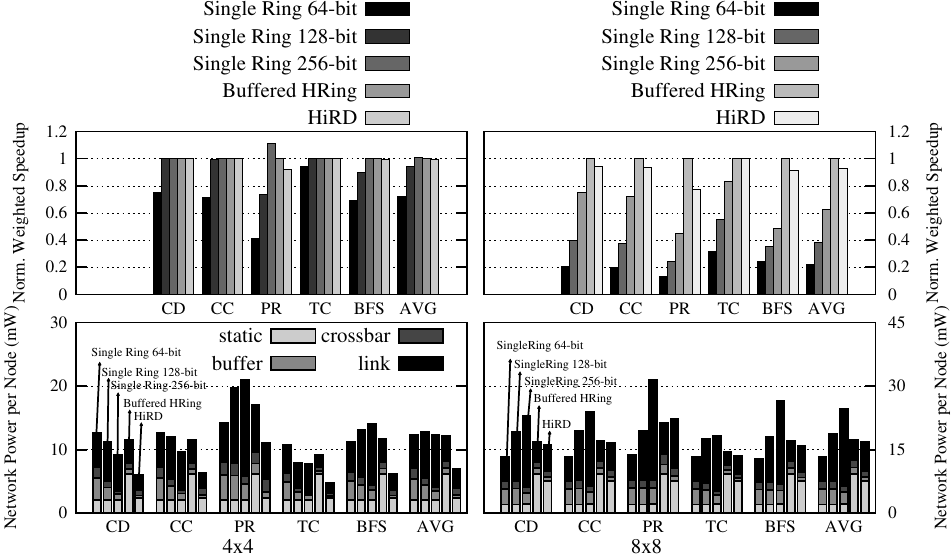}
\caption{HiRD as compared to buffered hierarchical rings and a single-ring network on multithreaded workloads. Reproduced from~\cite{hird-safari-tr}.}
\label{fig:hird-buf-hring-graph}
\end{figure*}

We provide a comprehensive evaluation of our proposed mechanism against other
ring baselines.  Since our goal is to provide a better ring design, our main
comparisons are to ring networks. However, we also provide sensitivity analyses
and comparisons to other network designs as well.



\subsection{Ring-based Network Designs}
\label{sec:eval-hrings}

\subsubsection{Multiprogrammed workloads}

Figure~\ref{fig:hird-buf-hring} shows performance (weighted speedup
normalized per node), power (total network power normalized per node),
and energy-efficiency (perf./power) for 16-node and 64-node HiRD and
buffered hierarchical rings in~\cite{ravindran97}, using identical topologies, as well as a single ring (with different bisection bandwidths).

1. A hierarchical topology yields significant performance advantages over a
single ring (i) when network load is high and/or (ii) when the network scales
to many nodes. As shown, the buffered hierarchical ring improves
performance by 7\% (and HiRD by 10\%) 
in high-load workloads at 16 nodes compared to a single ring with 128-bit links. The
hierarchical design also reduces power because hop count is reduced. Therefore,
link power reduces significantly with respect to a single ring.
On average, in the 8x8 configuration, the buffered hierarchical ring network
obtains 15.6\% better application performance than the single ring with 256-bit
links, while HiRD attains 18.2\% higher performance.


2. Compared to the buffered hierarchical ring, HiRD has significantly
lower network power and better performance. On average, HiRD reduces total network power
(links and routers) by 46.5\% (4x4) and 14.7\% (8x8) relative to this
baseline. This reduction in turn yields significantly better energy
efficiency (lower energy consumption for buffers and slightly higher for links).\footnote{Note 
that the low intensity workloads in the 8x8 network is an exception.
HiRD reduces energy efficiency for these as the static link power becomes dominant
for them.} Overall, HiRD is the most energy-efficient of the
ring-based designs evaluated in this paper for both 4x4 and 8x8 network
sizes. HiRD also performs better than Buffered HRing due to the reasons explained
in the next section (\S\ref{sec:eval-synth}).


3. While scaling the link bandwidth increases the performance of a single ring
network, the network power increases 25.9\% when the link bandwidth increases
from 64-bit to 128-bit and  15.7\% when the link bandwidth increases from
128-bit to 256-bit because of higher dynamic energy due to wider links. In addition, scaling the link bandwidth is not a scalable solution as a single ring network performs worse than the bufferred hierarchical ring baseline even when a 256-bit link is used.


We conclude that HiRD is effective in simplifying the design of the
hierarchical ring and making it more energy efficient, as we intended to as our design goal. We show that HiRD provides competitive performance compared
to the baseline buffered hierarchical ring design with equal or better energy
efficiency.

\subsubsection{Multithreaded workloads}

\indent Figure~\ref{fig:hird-buf-hring-graph} shows the performance and power of HiRD on
multithreaded applications compared to a buffered hierarchical ring and a
single-ring network for both 16-node and 64-node systems. On average, HiRD
performs 0.1\% (4x4) and 0.73\% (8x8) worse than the buffered hierarchical ring.
However, on average, HiRD consumes 43.8\% (4x4) and 3.1\% (8x8) less power,
leading to higher energy efficiency. This large reduction in energy comes from the
elimination of most buffers in HiRD.

Both the buffered hierarchical ring and HiRD outperform single ring networks,
and the performance improvement increases as we scale the size of the network.


Even though HiRD performs competitively with a buffered hierarchical ring
network in most cases, HiRD performs poorly on the Page Ranking application. We
observe that Page Ranking generates more non-local network traffic than other applications. As HiRD is
beneficial mainly at lowering the local-ring latency, it is unable to speed up
such non-local traffic, and is thus unable to help Page Ranking. In addition,
Page Ranking also has higher network traffic, causing more congestion in the
network (we observe 17.3\% higher average network latency for HiRD in an 8x8
network), and resulting in a performance drop for HiRD. However, it is
possible to use a different number of bridge routers as illustrated
in Figure~\ref{fig:topology}, to improve the performance of HiRD, which we will
analyze in Section~\ref{sec:eval-sensitivity}.  Additionally, it
is possible to apply a locality-aware cache mapping technique~\cite{CloudCache,ccraik-safari-tr}
in order to take advantage of lower local-ring latency in HiRD. 

We conclude that HiRD is effective in improving evergy efficiency significantly 
for both multiprogrammed and multithreaded applications.

\subsection{Synthetic-Traffic Network Behavior}
\label{sec:eval-synth}

\begin{figure}[h!]
\centering
\subfloat[Uniform Random, 4x4]{
\includegraphics[width=1.5in]{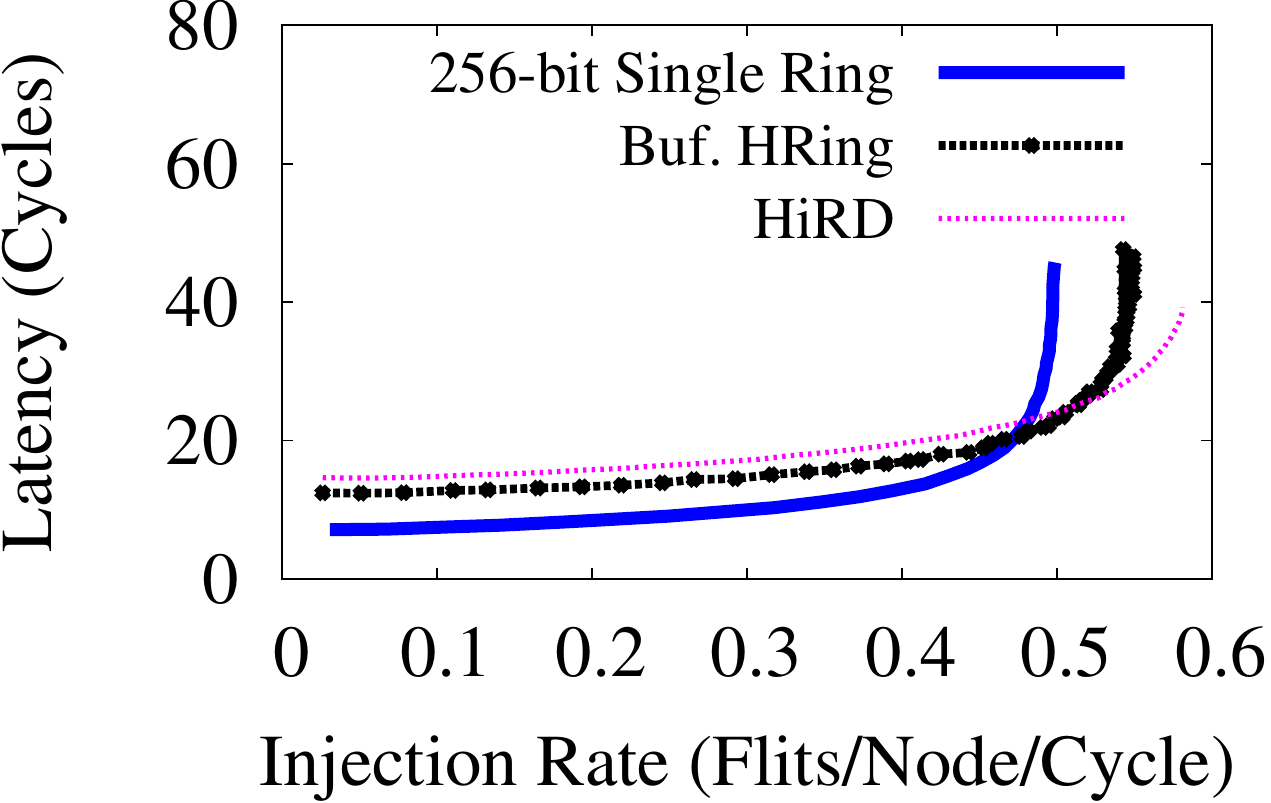}
\label{fig:synthUR}
}
\subfloat[Uniform Random, 8x8]{
\includegraphics[width=1.5in]{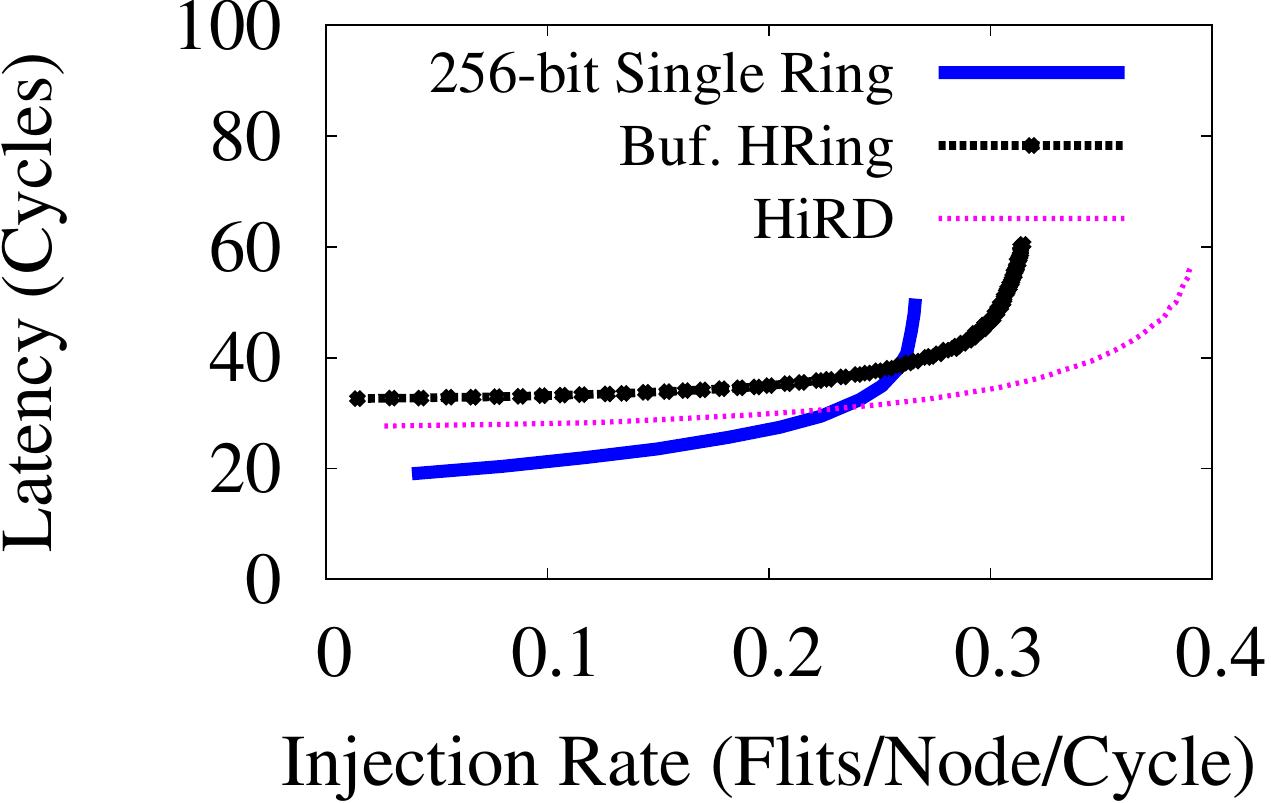}
\label{fig:synth8UR}
}
\\
\vspace{-0.1in}
\subfloat[Bit Complement, 4x4]{
\includegraphics[width=1.5in]{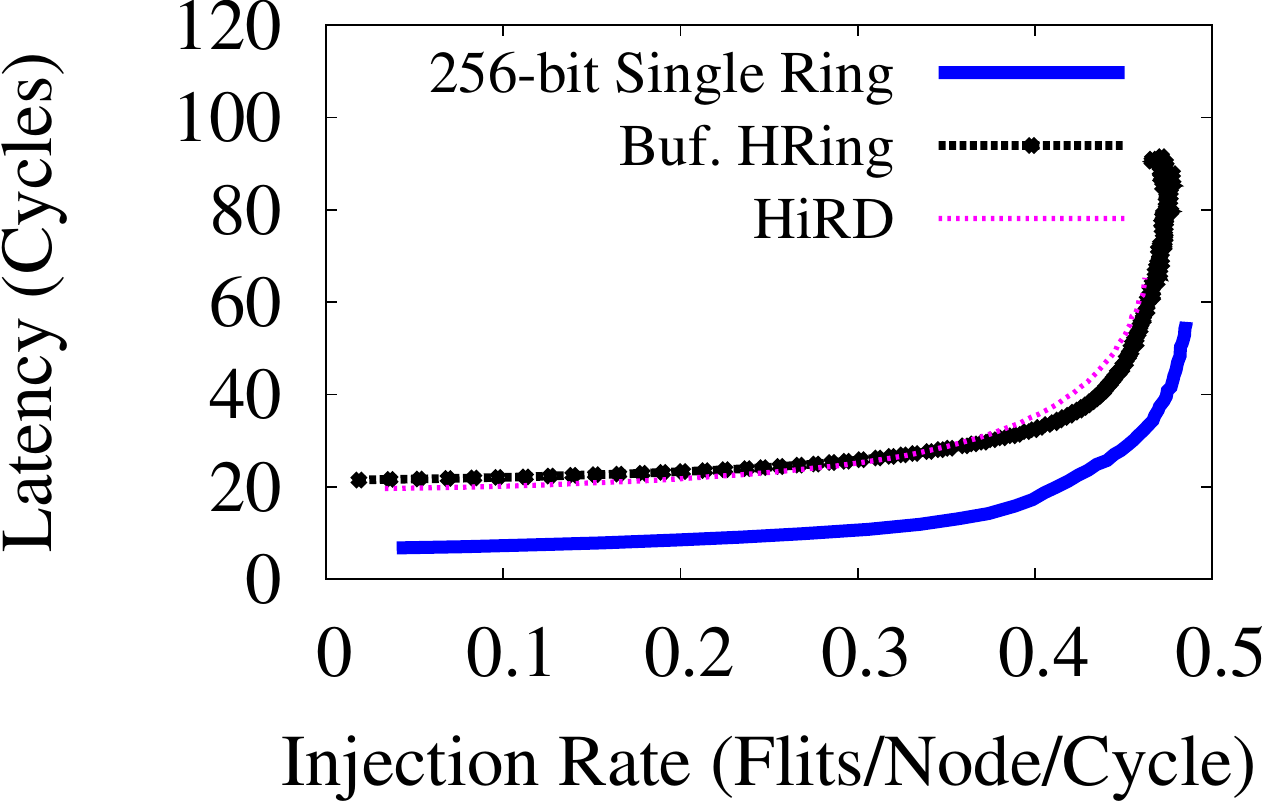}
\label{fig:synthBC}
}
\subfloat[Bit Complement, 8x8]{
\includegraphics[width=1.5in]{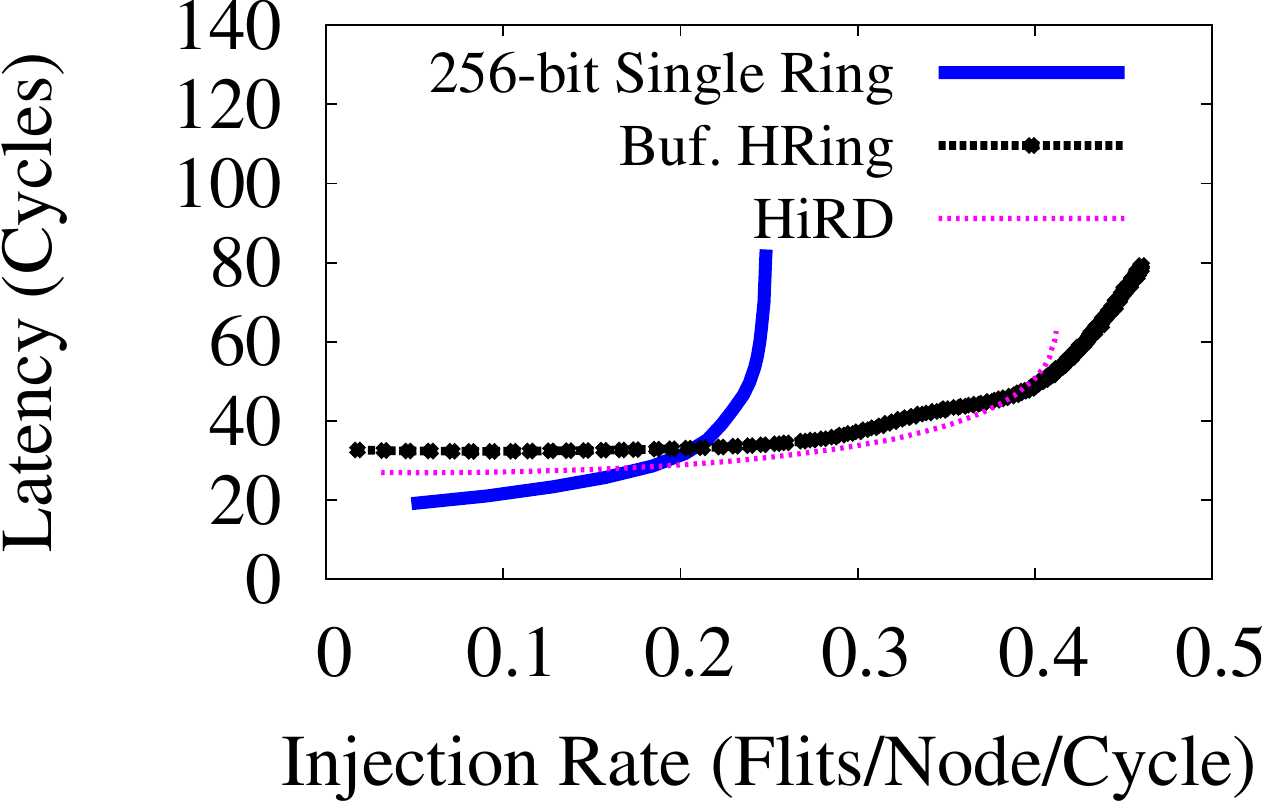}
\label{fig:synth8BC}
}
\\
\vspace{-0.1in}
\subfloat[Transpose, 4x4]{
\includegraphics[width=1.5in]{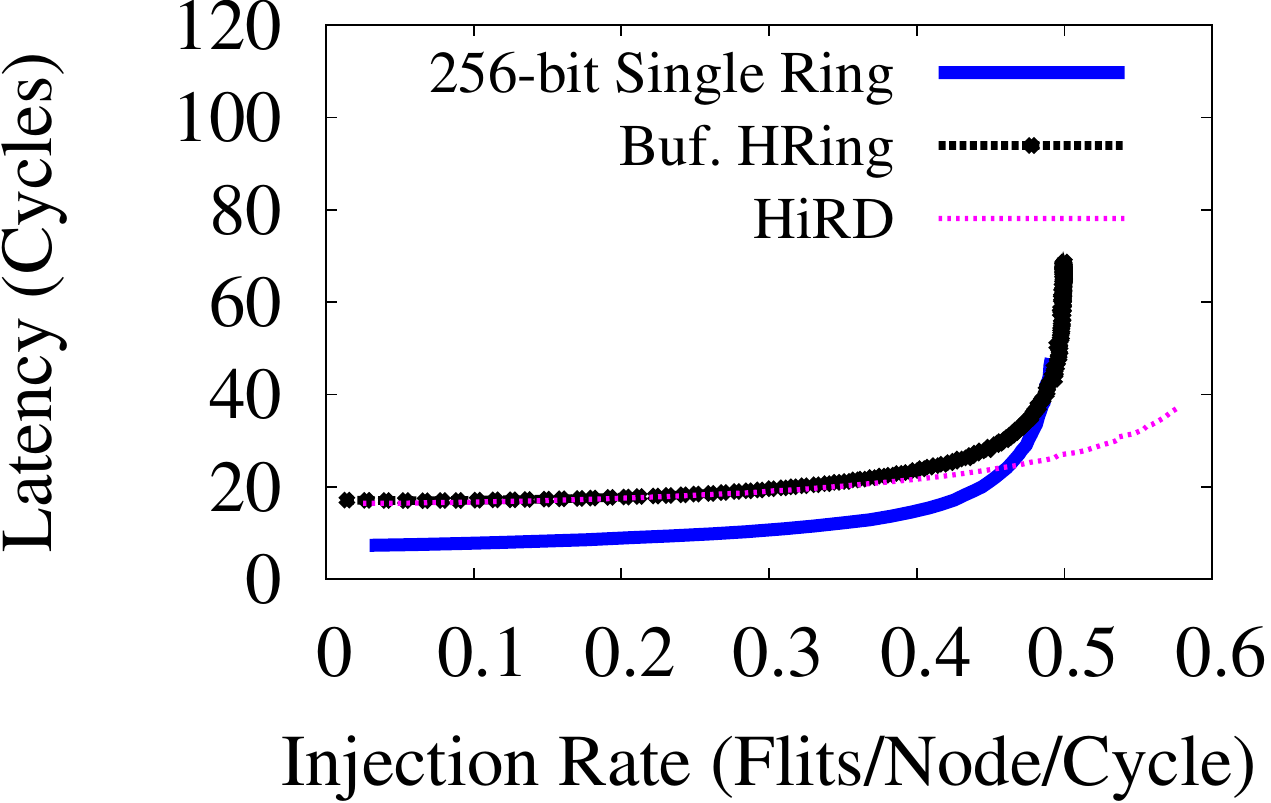}
\label{fig:synthTR}
}
\subfloat[Transpose, 8x8]{
\includegraphics[width=1.5in]{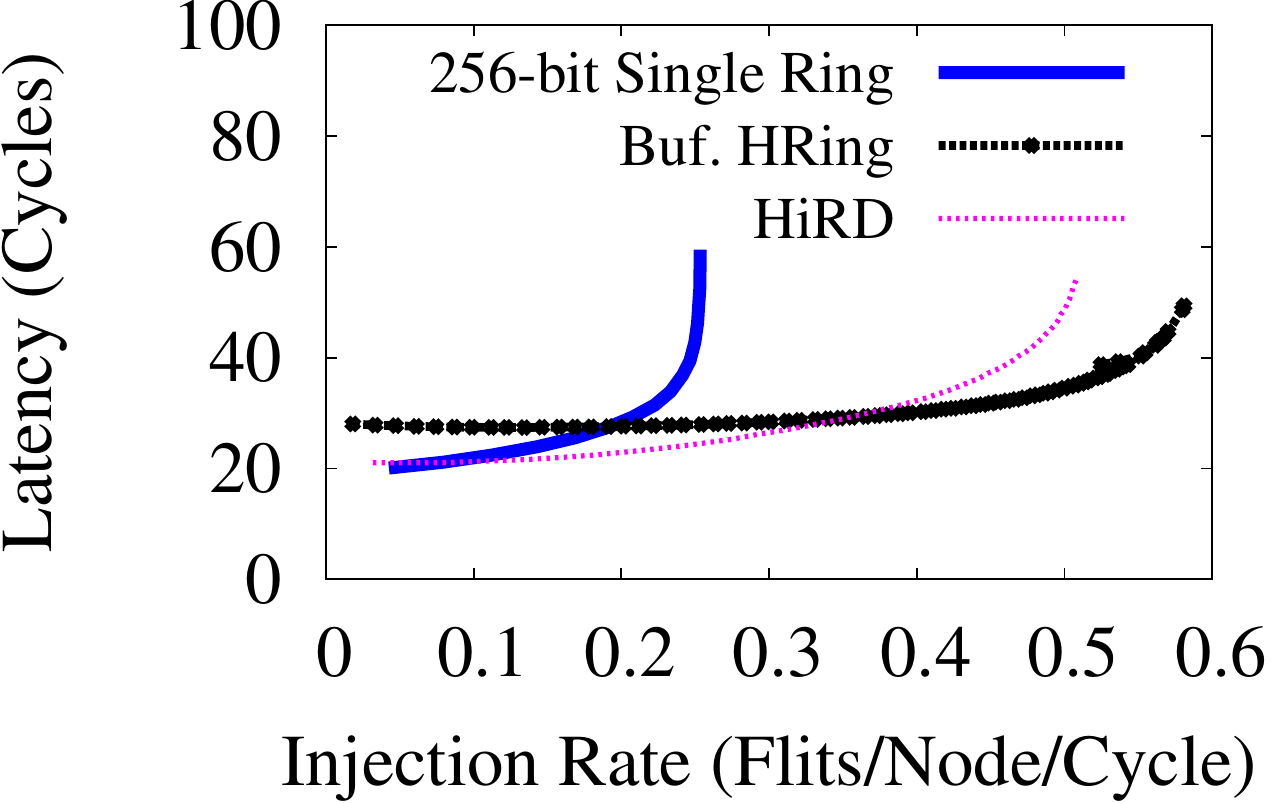}
\label{fig:synth8TR}
}
\vspace{-0.05in}
\caption{Synthetic-traffic evaluations for 4x4 and 8x8 networks. \ch{Reproduced from~\cite{hird-safari-tr}.}}
\label{fig:synth}
\vspace{-0.15in}
\end{figure}

Figure~\ref{fig:synth} shows the average packet latency as a function of injection rate for
buffered and bufferless mesh routers, a single-ring design, the buffered
hierarchical ring, and HiRD in 16 and 64-node systems. We show uniform random,
transpose and bit complement traffic patterns~\cite{principles}. Sweeps on injection rate
terminate at network saturation. The buffered hierarchical ring saturates at a
similar point to HiRD but maintains a slightly lower average latency because it
avoids transfer deflections. In contrast to these high-capacity designs, the
256-bit single ring saturates at a lower injection rate. 

\begin{table*}
\small
\centering
\begin{tabular}{|l||l|l|l||l|l|}
\hline
Configuration &  \multicolumn{3}{|c||}{\breakTable{Network Throughput\\(flits/node/cycle)}} & \breakTable{Transfer FIFO Wait\\(cycles)} & Deflections/Retries \\
\hline
 & Ring A & Ring B & Ring C & avg/max & avg/max \\
\hline
\hline
Without Guarantees & 0.164 & 0.000 & 0.163 & 2.5 / 299670 & 6.0 / 49983 \\
\hline
With Guarantees & 0.133 & 0.084 & 0.121 & 1.2 / 66 & 2.8 / 18 \\
\hline
\end{tabular}
\caption{Results of worst-case traffic pattern without and with
  injection/transfer guarantees enabled. Data reproduced from~\cite{hird-safari-tr}.}
\label{table:guarantee-results}
\end{table*}

\begin{table*}[h]
\small
\centering
\begin{tabular}{|l||l|l|}
\hline
Configuration & Transfer FIFO Wait time (cycles) & Deflections/Retries \\
\hline
 & (avg/max) & (avg/max) \\
\hline
\hline
Without guarantees & 3.3 / 169 & 3.7 / 19 \\
\hline
With guarantees & 0.76 / 72 & 0.7 / 8 \\
\hline
\end{tabular}
\caption{Effect of transfer guarantee mechanism on real workloads. Data reproduced from~\cite{hird-safari-tr}.}
\label{table:guarantee-real-results}
\end{table*}

As network size scales to 8x8, HiRD performs significantly better than the
256-bit single ring, because the hierarchy reduces the cross-chip latency while
preserving bisection bandwidth.  HiRD also performs better than Buffered HRing
because of two reasons. First, HiRD is able to allow higher peak utilization
(91\%) than Buffered HRing (71\%) on the global rings. We observed that when
flits have equal distance in a clock-wise and counter clock-wise direction,
Buffered HRing has to send flits to one direction in order to avoid deadlock
while deflections in HiRD allow flits to travel in both directions, leading to
better overall network utilization. Second, at high injection rates, the
transfer guarantee~\cite{hird} starts throttling the network, disallowing
future flits to be injected into the network until the existing flits arrive at
their destinations. This reduces congestion in the network and allows HiRD to
saturate at a higher injection rate than the buffered hierarchical ring design.

\subsection{Injection and Transfer Guarantees}
\label{sec:eval-guarantees}

In this subsection, we study HiRD's behavior under a worst-case
synthetic traffic pattern that triggers the injection and transfer
guarantees and demonstrates that they are necessary for correct
operation, and that they work as designed.

\noindent\textbf{Traffic Pattern.} In the worst-case traffic pattern, all nodes
on three rings in a two-level (16-node) hierarchy inject traffic (we call these
rings Ring A, Ring B, and Ring C). Rings A, B, and C have bridge routers
adjacent to each other, in that order, on the single global ring. All nodes in
Ring A continuously inject flits 
to nodes in Ring C, and all
nodes in Ring C likewise inject flits to nodes in Ring A. This creates heavy
traffic on the global ring across the point at which Ring B's bridge router
connects. All nodes on Ring B continuously inject flits (whenever they are
able) addressed to another ring elsewhere in the network. However, because
Rings A and C continuously inject flits, Ring B's bridge router will not be
able to transfer any flits to the global ring in the steady state (unless
another mechanism such as the throttling mechanism in~\cite{hird} intercedes).

\noindent\textbf{Results.} Table~\ref{table:guarantee-results} shows
three pertinent metrics on the network running the described traffic
pattern: average network throughput (flits/node/cycle) for nodes on
Rings A, B, and C, the maximum time (in cycles) spent by any one flit
at the head of a transfer FIFO, and the maximum number of times any
flit is deflected and has to circle a ring to try again. These metrics
are reported with the injection and transfer guarantee mechanisms
disabled and enabled. The experiment is run with the synthetic traffic
pattern for 300K cycles.

The results show that \emph{without} the injection and transfer guarantees,
Ring B is completely starved and cannot transfer any flits onto the
global ring. This is confirmed by the maximum transfer FIFO wait time,
which is almost the entire length of the simulation. In other words,
once steady state is reached, no flit ever transfers out of Ring
B. Once the transfer FIFO in Ring B's bridge router fills, the local
ring fills with more flits awaiting a transfer, and these flits are
continuously deflected. Hence, the maximum deflection count is very
high.  Without the injection or transfer guarantees, the network does
\emph{not} ensure forward progress for these flits. In contrast, when
the injection and transfer guarantees are enabled, (i) Ring B's bridge
router is able to inject flits into the global ring and (ii) Ring B's
bridge router fairly picks flits from its local ring to place into its
transfer FIFO. The maximum transfer FIFO wait time and maximum
deflection count are now bounded, and nodes on all rings receive
network throughput.  Thus, the guarantees are both necessary
and sufficient to ensure deterministic forward progress for all flits
in the network.

\begin{figure*}[h!]
\centering
\includegraphics[width=5in]{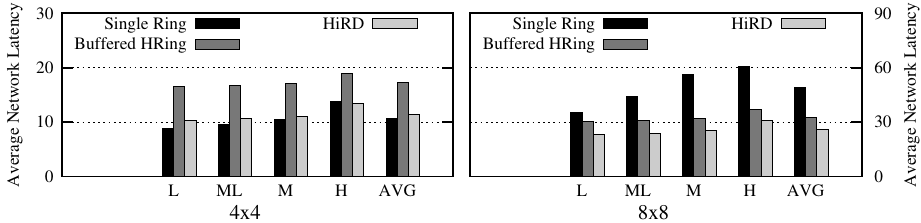}

\caption{Average network latency for 4x4 and 8x8 networks. \ch{Reproduced from~\cite{hird}.}}

\label{fig:avg_Latency}
\end{figure*}

\begin{figure*}[h!]
\centering
\includegraphics[width=5in]{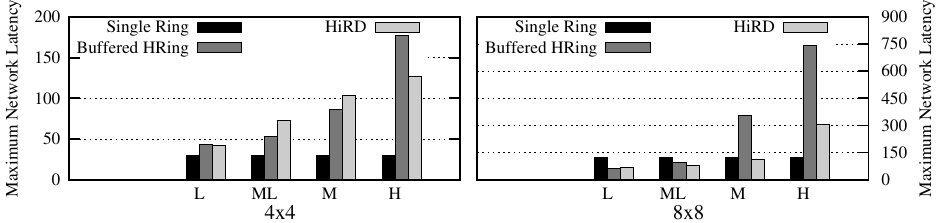}
\caption{Maximum network latency for 4x4 and 8x8 networks. \ch{Reproduced from~\cite{hird}.}}

\label{fig:max_Latency}
\end{figure*}

\begin{figure*}[h!]
\centering

\includegraphics[width=5in]{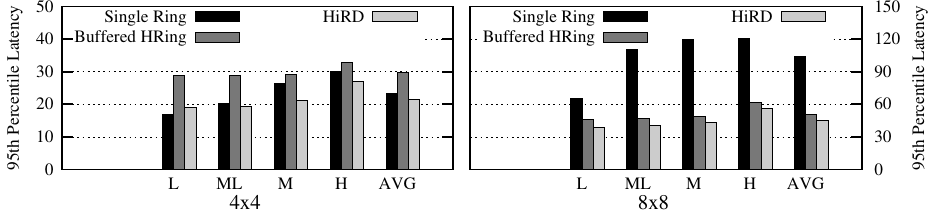}
\caption{95th percentile latency for 4x4 and 8x8 networks. \ch{Reproduced from~\cite{hird-safari-tr}.}}
\label{fig:95th_Latency}
\end{figure*}

%


\noindent\textbf{Real Applications.} Table~\ref{table:guarantee-real-results}
shows the effect of the transfer guarantee mechanism on real applications in a
4x4 network.  Average transfer FIFO wait time shows the average number of cycles
that a flit waits in the transfer FIFO across all 60 workloads.
Maximum transfer FIFO wait time shows the maximum observed flit wait time in the same FIFO across all
workloads.  As illustrated in Table~\ref{table:guarantee-real-results}, some number of
flits can experience very high wait times when there is no transfer guarantee.
Our transfer guarantee mechanism reduces both average and maximum FIFO
wait times\footnote{As the network scales to 64 nodes, we observe that the average
wait time in the transfer FIFO does not affect the overall performance significantly
(adding 1.5 cycles per flit).}. In addition, we observe that our transfer
guarantee mechanism not only provides livelock- and deadlock-freedom but also
provides lower maximum wait time in the transfer FIFO for each flit because
the guarantee provides a form of throttling when the network is congested. A
similar observation has been made in many previous network-on-chip works
that use source throttling to improve the performance of the
network~\cite{selftuned,baydal05,hotnets2010,hat-sbac-pad,sigcomm12}.



We conclude that our transfer guarantee mechanism is effective in eliminating
livelock and deadlock as well as reducing packet queuing delays in real workloads.

\subsection{Network Latency and Latency Distribution}



Figure~\ref{fig:avg_Latency} shows average network latency for our three
evaluated configurations: 256-bit single ring, buffered hierarchical ring and
HiRD.  This plot shows that our proposal can reduce the network latency by
having a faster local-ring hop latency compared to other ring-based designs.
Additionally, we found that, for all real workloads, the number of deflections
we observed is always less than 3\% of the total number of flits. Therefore,
the benefit of our deflection based router design outweighs the extra cost of
deflections compared to other ring-based router designs. Finally, in the
case of small networks such as a 4x4 network, a 1-cycle hop latency of a single
ring provides significant latency reduction compared to the buffered
hierarchical design. However, a faster local-ring hop latency in HiRD helps to
reduce the network latency of a hierarchical design and provides a
competitive network latency compared to a single ring design in small
networks.


In addition, Figure~\ref{fig:max_Latency} shows the maximum latency 
and Figure~\ref{fig:95th_Latency} shows the 95th percentile
latency for each network design. The 95th percentile latency shows the behavior of the network without
extreme outliers. These two figures provide quantitative evidence
that the network is deadlock-free and livelock-free. Several
observations are in order:


1. HiRD provides lower latency at the 95th percentile and the lowest average
latency observed in the network. This lower latency comes from our transfer
guarantee mechanism, which is triggered when flits spend more than 100 cycles in
each local ring, draining all flits in the network to their destination. This
also means that HiRD improves the worst-case latency that a flit can
experience because none of the flits are severely delayed.

2. While both HiRD and the buffered hierarchical ring have higher 95th
percentile and maximum flit latency compared to a 64-bit single ring network,
both hierarchical designs have 60.1\% (buffered hierarchical ring) and 53.9\%
(HiRD) lower average network latency in an 8x8 network because a hierarchical design
provides better scalability on average. 

3. Maximum latency in the single ring is low because contention happens only at
injection and ejection, as opposed to hierarchical designs where contention can
also happen when flits travel through different level of the hierarchy.

4. The transfer guarantee in HiRD also helps to significantly reduce the maximum
latency observed by some flits compared to a buffered design because the
guarantee enables the throttling of the network, thereby alleviating congestion.
Reduced congestion leads to reduced maximum latency. This observation is confirmed by our synthetic
traffic results shown in Section~\ref{sec:eval-synth}.

\begin{figure*}[h!]
\centering
\includegraphics[width=5in]{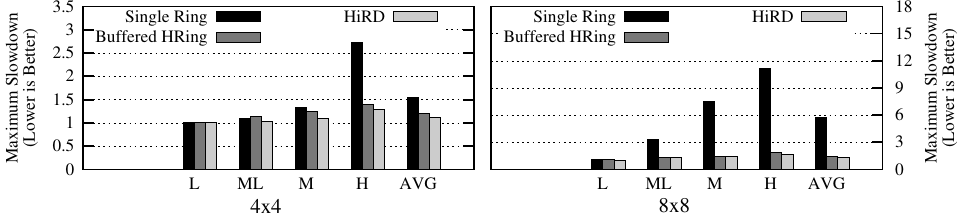}
\caption{Unfairness for 4x4 and 8x8 networks. \ch{Reproduced from~\cite{hird-safari-tr}.}}
\label{fig:hird-fairness}
\end{figure*}


\subsection{Fairness}

Figure~\ref{fig:hird-fairness} shows the fairness, measured by the maximum
slowdown metric, for our three evaluated configurations. Compared to a buffered
hierarchical ring design HiRD, is 8.3\% (5.1\%) more fair on a 4x4 (8x8)
network. Compared to a single ring design, HiRD is 40.0\% (296.4\%) more fair
on a 4x4 (8x8) network. In addition, we provide several observations:



1. HiRD is the most fair design compared to the buffered hierarchical ring and
the single ring designs. Compared to a single ring design, hierarchical designs
are more fair because the global ring in the hierarchical designs allows flits
to arrive at the destination faster. Compared to the buffered hierarchical ring
design, HiRD is more fair because HiRD has lower average network latency. HiRD
is much more fair for medium and high intensity workloads, where the throttling
mechanism in HiRD lowers average network latency. 

2. Global rings allow both hierarchical designs to provide better fairness
compared to the single ring design as the size of the network gets bigger from
4x4 to 8x8. 

3. We conclude that HiRD is the most fair ring design among all evaluated
designs due to its overall lower packet latencies and reduced congestion across
all applications.

\begin{figure*}[h!]
\centering
\includegraphics[width=5in]{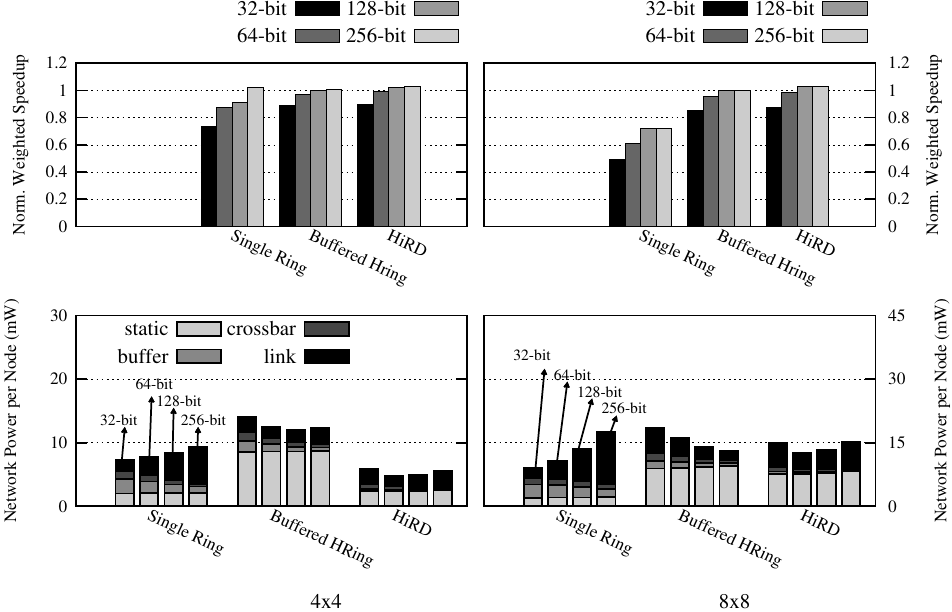}
\caption{Sensitivity to different link bandwidth for 4x4 and 8x8 networks. \ch{Reproduced from~\cite{hird-safari-tr}.}}
\vspace{-.1in}
\label{fig:bw_sweep}
\end{figure*}

\subsection{Router Area and Timing}
\label{sec:eval-area}

We show both critical path length and normalized die area for single-ring,
buffered hierarchical ring, and HiRD, in Table~\ref{table:timing_area}. Area
results are normalized to the buffered hierarchical ring baseline, and are
reported for all routers required by a 16-node network (e.g., for HiRD, 16 node
routers and 8 bridge routers).

\linespread{1}
\begin{table}[h!]
\centering
\vspace{-0.1in}
\footnotesize{
\begin{tabular}{|p{1.9cm}||p{15mm}|p{15mm}|p{10mm}|}
\hline
Metric & Single-Ring & Buffered & HiRD \\
 &  &  HRing &  \\
\hline
\hline
Critical path (ns) & 0.33 & 0.87 & 0.61 \\
\hline
Normalized area & 0.281 & 1 & 0.497 \\
\hline
\end{tabular}
}
\vspace{-0.05in}
\caption{Total router area (16-node network) and critical path. Data reproduced from~\cite{hird-safari-tr}.}
\label{table:timing_area}
\vspace{-0.15in}
\end{table}

\linespread{0.85}

\begin{figure*}[h]
\centering
\subfloat[\scriptsize Number of bridge routers]{
\includegraphics[width=1.25in]{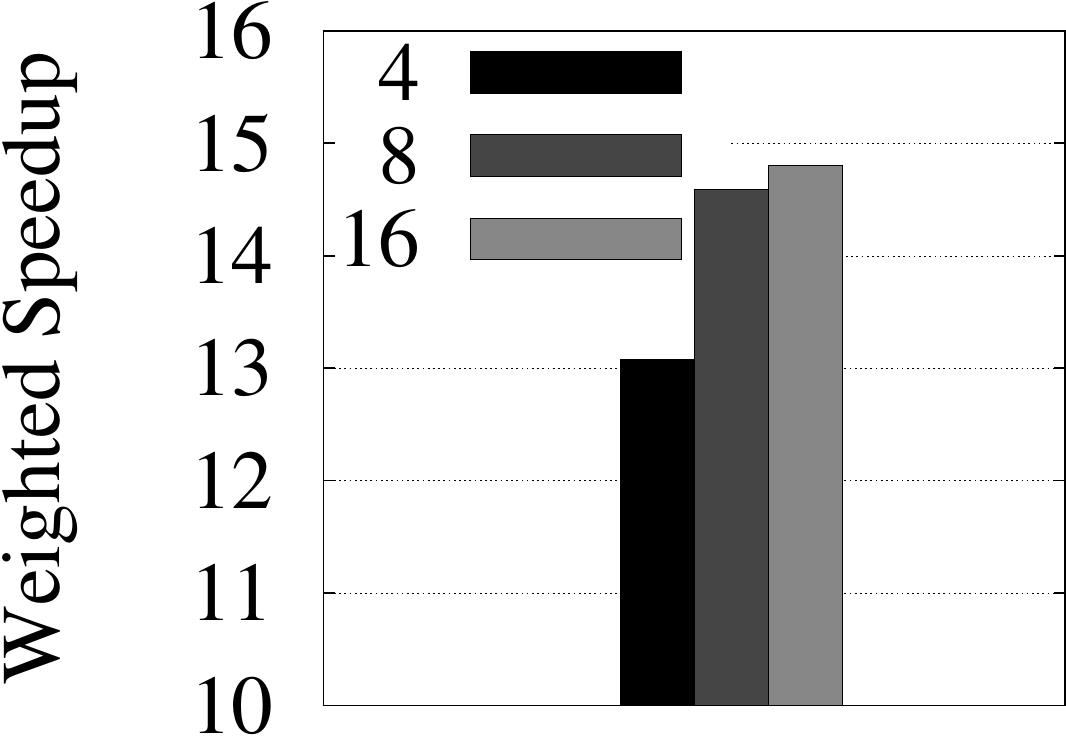}
\label{fig:bridges}
}
\subfloat[\scriptsize Local-to-global bridge buffer]{
\includegraphics[width=1.25in]{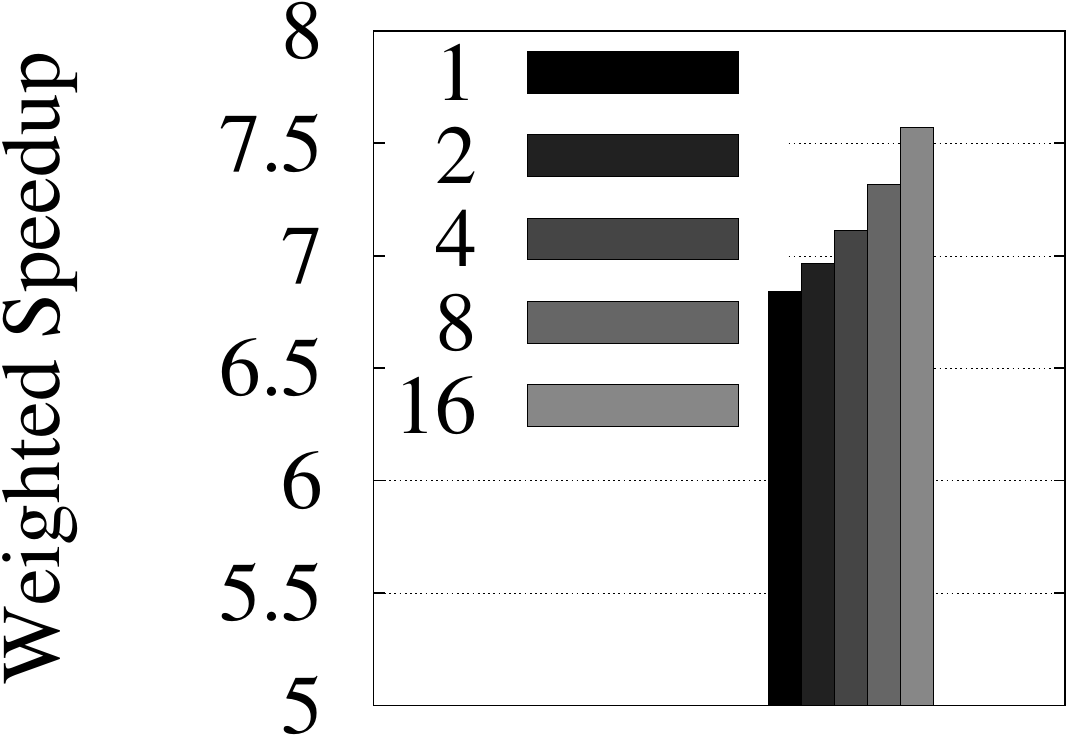}
\label{fig:L2GBuf}
}
\subfloat[\scriptsize Global-to-local bridge buffer]{
\includegraphics[width=1.25in]{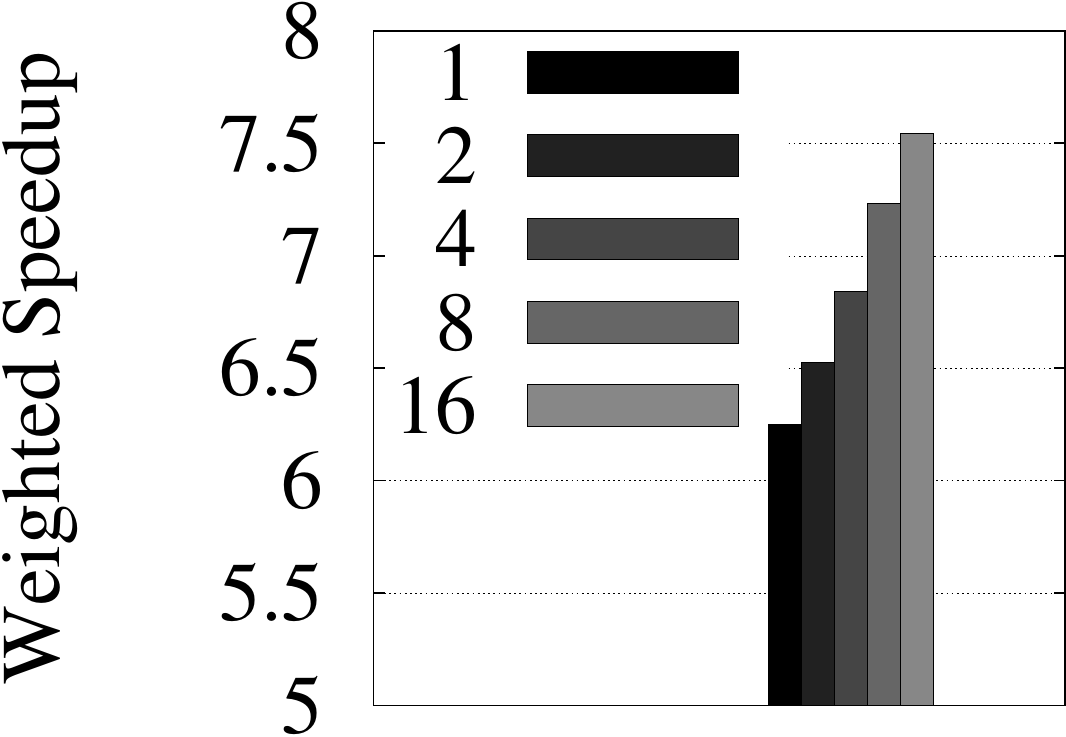}
\label{fig:G2LBuf}
}
\subfloat[\scriptsize Global-Local ring B/W ratio]{
\includegraphics[width=1.25in]{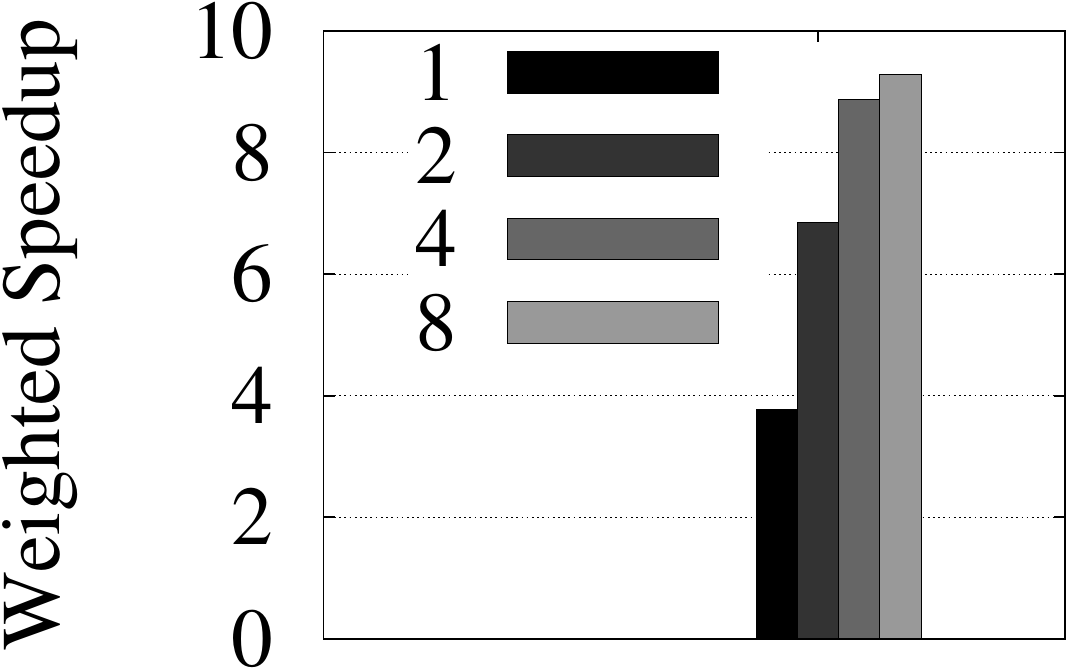}
\label{fig:GWidth}
}
\caption{Performance sensitivity to buffer sizes and the global ring
  bandwidth in a 4x4 network. \ch{Reproduced from~\cite{hird-safari-tr}.}}
\end{figure*}

Two observations are in order. First, HiRD reduces area relative to the
buffered hierarchical ring routers, because the node router required at each
network node is much simpler and does not require complex flow control logic.
HiRD reduces total router area by 50.3\% vs. the buffered hierarchical ring.
Its area is higher than a single ring router because it contains buffers in
bridge routers. However, the energy efficiency of HiRD and its performance at
high load make up for this shortcoming. Second, the buffered
hierarchical ring router's critical path is 42.6\% longer than HiRD because its control
logic must also handle flow control (it must check whether credits are
available for a downstream buffer).  The single-ring network has a higher
operating frequency than HiRD because it does not need to accommodate ring
transfers (but recall that this simplicity comes at the cost of poor
performance at high load for the single ring).

\subsection{Sensitivity to Link Bandwidth}


The bandwidth of each link also has an effect on the performance of different
network designs. We evaluate the effect of different link
bandwidths on several ring-based networks by using 32-, 64- and 128-bit links on all
network designs. Figure~\ref{fig:bw_sweep} shows the performance and power
consumption of each network design. As links get
wider, the performance of each design increases.  According to the evaluation results, HiRD performs
slightly better than a buffered hierarchical ring design for almost all link
bandwidths while maintaining much lower power consumption on a 4x4 network, and
slightly lower power consumption on an 8x8 network. 


Additionally, we observe that increasing link bandwidth can decrease the network
power in a hierarchical design because lower link bandwidth causes more
congestion in the network and leads to more dynamic buffer, crossbar and link power
consumption due to additional deflections at the buffers. 
As the link bandwidth increases, congestion reduces, lowering
dynamic power. However, we observe that past a certain link bandwidth (e.g., 128 bits for
buffered hierarchical ring and HiRD), congestion no longer reduces, because
deflections at the buffers become the bottleneck instead. This leads to
diminishing returns in performance yet increased dynamic power.

\subsection{Sensitivity to Configuration Parameters}
\label{sec:eval-sensitivity}

\noindent\textbf{Bridge Router Organization.} The number of bridge
routers connecting the global ring(s) to the local rings has an
important effect on system performance because the connection between
local and global rings can limit bisection bandwidth. In
Figure~\ref{fig:topology}, we showed three alternative arrangements for
a 16-node network, with 4, 8, and 16 bridge routers. So far, we have
assumed an 8-bridge design in 4x4-node systems, and a system with 8
bridge routers at each level in 8x8-node networks
(Figure~\ref{fig:scale}). In Figure~\ref{fig:bridges}, we show average
performance across all workloads for a 4x4-node system with 4, 8, and
16 bridge routers. Buffer capacity is held constant. As shown,
significant performance is lost if only 4 bridge routers are used
(10.4\% on average). However, doubling from 8 to 16 bridge routers
gains only 1.4\% performance on average. Thus, the 8-bridge design
provides the best tradeoff of performance and network cost (power and
area) overall in our evaluations.



\noindent\textbf{Bridge Router Buffer Size.} The size of the FIFO queues used
to transfer flits between local and global rings can have an impact on
performance if they are too small (and hence are often full, leading them to
deflect transferring flits) or too large (and hence increase bridge router
power and die area). We show the effect of local-to-global and global-to-local
FIFO sizes in Figures~\ref{fig:L2GBuf}~and~\ref{fig:G2LBuf}, respectively, for
the 8-bridge 4x4-node design. In both cases, increased buffer size leads to
increased performance. However, performance is more sensitive to
global-to-local buffer size (20.7\% gain from 1-flit to 16-flit buffer size)
than to local-to-global size (10.7\% performance gain from 1 to 16 flits),
because in the 8-bridge configuration, the whole-loop latency around the global
ring is slightly higher than the loop latency in each of the local ring, making
a global-to-local transfer retry more expensive than a local-to-global one.


For our evaluations, we use a 4-flit global-to-local and 1-flit
local-to-global buffer per bridge router, which results in transfer deflection
rates of 28.2\% (global-to-local) and 34\% (local-to-global) on average for
multiprogrammed workloads. These deflection rates are less than 1\% for
all of our multithreaded workloads. The deflection rate is much lower in multithreaded
workloads because these workloads are less memory-intensive and hence the contention
in the on-chip interconnect is low for them.

\begin{figure*}[h!]
\vspace{-.35in}
\centering
\hbox{
\includegraphics[width=5.5in]{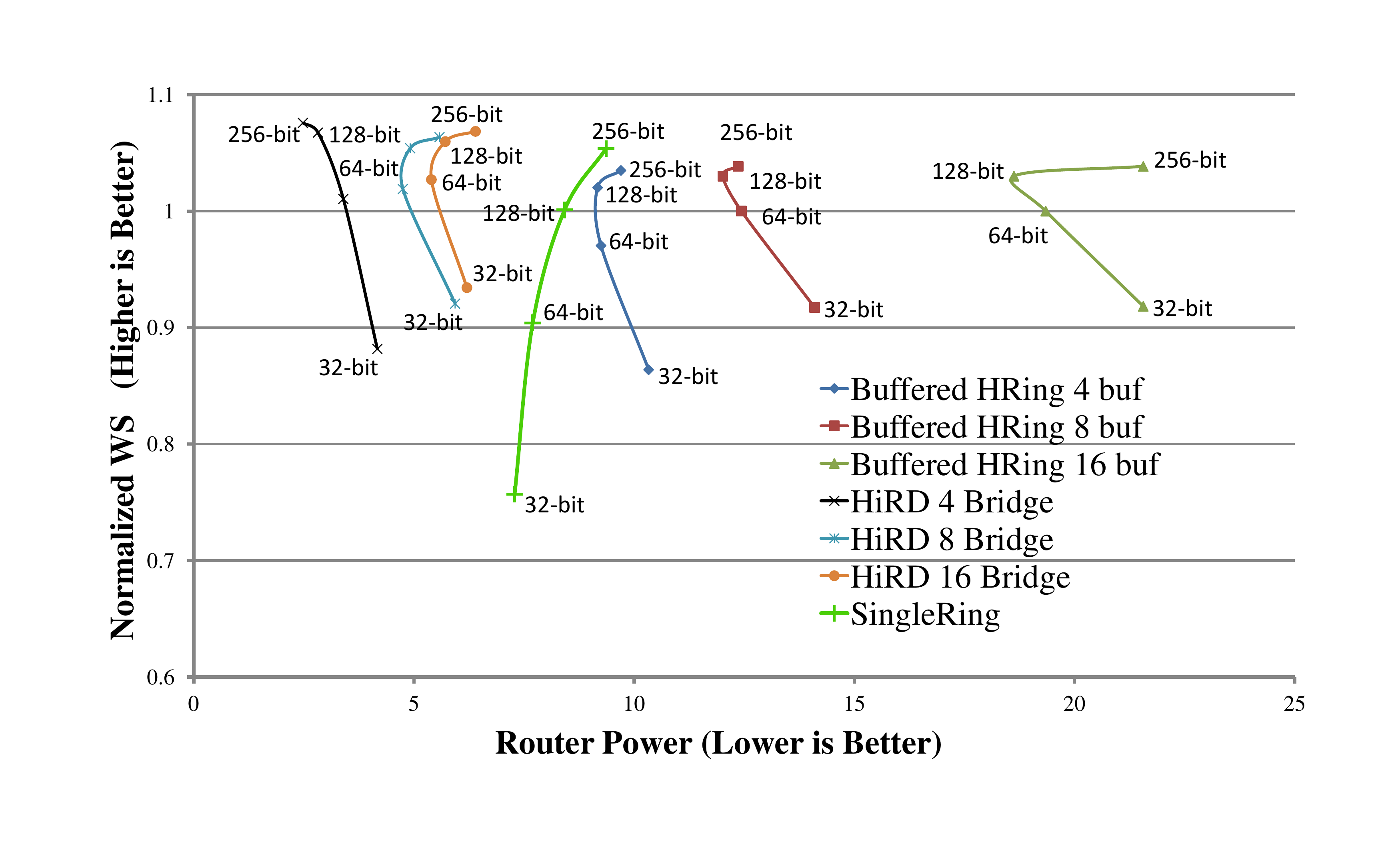}
}
\vspace{-.3in}
\caption{Weighted speedup (Y) vs. power (X) for 4x4 networks. \ch{Reproduced from~\cite{hird-safari-tr}.}}
\label{fig:WS_power}
\end{figure*}


\noindent\textbf{Global Ring Bandwidth.} Previous work on
hierarchical ring designs does not examine the impact of global ring
bandwidth on performance but instead assume equal bandwidth in
local and global rings~\cite{ravindran98}. In Figure~\ref{fig:GWidth},
we examine the sensitivity of system performance to global ring
bandwidth relative to local ring bandwidth, for the all-High category
of workloads (in order to stress check bisection bandwidth). Each point in
the plot is described by this global-to-local ring bandwidth ratio. The local ring design
is held constant while the width of the global ring is adjusted. If a
ratio of 1:1 is assumed (leftmost bar), performance is significantly
worse than the best possible design. Our main evaluations in 4x4
networks use a ratio of 2:1 (global:local) in order to provide
equivalent bisection bandwidth to a 4x4 mesh baseline. Performance
increases by 81.3\% from a 1:1 ratio to the 2:1 ratio that we
use. After a certain point, the global ring becomes less of a
bottleneck, and further global-ring bandwidth increases have massively smaller
effects. 

\noindent\textbf{Delivery Guarantee Parameters.}
We introduced injection guarantee and ejection guarantee mechanisms to ensure
every flit is eventually delivered to its destination. These guarantees are clearly described in detail in our original work~\cite{hird}. The
injection guarantee mechanism takes a threshold parameter that specifies how
long an injection can be blocked before action is taken.  Setting this
parameter too low can have an adverse impact on performance, because the system
throttles nodes too aggressively and thus underutilizes the network. Our main
evaluations use a $100$-cycle threshold. For high-intensity workloads,
performance drops by 21.3\% when using an aggressive threshold of only $1$ cycle. From $10$
cycles upward, variation in performance is at most 0.6\%: the mechanism is
invoked rarely enough that the exact threshold does not matter, only that it is
finite (is required for correctness guarantees). In fact, for a 100-cycle threshold, the injection
guarantee mechanism is \emph{never} triggered in our real applications. Hence, the
mechanism is necessary only for corner-case correctness. In addition, we
evaluate the impact of communication latency between routers and the
coordinator. We find less than 0.1\% variation in performance for latencies
ranging from $1$ to $30$ cycles (when parameters are set so that the mechanism becomes
active); thus, slow, low-cost wires may be used for this mechanism.

The ejection guarantee takes a single threshold parameter: the number
of times a flit is allowed to circle around a ring before action is
taken. We find less than 0.4\% variation in performance when sweeping the
threshold from $1$ to $16$.  Thus, the mechanism provides correctness
in corner cases but is unimportant for performance in the common case.

\subsection{Comparison Against Other Ring Configurations}
\label{sec:eval-ws-power}

Figure~\ref{fig:WS_power} highlights the energy-efficiency comparison of
different ring-based design configurations by showing weighted speedup (Y axis)
against power (X axis) for all evaluated 4x4 networks. HiRD is shown with the
three different bridge-router configurations (described in
\S\ref{sec:bridge_router}). Every ring design is evaluated at various
link bandwidths (32-, 64-, 128- and 256-bit links). The top-left is the ideal
corner (high performance, low power). As the results show, at the same link
bandwidth, all three configurations of HiRD are more energy efficient than the
evaluated buffered hierarchical ring baseline designs at this network size.

\begin{table*}[h!]
\small
\centering
\begin{tabular}{|l||l|l||l|l|}
\hline
Topologies & \multicolumn{2}{c||}{4x4} & \multicolumn{2}{|c|}{8x8} \\
\hline
 & Norm. WS & Power (mWatts) & Norm. WS & Power (mWatts) \\
\hline
\hline
Single Ring & 0.904 & 7.696 & 0.782 & 13.603 \\
\hline
Buffered HRing & 1 & 12.433 & 1 & 16.188 \\
\hline
Buffered Mesh & 1.025 & 11.947 & 1.091 & 13.454 \\
\hline
CHIPPER & 0.986 & 4.631 & 1.013 & 7.275 \\
\hline
Flattened Butterfly & 1.037 & 10.760 & 1.211 & 30.434 \\
\hline
HiRD & 1.020 & 4.746 & 1.066 & 12.480 \\
\hline
\end{tabular}
\caption{Evaluation for 4x4 and 8x8 networks against different network designs. Data reproduced from~\cite{hird-safari-tr}.}
\label{table:results-topo}
\end{table*}

We also observe that increasing link bandwidth can sometimes decrease router
power as it reduces deflections in HiRD or lowers contention at the buffers in
a buffered hierarchical ring design. However, once links are wide enough, this
benefit diminishes for two reasons: 1) 
links and crossbars consume more energy, 2)
packets arrive at the destination faster, leading to higher power as more energy
is consumed in less time.

\subsection{Comparison Against Other Network Designs}
\label{sec:eval-topos}

For completeness, Table~\ref{table:results-topo} compares HiRD against several other network
designs on 4x4 and 8x8 networks using the multiprogrammed workloads described in
Section~\ref{sec:meth}.

We compare our mechanism against a buffered mesh design with buffer bypassing~\cite{wang03,michelog10}.
We configure the buffered mesh to have 4 virtual channels (VCs) per port with 8
buffers per VC.  We also compare our mechanism against CHIPPER~\cite{chipper},
a low-complexity bufferless mesh network. We use 128-bit links for both designs.
Additionally, we compare our mechanism against a flattened
butterfly~\cite{flattened_bfly} with 4 VCs per output port, 8 buffers per VC,
and 64-bit links. Our main conclusions are as follows:

\indent 1. Against designs using the mesh topology, we observe that HiRD performs very
closely to the buffered mesh design both for 4x4 and 8x8 network sizes, while a
buffered hierarchical ring design performs slightly worse compared to HiRD and
buffered mesh designs. Additionally, HiRD performs better than CHIPPER in both
4x4 and 8x8 networks, though CHIPPER consumes less power in an 8x8 design as there
is no buffer in CHIPPER. 

2. Compared to a flattened butterfly design, we observe that HiRD performs
competitively with a flattened butterfly in a 4x4 network, but consumes lower
router power. In an 8x8 network, HiRD does not scale as well as a flattened
butterfly network and performs 11\% worse than a flattened butterfly network;
however, HiRD consumes 59\% less power than the flattened butterfly design. 

3. Overall, we conclude that HiRD is competitive in performance with the
highest performing designs while having much lower power consumption.

%% file: main.bbl
\begin{thebibliography}{100}
\expandafter\ifx\csname url\endcsname\relax
  \def\url#1{\texttt{#1}}\fi
\expandafter\ifx\csname urlprefix\endcsname\relax\def\urlprefix{URL }\fi
\expandafter\ifx\csname href\endcsname\relax
  \def\href#1#2{#2} \def\path#1{#1}\fi

\bibitem{principles}
W.~Dally, B.~Towles, Principles and Practices of Interconnection Networks,
  {Morgan Kaufmann}, 2004.

\bibitem{casebufferless}
T.~Moscibroda, O.~Mutlu, A case for bufferless routing in on-chip networks,
  {ISCA} (2009).

\bibitem{minbd}
C.~Fallin, et~al., {MinBD}: Minimally-buffered deflection routing for
  energy-efficient interconnect, NOCS (2012).

\bibitem{chipper}
C.~Fallin, et~al., {CHIPPER}: A low-complexity bufferless deflection router,
  {HPCA} (2011).

\bibitem{hat-sbac-pad}
K.~K.-W. Chang, R.~Ausavarungnirun, C.~Fallin, O.~Mutlu, {HAT: Heterogeneous
  Adaptive Throttling for On-Chip Networks}, in: SBAC-PAD, 2012.

\bibitem{carpool}
X.~Xiang, W.~Shi, S.~Ghose, L.~Peng, O.~Mutlu, N.-F. Tzeng, {Carpool: A
  Bufferless on-Chip Network Supporting Adaptive Multicast and Hotspot
  Alleviation}, in: ICS, 2017.

\bibitem{xiyue-iccd16}
X.~Xiang, S.~Ghose, O.~Mutlu, N.-F. Tzeng, {A model for Application Slowdown
  Estimation in on-chip networks and its use for improving system fairness and
  performance}, in: ICCD, 2016.

\bibitem{hotnets2010}
G.~P. Nychis, et~al., Next generation on-chip networks: What kind of congestion
  control do we need?, Hotnets (2010).

\bibitem{sigcomm12}
G.~P. Nychis, et~al., On-chip networks from a networking perspective:
  Congestion and scalability in many-core interconnects, SIGCOMM (2012).

\bibitem{grot09}
B.~Grot, et~al., Express cube topologies for on-chip interconnects, HPCA
  (2009).

\bibitem{pvc}
B.~Grot, et~al., {Preemptive Virtual Clock: A Flexible, Efficient, and
  Cost-effective QOS Scheme for Networks-on-Chip}, MICRO (2009).

\bibitem{grot2010}
B.~Grot, et~al., Topology-aware quality-of-service support in highly integrated
  chip multiprocessors, in: WIOSCA, 2010.

\bibitem{kilonocs}
B.~Grot, et~al., {Kilo-NOC}: A heterogeneous network-on-chip architecture for
  scalability and service guarantees, ISCA (2011).

\bibitem{Nicopoulos06}
C.~Nicopoulos, et~al., {ViChaR}: A dynamic virtual channel regulator for
  on-chip networks, MICRO (2006).

\bibitem{stc}
R.~Das, et~al., Application-aware prioritization mechanisms for on-chip
  networks, MICRO (2009).

\bibitem{aergia}
R.~Das, et~al., A{\'e}rgia: exploiting packet latency slack in on-chip
  networks, ISCA (2010).

\bibitem{a2c}
R.~Das, et~al., {Application-to-core mapping policies to reduce memory system
  interference in multi-core systems}, in: HPCA, 2013.

\bibitem{flattened_bfly}
J.~Kim, W.~Dally, Flattened butterfly: A cost-efficient topology for high-radix
  networks, ISCA (2007).

\bibitem{tera}
R.~Alverson, et~al., The {Tera} computer system, ICS (1990).

\bibitem{taylor02}
M.~Taylor, J.~Kim, J.~Miller, D.~Wentzlaff, The raw microprocessor: A
  computational fabric for software circuits and general-purpose programs, IEEE
  Micro (Mar 2002).

\bibitem{intel-scc}
{Intel Corporation}, Single-chip cloud computer,
  \url{http://techresearch.intel.com/articles/Tera-Scale/1826.htm}.

\bibitem{slimnoc}
M.~Besta, S.~M. Hassan, S.~Yalamanchili, R.~Ausavarungnirun, O.~Mutlu,
  T.~Hoefler, {Slim NoC: A Low-Diameter On-Chip Network Topology for High
  Energy Efficiency and Scalability}, in: ASPLOS, 2018.

\bibitem{hird}
R.~Ausavarungnirun, et~al., Design and evaluation of hierarchical rings with
  deflection routing, in: SBAC-PAD, 2014.

\bibitem{hird-parco}
R.~Ausavarungnirun, C.~Fallin, X.~Yu, K.~K.-W. Chang, G.~Nazario, R.~Das, G.~H.
  Loh, O.~Mutlu, {A Case for Hierarchical Rings with Deflection Routing},
  Parallel Comput. (May 2016) 54~(C) (2016) 29–45.

\bibitem{fattah-maze}
M.~Fattah, A.~Airola, R.~Ausavarungnirun, N.~Mirzaei, P.~Liljeberg, J.~Plosila,
  S.~Mohammadi, T.~Pahikkala, O.~Mutlu, H.~Tenhunen, {A Low-Overhead,
  Fully-Distributed, Guaranteed-Delivery Routing Algorithm for Faulty
  Network-on-Chips}, in: NOCS, 2015.

\bibitem{gratz-iccd06}
P.~{Gratz}, C.~{Kim}, R.~{McDonald}, S.~W. {Keckler}, D.~{Burger},
  {Implementation and Evaluation of On-Chip Network Architectures}, in: ICCD,
  2006.

\bibitem{michelog10}
G.~Michelogiannakis, et~al., Evaluating bufferless flow-control for on-chip
  networks, NOCS (2010).

\bibitem{wang03}
H.~Wang, et~al., Power-driven design of router microarchitectures in on-chip
  networks, MICRO (2003).

\bibitem{hotpotato}
P.~Baran, On distributed communications networks, {IEEE Trans. on Comm.}
  (1964).

\bibitem{borkar12}
S.~Borkar, {NoCs}: What's the point?, NSF Workshop on Emerging Tech. for
  Interconnects (WETI), Feb. 2012.

\bibitem{mesh5ghz}
Y.~Hoskote, et~al., A 5-{GHz} mesh interconnect for a teraflops processor, IEEE
  Micro (2007).

\bibitem{gratz06}
P.~Gratz, C.~Kim, R.~McDonald, S.~Keckler, Implementation and evaluation of
  on-chip network architectures, ICCD (2006).

\bibitem{tile64}
D.~Wentzlaff, et~al., On-chip interconnection architecture of the tile
  processor, IEEE Micro (2007) 27~(5) (2007) 15--31.

\bibitem{tile100}
{Tilera Corporation}, Tilera announces the world's first 100-core processor
  with the new {TILE-Gx} family,
  \url{http://www.tilera.com/news_&_events/press_release_091026.php}.

\bibitem{intel-skylake}
{6th Generation Intel Core Processor Family Datasheet},
  \url{http://www.intel.com/content/www/us/en/processors/core/desktop-6th-gen-core-family-datasheet-vol-1.html}.

\bibitem{intel-cascade-lake}
{Second Generation Intel Xeon Scalable Datasheet},
  \url{https://www.intel.com/content/www/us/en/products/docs/processors/xeon/2nd-gen-xeon-scalable-datasheet-vol-1.html}.

\bibitem{intel-ice-lake}
{10th Generation Intel Core Processor Families Datasheet},
  \url{https://www.intel.com/content/dam/www/public/us/en/documents/datasheets/10th-gen-core-families-datasheet-vol-1-datasheet.pdf}.

\bibitem{larrabee}
L.~Seiler, et~al., Larrabee: a many-core x86 architecture for visual computing,
  SIGGRAPH (2008).

\bibitem{cell}
D.~Pham, et~al., Overview of the architecture, circuit design, and physical
  implementation of a first-generation {CELL} processor, JSSC (2006).

\bibitem{sandybridge}
{Intel Corporation}, Intel details 2011 processor features, \newline
  \url{http://newsroom.intel.com/community/intel_newsroom/blog/2010/09/13/intel-details-2011-processor-features-offers-stunning-visuals-built-in}
  (2011).

\bibitem{intel-coffee-lake}
{8th and 9th Generation Intel Core Processor Family and Intel Xeon E Processor
  Families Datasheet},
  \url{https://www.intel.com/content/dam/www/public/us/en/documents/datasheets/8th-gen-core-family-datasheet-vol-1.pdf}.

\bibitem{dgx1}
{NVIDIA DGX-1 With Tesla V100 System Architecture},
  \url{https://images.nvidia.com/content/pdf/dgx1-v100-system-architecture-whitepaper.pdf}.

\bibitem{kim09nocarc}
J.~Kim, H.~Kim, Router microarchitecture and scalability of ring topology in
  on-chip networks, {NoCArc} (2009).

\bibitem{farkas-sc92}
K.~Farkas, Z.~Vranesic, M.~Stumm, {Cache Consistency in Hierarchical-Ring-Based
  Multiprocessors}, in: SC, 1992.

\bibitem{hector}
Z.~G. Vranesic, M.~Stumm, D.~M. Lewis, R.~White,
  \href{https://doi.org/10.1109/2.67196}{{Hector: A Hierarchically Structured
  Shared-Memory Multiprocessor}}, Computer (Jan. 1991) 24~(1) (1991) 72–79.
\newblock \href {http://dx.doi.org/10.1109/2.67196}
  {\path{doi:10.1109/2.67196}}.
\newline\urlprefix\url{https://doi.org/10.1109/2.67196}

\bibitem{ravindran97}
G.~Ravindran, M.~Stumm, A performance comparison of hierarchical ring- and
  mesh-connected multiprocessor networks, HPCA (1997).

\bibitem{xiangdong95}
X.~Zhang, Y.~Yan, Comparative modeling and evaluation of {CC-NUMA} and {COMA}
  on hierarchical ring architectures, IEEE TPDS (1995).

\bibitem{hr-model}
V.~C. Hamacher, H.~Jiang, Hierarchical ring network configuration and
  performance modeling, IEEE Transaction on Computers (2001).

\bibitem{ravindran98}
G.~Ravindran, M.~Stumm, On topology and bisection bandwidth for
  hierarchical-ring networks for shared memory multiprocessors, HPCA (1998).

\bibitem{numachine}
R.~Grindley, et~al., The {NUMA}chine multiprocessor, ICPP (2000).

\bibitem{holliday94}
M.~Holliday, M.~Stumm, {Performance Evaluation of Hierarchical Ring-based
  Shared Memory Multiprocessors}, IEEE Transactions on Computers (1994) 43~(1)
  (1994) 52--67.
\newblock \href {http://dx.doi.org/10.1109/12.250609}
  {\path{doi:10.1109/12.250609}}.

\bibitem{hird-safari-tr}
R.~Ausavarungnirun, et~al., Improving energy efficiency of hierarchical rings
  via deflection routing, {SAFARI} Technical Report {TR-2014-002}:
  \url{http://safari.ece.cmu.edu/tr.html} (Apr 2014).

\bibitem{cm}
W.~Hillis, The Connection Machine, MIT Press, 1989.

\bibitem{hep}
B.~Smith, Architecture and applications of the {HEP} multiprocessor computer
  system, SPIE (1981).

\bibitem{scarab}
M.~Hayenga, et~al., {SCARAB}: A single cycle adaptive routing and bufferless
  network, MICRO (2009).

\bibitem{gomez08}
C.~G{\'o}mez, et~al., Reducing packet dropping in a bufferless {NoC}, EuroPar
  (2008).

\bibitem{glsvlsi}
S.~Tota, et~al., Implementation analysis of {NoC}: a {MPSoC} trace-driven
  approach, GLSVLSI (2006).

\bibitem{deflection_routing}
Z.~Lu, M.~Zhong, A.~Jantsch, Evaluation of on-chip networks using deflection
  routing, GLSVLSI (2006).

\bibitem{cai-isqed}
Y.~Cai, K.~Mai, O.~Mutlu, {Comparative evaluation of FPGA and ASIC
  implementations of bufferless and buffered routing algorithms for on-chip
  networks}, in: ISQED, 2015, pp. 475--484.
\newblock \href {http://dx.doi.org/10.1109/ISQED.2015.7085472}
  {\path{doi:10.1109/ISQED.2015.7085472}}.

\bibitem{chich01}
T.~Chich, P.~Fraigniaud, J.~Cohen, Unslotted deflection routing: a practical
  and efficient protocol for multihop optical networks, IEEE/ACM Transactions
  on Networking (2001).

\bibitem{greenberg92}
A.~Greenberg, B.~Hajek, Deflection routing in hypercube networks, IEEE Trans.
  on Comm. (1992).

\bibitem{barnoy93}
A.~Bar-Noy, P.~Raghavan, B.~Schieber, Fast deflection routing for packets and
  worms, PODC-12 (1993).

\bibitem{busch07}
C.~Busch, M.~Magdon-Ismail, M.~Mavronicolas, Efficient bufferless packet
  switching on trees and leveled networks, J. of Par. and Dist. Comp. (2007).

\bibitem{cfc-cal}
H.~Kim, et~al., Clumsy flow control for high-throughput bufferless on-chip
  networks, IEEE CAL (2013).

\bibitem{bless_switching}
C.~G{\'o}mez, et~al., {BPS:} a bufferless switching technique for {NoCs}, Wina
  (2008).

\bibitem{busch00}
C.~Busch, M.~Herlihy, R.~Wattenhofer, Hard-potato routing, STOC (2000).

\bibitem{chaosrouter}
S.~Konstantinidou, L.~Snyder, Chaos router: architecture and performance, ISCA
  (1991).

\bibitem{mshr-isca81}
D.~Kroft, {Lockup-Free Instruction Fetch/Prefetch Cache Organization}, in:
  ISCA, 1981.

\bibitem{mshr-isca94}
K.~I. Farkas, N.~P. Jouppi, {Complexity/Performance Tradeoffs with Non-Blocking
  Loads}, in: ISCA, 1994.

\bibitem{mshr-micro06}
J.~{Tuck}, L.~{Ceze}, J.~{Torrellas}, {Scalable Cache Miss Handling for High
  Memory-Level Parallelism}, in: MICRO, 2006.

\bibitem{afc}
S.~Jafri, et~al., Adaptive flow control for robust performance and energy,
  MICRO (2010).

\bibitem{minbd-tr}
C.~Fallin, et~al., {MinBD}: Minimally-buffered deflection routing for
  energy-efficient interconnect, {SAFARI} Technical Report {TR-2011-008}:
  \url{http://safari.ece.cmu.edu/tr.html} (Sep 2011).

\bibitem{minbd-book}
C.~Fallin, et~al., Bufferless and Minimally-Buffered Deflection Routing, in
  Routing Algorithms in Networks-on-Chip, Springer New York, New York, NY,
  2014, pp. 241--275.

\bibitem{udipi10}
A.~N. Udipi, et~al., Towards scalable, energy-efficient, bus-based on-chip
  networks, HPCA (2010).

\bibitem{das09}
R.~Das, et~al., Design and evaluation of hierarchical on-chip network
  topologies for next generation {CMPs}, HPCA (2009).

\bibitem{dsent}
C.~Sun, et~al., {DSENT} - a tool connecting emerging photonics with electronics
  for opto-electronic networks-on-chip modeling, NOCS (2012).

\bibitem{kroft81}
D.~Kroft, Lockup-free instruction fetch/prefetch cache organization, ISCA
  (1981).

\bibitem{hansson07}
A.~Hansson, K.~Goossens, A.~Radulescu, Avoiding message-dependent deadlock in
  network-based systems-on-chip, VLSI Design (2007).

\bibitem{kim09}
J.~Kim, Low-cost router microarchitecture for on-chip networks, MICRO (2009).

\bibitem{laudon97}
J.~Laudon, D.~Lenoski, The {SGI} {Origin}: a {ccNUMA} highly scalable server,
  ISCA (1997).

\bibitem{bubbleflow}
C.~Carri\'{o}n, et~al., A flow control mechanism to avoid message deadlock in
  k-ary n-cube networks, HiPC (1997).

\bibitem{NOCulator}
R.~Ausavarungnirun, et~al., {NOCulator}, in:
  \url{https://github.com/CMU-SAFARI/NOCulator}.

\bibitem{pin}
C.-K. Luk, et~al., Pin: building customized program analysis tools with dynamic
  instrumentation, PLDI (2005).

\bibitem{CloudCache}
H.~Lee, et~al., Cloudcache: Expanding and shrinking private caches, HPCA
  (2011).

\bibitem{ccraik-safari-tr}
C.~Craik, O.~Mutlu, Investigating the viability of bufferless {NoCs} in modern
  chip multi-processor systems, {SAFARI} Technical Report {TR-2011-004}:
  \url{http://safari.ece.cmu.edu/tr.html} (Aug 2011).

\bibitem{graphchi}
A.~Kyrola, et~al., {GraphChi}: Large-scale graph computation on just a {PC},
  2012.

\bibitem{graphlab}
Y.~Low, et~al., {GraphLab}: A new parallel framework for machine learning,
  2010.

\bibitem{twitter_rv}
H.~Kwak, et~al., {W}hat is {T}witter, a social network or a news media?, 2010.

\bibitem{weighted_speedup}
A.~Snavely, D.~M. Tullsen, Symbiotic jobscheduling for a simultaneous
  multithreaded processor, ASPLOS (2000).

\bibitem{harmonic_speedup}
S.~Eyerman, L.~Eeckhout, System-level performance metrics for multiprogram
  workloads, IEEE Micro (2008).

\bibitem{atlas}
Y.~Kim, et~al., {ATLAS: a scalable and high-performance scheduling algorithm
  for multiple memory controllers}, HPCA (2010).

\bibitem{tcm}
Y.~Kim, et~al., Thread cluster memory scheduling: Exploiting differences in
  memory access behavior, MICRO (2010).

\bibitem{vandierendonck}
H.~Vandierendonck, A.~Seznec, Fairness metrics for multi-threaded processors,
  IEEE Computer Architecture Letters (Feb 2011).

\bibitem{fst}
E.~Ebrahimi, et~al., Fairness via source throttling: a configurable and
  high-performance fairness substrate for multi-core memory systems, ASPLOS
  (2010).

\bibitem{eiman-isca11}
E.~Ebrahimi, et~al., Prefetch-aware shared resource management for multi-core
  systems, in: ISCA, 2011.

\bibitem{eiman-micro09}
E.~Ebrahimi, et~al., Coordinated control of multiple prefetchers in multi-core
  systems, in: MICRO, 2009.

\bibitem{mcp}
S.~P. Muralidhara, et~al., Reducing memory interference in multicore systems
  via application-aware memory channel partitioning, MICRO (2011).

\bibitem{asm-micro15}
L.~Subramanian, et~al., {The application slowdown model: Quantifying and
  controlling the impact of inter-application interference at shared caches and
  main memory}, in: MICRO, 2015.

\bibitem{mise-hpca13}
L.~Subramanian, et~al., {MISE: Providing performance predictability and
  improving fairness in shared main memory systems}, in: HPCA, 2013.

\bibitem{dash-taco16}
H.~Usui, et~al., {DASH: Deadline-Aware High-Performance Memory Scheduler for
  Heterogeneous Systems with Hardware Accelerators}, ACM TACO (2016).

\bibitem{sms}
R.~Ausavarungnirun, et~al., {Staged Memory Scheduling: Achieving High
  Performance and Scalability in Heterogeneous Systems}, in: ISCA, 2012.

\bibitem{bliss}
L.~Subramanian, et~al., The blacklisting memory scheduler: Achieving high
  performance and fairness at low cost, in: ICCD, 2014.

\bibitem{bliss-tpds}
L.~Subramanian, et~al., The blacklisting memory scheduler: Balancing
  performance, fairness and complexity, in: TPDS, 2016.

\bibitem{selftuned}
M.~Thottethodi, et~al., Self-tuned congestion control for multiprocessor
  networks, HPCA (2001).

\bibitem{baydal05}
E.~Baydal, et~al., A family of mechanisms for congestion control in wormhole
  networks, IEEE Trans. on Par. and Dist. Sys. (2005) 16.

\bibitem{kim-hpca14}
H.~Kim, et~al., Transportation-network-inspired network-on-chip, HPCA (2014).

\bibitem{sci}
D.~Gustavson, The scalable coherent interface and related standards projects,
  {IEEE} Micro (1992).

\bibitem{ksr}
T.~H. Dunigan, Kendall square multiprocessor: Early experiences and
  performance, in: of the Intel Paragon, ORNL/TM-12194, 1994.

\bibitem{gemini}
R.~{Alverson}, D.~{Roweth}, L.~{Kaplan}, {The Gemini System Interconnect}, in:
  HOTL, 2010.

\bibitem{balfour06}
J.~Balfour, W.~J. Dally, Design tradeoffs for tiled {CMP} on-chip networks, ICS
  (2006).

\bibitem{kilo-noc-toppick}
B.~Grot, J.~Hestness, S.~Keckler, O.~Mutlu, {A QoS-Enabled On-Die Interconnect
  Fabric for Kilo-Node Chips}, IEEE Micro (2012) 32~(3) (2012) 17--25.

\bibitem{hyperx}
J.~H. {Ahn}, N.~{Binkert}, A.~{Davis}, M.~{McLaren}, R.~S. {Schreiber},
  {HyperX: topology, routing, and packaging of efficient large-scale networks},
  in: SC, 2009.

\bibitem{slimfly}
M.~{Besta}, T.~{Hoefler}, {Slim Fly: A Cost Effective Low-Diameter Network
  Topology}, in: SC, 2014.

\bibitem{lionel-isca92-turn-model}
C.~J. Glass, L.~M. Ni, {The Turn Model for Adaptive Routing}, in: ISCA, 1992.

\bibitem{lionel-isca98-turn-model}
L.~Ni, {Retrospective: The Turn Model for Adaptive Routing}, in: ISCA, 1998.

\bibitem{mullins04}
R.~Mullins, et~al., Low-latency virtual-channel routers for on-chip networks,
  ISCA (2004).

\bibitem{rotary-router}
P.~Abad, et~al., {Rotary router}: an efficient architecture for {CMP}
  interconnection networks, ISCA (2007).

\bibitem{Kodi08}
A.~Kodi, et~al., {iDEAL}: Inter-router dual-function energy and area-efficient
  links for network-on-chip ({NoC}) architectures, ISCA (2008).

\bibitem{chiplet-isca18}
J.~{Yin}, Z.~{Lin}, O.~{Kayiran}, M.~{Poremba}, M.~{Shoaib Bin Altaf},
  N.~{Enright Jerger}, G.~H. {Loh}, {Modular Routing Design for Chiplet-Based
  Systems}, in: ISCA, 2018.

\bibitem{centaur}
R.~Hwang, T.~Kim, Y.~Kwon, M.~Rhu, {Centaur: A Chiplet-Based, Hybrid
  Sparse-Dense Accelerator for Personalized Recommendations}, in: ISCA, 2020.

\bibitem{network-on-memory}
S.~H.~S. Rezaei, M.~Modarressi, R.~Ausavarungnirun, M.~Sadrosadati, O.~Mutlu,
  M.~Daneshtalab, {NoM: Network-on-Memory for Inter-Bank Data Transfer in
  Highly-Banked Memories}, IEEE CAL (2020).

\bibitem{stringfigure}
M.~Ogleari, Y.~Yu, C.~Qian, E.~Miller, J.~Zhao, {String Figure: A Scalable and
  Elastic Memory Network Architecture}, in: HPCA, 2019.

\bibitem{memory-centric-gpu}
Y.~Kwon, M.~Rhu, {Beyond the Memory Wall: A Case for Memory-centric HPC System
  for Deep Learning}, in: MICRO, 2018.

\bibitem{ahn.tesseract.isca15}
J.~Ahn, S.~Hong, S.~Yoo, O.~Mutlu, K.~Choi, {A Scalable Processing-in-Memory
  Accelerator for Parallel Graph Processing}, in: ISCA, 2015.

\bibitem{graphp}
M.~Zhang, Y.~Zhuo, C.~Wang, M.~Gao, Y.~Wu, K.~Chen, C.~Kozyrakis, X.~Qian,
  {GraphP: Reducing Communication for PIM-Based Graph Processing with Efficient
  Data Partition}, in: HPCA, 2018.

\bibitem{mutlu2020modern}
O.~Mutlu, S.~Ghose, J.~Gómez-Luna, R.~Ausavarungnirun, A modern primer on
  processing in memory (2020).
\newblock \href {http://arxiv.org/abs/2012.03112} {\path{arXiv:2012.03112}}.

\bibitem{micpro2019}
O.~Mutlu, et~al., {Processing Data Where It Makes Sense: {E}nabling In-Memory
  Computation}, MicPro (2019).

\bibitem{zhu2013accelerating}
Q.~Zhu, T.~Graf, H.~E. Sumbul, L.~Pileggi, F.~Franchetti, {Accelerating Sparse
  Matrix-Matrix Multiplication with 3D-Stacked Logic-in-Memory Hardware}, in:
  HPEC, 2013.

\bibitem{pugsley2014ndc}
S.~H. Pugsley, J.~Jestes, H.~Zhang, R.~Balasubramonian, V.~Srinivasan,
  A.~Buyuktosunoglu, A.~Davis, F.~Li, {{NDC: Analyzing the Impact of 3D-Stacked
  Memory+Logic Devices on MapReduce Workloads}}, in: ISPASS, 2014.

\bibitem{zhang.hpdc14}
D.~P. Zhang, N.~Jayasena, A.~Lyashevsky, J.~L. Greathouse, L.~Xu,
  M.~Ignatowski, {TOP-PIM: Throughput-Oriented Programmable Processing in
  Memory}, in: HPDC, 2014.

\bibitem{farmahini2015nda}
A.~Farmahini-Farahani, J.~H. Ahn, K.~Morrow, N.~S. Kim, {NDA: Near-DRAM
  acceleration architecture leveraging commodity DRAM devices and standard
  memory modules}, in: HPCA, 2015.

\bibitem{cali2020genasm}
D.~S. Cali, G.~S. Kalsi, Z.~Bing{\"o}l, C.~Firtina, L.~Subramanian, J.~S. Kim,
  R.~Ausavarungnirun, M.~Alser, J.~Gomez-Luna, A.~Boroumand, et~al., {GenASM: A
  High-Performance, Low-Power Approximate String Matching Acceleration
  Framework for Genome Sequence Analysis}, in: MICRO, 2020.

\bibitem{ahn.pei.isca15}
J.~Ahn, S.~Yoo, O.~Mutlu, K.~Choi, {PIM-Enabled Instructions: A Low-Overhead,
  Locality-Aware Processing-in-Memory Architecture}, in: ISCA, 2015.

\bibitem{loh2013processing}
G.~H. Loh, N.~Jayasena, M.~Oskin, M.~Nutter, D.~Roberts, M.~Meswani, D.~P.
  Zhang, M.~Ignatowski, {A Processing in Memory Taxonomy and a Case for
  Studying Fixed-Function PIM}, in: WoNDP, 2013.

\bibitem{hsieh.isca16}
K.~Hsieh, E.~Ebrahimi, G.~Kim, N.~Chatterjee, M.~O'Conner, N.~Vijaykumar,
  O.~Mutlu, S.~Keckler, {Transparent Offloading and Mapping (TOM): Enabling
  Programmer-Transparent Near-Data Processing in GPU Systems}, in: ISCA, 2016.

\bibitem{impica}
K.~Hsieh, S.~Khan, N.~Vijaykumar, K.~K. Chang, A.~Boroumand, S.~Ghose,
  O.~Mutlu, {Accelerating Pointer Chasing in 3D-Stacked Memory: Challenges,
  Mechanisms, Evaluation}, in: ICCD, 2016.

\bibitem{DBLP:conf/sigmod/BabarinsaI15}
O.~O. Babarinsa, S.~Idreos, {JAFAR: Near-Data Processing for Databases}, in:
  SIGMOD, 2015.

\bibitem{DBLP:conf/IEEEpact/LeeSK15}
J.~H. Lee, J.~Sim, H.~Kim, {BSSync: Processing Near Memory for Machine Learning
  Workloads with Bounded Staleness Consistency Models}, in: PACT, 2015.

\bibitem{DBLP:conf/hpca/GaoK16}
M.~Gao, C.~Kozyrakis, {HRL: Efficient and Flexible Reconfigurable Logic for
  Near-Data Processing}, in: HPCA, 2016.

\bibitem{chi2016prime}
P.~Chi, S.~Li, C.~Xu, T.~Zhang, J.~Zhao, Y.~Liu, Y.~Wang, Y.~Xie, {PRIME: A
  Novel Processing-In-Memory Architecture for Neural Network Computation In
  ReRAM-Based Main Memory}, in: ISCA, 2016.

\bibitem{gu.isca16}
B.~Gu, A.~S. Yoon, D.-H. Bae, I.~Jo, J.~Lee, J.~Yoon, J.-U. Kang, M.~Kwon,
  C.~Yoon, S.~Cho, J.~Jeong, D.~Chang, {Biscuit: {A} Framework for Near-Data
  Processing of Big Data Workloads}, in: ISCA, 2016.

\bibitem{kim.isca16}
D.~Kim, J.~Kung, S.~Chai, S.~Yalamanchili, S.~Mukhopadhyay, {Neurocube: {A}
  Programmable Digital Neuromorphic Architecture with High-Density {3D}
  Memory}, in: ISCA, 2016.

\bibitem{asghari-moghaddam.micro16}
H.~Asghari-Moghaddam, Y.~H. Son, J.~H. Ahn, N.~S. Kim, {Chameleon: Versatile
  and Practical Near-DRAM Acceleration Architecture for Large Memory Systems},
  in: MICRO, 2016.

\bibitem{boroumand2016pim}
A.~Boroumand, S.~Ghose, M.~Patel, H.~Hassan, B.~Lucia, K.~Hsieh, K.~T. Malladi,
  H.~Zheng, O.~Mutlu, {LazyPIM: An Efficient Cache Coherence Mechanism for
  Processing-in-Memory}, CAL (2016).

\bibitem{hashemi.isca16}
M.~Hashemi, Khubaib, E.~Ebrahimi, O.~Mutlu, Y.~N. Patt, {Accelerating Dependent
  Cache Misses with an Enhanced Memory Controller}, in: ISCA, 2016.

\bibitem{gao.pact15}
M.~Gao, G.~Ayers, C.~Kozyrakis, {Practical Near-Data Processing for In-Memory
  Analytics Frameworks}, in: PACT, 2015.

\bibitem{guo2014wondp}
Q.~Guo, N.~Alachiotis, B.~Akin, F.~Sadi, G.~Xu, T.~M. Low, L.~Pileggi, J.~C.
  Hoe, F.~Franchetti, {3D-Stacked Memory-Side Acceleration: Accelerator and
  System Design}, in: WoNDP, 2014.

\bibitem{sura.cf15}
Z.~Sura, A.~Jacob, T.~Chen, B.~Rosenburg, O.~Sallenave, C.~Bertolli, S.~Antao,
  J.~Brunheroto, Y.~Park, K.~O'Brien, R.~Nair, {Data Access Optimization in a
  Processing-in-Memory System}, in: CF, 2015.

\bibitem{morad.taco15}
A.~Morad, L.~Yavits, R.~Ginosar, {GP-SIMD Processing-in-Memory}, ACM TACO
  (2015).

\bibitem{hassan.memsys15}
S.~M. Hassan, S.~Yalamanchili, S.~Mukhopadhyay, {Near Data Processing: Impact
  and Optimization of 3D Memory System Architecture on the Uncore}, in: MEMSYS,
  2015.

\bibitem{li.dac16}
S.~Li, C.~Xu, Q.~Zou, J.~Zhao, Y.~Lu, Y.~Xie, {Pinatubo: A Processing-in-Memory
  Architecture for Bulk Bitwise Operations in Emerging Non-Volatile Memories},
  in: DAC, 2016.

\bibitem{kang.icassp14}
M.~Kang, M.-S. Keel, N.~R. Shanbhag, S.~Eilert, K.~Curewitz, {An
  Energy-Efficient VLSI Architecture for Pattern Recognition via Deep Embedding
  of Computation in SRAM}, in: ICASSP, 2014.

\bibitem{aga.hpca17}
S.~Aga, S.~Jeloka, A.~Subramaniyan, S.~Narayanasamy, D.~Blaauw, R.~Das,
  {Compute Caches}, in: HPCA, 2017.

\bibitem{shafiee2016isaac}
A.~Shafiee, A.~Nag, N.~Muralimanohar, et~al., {ISAAC: A Convolutional Neural
  Network Accelerator with In-situ Analog Arithmetic in Crossbars}, in: ISCA,
  2016.

\bibitem{nai2017graphpim}
L.~Nai, R.~Hadidi, J.~Sim, H.~Kim, P.~Kumar, H.~Kim, {GraphPIM: Enabling
  Instruction-Level PIM Offloading in Graph Computing Frameworks}, in: HPCA,
  2017.

\bibitem{kim.arxiv17}
J.~S. Kim, D.~Senol, H.~Xin, D.~Lee, S.~Ghose, M.~Alser, H.~Hassan, O.~Ergin,
  C.~Alkan, O.~Mutlu, {GRIM-Filter: Fast Seed Filtering in Read Mapping Using
  Emerging Memory Technologies}, arXiv:1708.04329 [q-bio.GN] (2017).

\bibitem{kim.bmc18}
J.~S. Kim, D.~Senol, H.~Xin, D.~Lee, S.~Ghose, M.~Alser, H.~Hassan, O.~Ergin,
  C.~Alkan, O.~Mutlu, {GRIM-Filter: Fast Seed Location Filtering in DNA Read
  Mapping Using Processing-in-Memory Technologies}, BMC Genomics (2018).

\bibitem{li.micro17}
S.~Li, D.~Niu, K.~T. Malladi, H.~Zheng, B.~Brennan, Y.~Xie, {DRISA: A
  DRAM-Based Reconfigurable In-Situ Accelerator}, in: MICRO, 2017.

\bibitem{kim.sc17}
G.~Kim, N.~Chatterjee, M.~O'Connor, K.~Hsieh, {Toward Standardized Near-Data
  Processing with Unrestricted Data Placement for GPUs}, in: SC, 2017.

\bibitem{boroumand.asplos18}
A.~Boroumand, S.~Ghose, Y.~Kim, R.~Ausavarungnirun, E.~Shiu, R.~Thakur, D.~Kim,
  A.~Kuusela, A.~Knies, P.~Ranganathan, O.~Mutlu, {Google Workloads for
  Consumer Devices: Mitigating Data Movement Bottlenecks}, in: ASPLOS, 2018.

\bibitem{fernandez2020natsa}
I.~Fernandez, R.~Quislant, C.~Giannoula, M.~Alser, J.~Gomez-Luna, E.~Gutierrez,
  O.~Plata, O.~Mutlu, {NATSA: A Near-Data Processing Accelerator for Time
  Series Analysis}, in: ICCD, 2020.

\bibitem{singh2019napel}
G.~Singh, J.~Gomez-Luna, G.~Mariani, G.~F. Oliveira, S.~Corda, S.~Stujik,
  O.~Mutlu, H.~Corporaal, {NAPEL: Near-memory Computing Application Performance
  Prediction via Ensemble Learning}, in: DAC, 2019.

\bibitem{rezaei2020nom}
S.~H.~S. {Rezaei}, M.~{Modarressi}, R.~{Ausavarungnirun}, M.~{Sadrosadati},
  O.~{Mutlu}, M.~{Daneshtalab}, {NoM: Network-on-Memory for Inter-Bank Data
  Transfer in Highly-Banked Memories}, CAL (2020).

\bibitem{sisa}
M.~Besta, R.~Kanakagiri, G.~Kwasniewski, R.~Ausavarungnirun, J.~Ber\'{a}nek,
  K.~Kanellopoulos, K.~Janda, Z.~Vonarburg-Shmaria, L.~Gianinazzi, I.~Stefan,
  J.~G. Luna, J.~Golinowski, M.~Copik, L.~Kapp-Schwoerer, S.~Di~Girolamo,
  N.~Blach, M.~Konieczny, O.~Mutlu, T.~Hoefler, {SISA: Set-Centric Instruction
  Set Architecture for Graph Mining on Processing-in-Memory Systems}, in:
  MICRO, 2021.

\bibitem{singh2020nero}
G.~Singh, D.~Diamantopoulos, C.~Hagleitner, J.~Gomez-Luna, S.~Stuijk, O.~Mutlu,
  H.~Corporaal, {NERO: A Near High-Bandwidth Memory Stencil Accelerator for
  Weather Prediction Modeling}, in: FPL, 2020.

\bibitem{wang2020figaro}
Y.~Wang, L.~Orosa, X.~Peng, Y.~Guo, S.~Ghose, M.~Patel, J.~S. Kim, J.~G. Luna,
  M.~Sadrosadati, N.~M. Ghiasi, et~al., {FIGARO: Improving System Performance
  via Fine-Grained In-DRAM Data Relocation and Caching}, in: MICRO, 2020.

\bibitem{synchron}
C.~Giannoula, N.~Vijaykumar, N.~Papadopoulou, V.~Karakostas, I.~Fernandez,
  J.~Gómez-Luna, L.~Orosa, N.~Koziris, G.~Goumas, O.~Mutlu, {SynCron:
  Efficient Synchronization Support for Near-Data-Processing Architectures},
  in: HPCA, 2021.

\bibitem{gomezluna2021benchmarking}
J.~Gómez-Luna, I.~E. Hajj, I.~Fernandez, C.~Giannoula, G.~F. Oliveira,
  O.~Mutlu, Benchmarking a new paradigm: An experimental analysis of a real
  processing-in-memory architecture (2021).
\newblock \href {http://arxiv.org/abs/2105.03814} {\path{arXiv:2105.03814}}.

\bibitem{boroumand-pact21}
A.~Boroumand, S.~Ghose, B.~Akin, R.~Narayanaswami, G.~F. Oliveira, X.~Ma,
  E.~Shiu, O.~Mutlu, {Google Neural Network Models for Edge Devices: Analyzing
  and Mitigating Machine Learning Inference Bottlenecks}, in: PACT, 2021.

\bibitem{damov}
G.~F. Oliveira, J.~Gómez-Luna, L.~Orosa, S.~Ghose, N.~Vijaykumar,
  I.~Fernandez, M.~Sadrosadati, O.~Mutlu, {DAMOV: A New Methodology and
  Benchmark Suite for Evaluating Data Movement Bottlenecks}, IEEE Access
  (2021).

\bibitem{fpga-pim-2021}
G.~Singh, M.~Alser, D.~Cali, D.~Diamantopoulos, J.~Gomez-Luna, H.~Corporaal,
  O.~Mutlu, Fpga-based near-memory acceleration of modern data-intensive
  applications, IEEE Micro (jul 2021) 41~(04) (2021) 39--48.

\bibitem{mutlu-date21}
O.~Mutlu, {Intelligent Architectures for Intelligent Computing Systems}, in:
  DATE, 2021.

\bibitem{fimdram-isscc21}
Y.-C. Kwon, S.~H. Lee, J.~Lee, S.-H. Kwon, J.~M. Ryu, J.-P. Son, O.~Seongil,
  H.-S. Yu, H.~Lee, S.~Y. Kim, Y.~Cho, J.~G. Kim, J.~Choi, H.-S. Shin, J.~Kim,
  B.~Phuah, H.~Kim, M.~J. Song, A.~Choi, D.~Kim, S.~Kim, E.-B. Kim, D.~Wang,
  S.~Kang, Y.~Ro, S.~Seo, J.~Song, J.~Youn, K.~Sohn, N.~S. Kim, {25.4 A 20nm
  6GB Function-In-Memory DRAM, Based on HBM2 with a 1.2TFLOPS Programmable
  Computing Unit Using Bank-Level Parallelism, for Machine Learning
  Applications}, in: ISSCC, 2021.

\bibitem{fimdram-isca21}
S.~Lee, S.-h. Kang, J.~Lee, H.~Kim, E.~Lee, S.~Seo, H.~Yoon, S.~Lee, K.~Lim,
  H.~Shin, J.~Kim, O.~Seongil, A.~Iyer, D.~Wang, K.~Sohn, N.~S. Kim, {Hardware
  Architecture and Software Stack for PIM Based on Commercial DRAM Technology :
  Industrial Product}, in: ISCA, 2021.

\bibitem{upmem-hotchip}
J.~G. Luna, I.~E. Hajj, I.~Fernandez, C.~Giannoula, G.~F. Oliveira, O.~Mutlu,
  {Understanding a Modern Processing-in-Memory Architecture: Benchmarking and
  Experimental Characterization}, in: HotChips Presentation, 2019.

\bibitem{simdram}
N.~Hajinazar, G.~F. Oliveira, S.~Gregorio, J.~a.~D. Ferreira, N.~M. Ghiasi,
  M.~Patel, M.~Alser, S.~Ghose, J.~G\'{o}mez-Luna, O.~Mutlu, {SIMDRAM: A
  Framework for Bit-Serial SIMD Processing Using DRAM}, in: ASPLOS, 2021.

\bibitem{Seshadri:2015:ANDOR}
V.~Seshadri, K.~Hsieh, A.~Boroumand, D.~Lee, M.~A. Kozuch, O.~Mutlu, P.~B.
  Gibbons, T.~C. Mowry, {Fast Bulk Bitwise AND and OR in DRAM}, CAL (2015).

\bibitem{chang.hpca16}
K.~K. Chang, P.~J. Nair, D.~Lee, S.~Ghose, M.~K. Qureshi, O.~Mutlu, {Low-Cost
  Inter-Linked Subarrays (LISA): Enabling Fast Inter-Subarray Data Movement in
  DRAM}, in: HPCA, 2016.

\bibitem{seshadri.micro17}
V.~Seshadri, D.~Lee, T.~Mullins, H.~Hassan, A.~Boroumand, J.~Kim, M.~A. Kozuch,
  O.~Mutlu, P.~B. Gibbons, T.~C. Mowry, {Ambit: In-Memory Accelerator for Bulk
  Bitwise Operations Using Commodity DRAM Technology}, in: MICRO, 2017.

\end{thebibliography}
